%% file: CC_NSI_at_Daya_Bay.tex
\documentclass[a4paper,11pt]{article}
\pdfoutput=1 

\usepackage{jheppub} 

\usepackage[T1]{fontenc} 
\usepackage[latin9]{inputenc}
\usepackage{xcolor}
\usepackage{babel}
\usepackage{booktabs}
\usepackage{mathrsfs}
\usepackage{multirow}
\usepackage{amsmath}
\usepackage{amsthm}
\usepackage{amssymb}
\usepackage{subfigure}
\usepackage{graphicx}

\title{\boldmath Charged-current non-standard neutrino interactions at Daya Bay}

\input{name_JHEP.tex}

\abstract{The full data set of the Daya Bay reactor neutrino experiment is used to probe the effect of the charged current non-standard interactions (CC-NSI) on neutrino oscillation experiments. Two different approaches are applied and constraints on the corresponding CC-NSI parameters are obtained with the neutrino flux taken from the Huber-Mueller model with a $5\%$ uncertainty. For the quantum mechanics-based approach (QM-NSI), the constraints on the CC-NSI parameters $\epsilon_{e\alpha}$ and $\epsilon_{e\alpha}^{s}$ are extracted with and without the assumption that the effects of the new physics are the same in the production and detection processes, respectively. The approach based on the weak effective field theory (WEFT-NSI) deals with four types of CC-NSI represented by the parameters $[\varepsilon_{X}]_{e\alpha}$. For both approaches, the results for the CC-NSI parameters are shown for cases with various fixed values of the CC-NSI and the Dirac CP-violating phases, and when they are allowed to vary freely. We find that constraints on the QM-NSI parameters $\epsilon_{e\alpha}$ and $\epsilon_{e\alpha}^{s}$ from the Daya Bay experiment alone can reach the order $\mathcal{O}(0.01)$ for the former and $\mathcal{O}(0.1)$ for the latter, while for WEFT-NSI parameters $[\varepsilon_{X}]_{e\alpha}$, we obtain $\mathcal{O}(0.1)$ for both cases.}

\begin{document}
\maketitle
\flushbottom

\section{Introduction \label{sec:Introduction}}

Neutrino oscillation has been observed for more than two decades. Most results of the oscillation experiments can be explained with good accuracy in the standard three-flavor neutrino oscillation framework which is parameterized with three mixing angles $\theta_{12}$, $\theta_{23}$ and $\theta_{13}$, one Dirac CP-violating phase $\delta_{\text{CP}}$ and two mass squared differences $\Delta m_{21}^{2}\equiv m_{2}^{2}-m_{1}^{2}$
and $\Delta m_{32}^{2}\equiv m_{3}^{2}-m_{2}^{2}$ (and thus $\Delta m_{31}^{2}=\Delta m_{32}^{2}+\Delta m_{21}^{2}$). Although the values of most of the parameters have been measured at the percent level, the mass ordering ($\left|\Delta m_{31}^{2}\right|=\left|\Delta m_{32}^{2}\right|\pm\left|\Delta m_{21}^{2}\right|$ where the sign + ($-$) is for the normal (inverted) ordering), the value of $\delta_{\text{CP}}$ and the octant of $\theta_{23}$ are still unknown. Together with other undetermined neutrino properties, e.g., the nature of neutrinos (whether Dirac or Majorana), these unknowns about neutrinos are the goals of the current and future neutrino experiments \cite{Workman:2022ynf}.

The phenomena of neutrino oscillations indicate that neutrinos are massive particles, as opposed to the hypothesis of the Standard Model (SM) of particle physics. The source of the neutrino masses is expected to originate from new physics (NP) beyond the SM. The NP not only gives rise to the neutrino masses and mixing but may also modify neutrino interactions. In the case that the scale of the NP is much larger than the typical energy scale of the experiment of interest, the effect of the NP can be approximated by an effective four-fermion Lagrangian \citep{Proceedings:2019qno}. Such new interactions  are referred to as the non-standard interactions (NSI) \citep{PhysRevD.17.2369,Guzzo:1991hi,Biggio:2009nt,ANTUSCH2009369,Ohlsson:2012kf,Miranda:2015dra,Farzan:2017xzy,Esteban:2018ppq}. NSI involving neutrinos can have charged current (CC) and neutral current (NC) types and can be written as 
\begin{align}
\mathscr{L}_{\textrm{CC-NSI}} & =-2\sqrt{2}G_{F}\sum_{f,f',\alpha,\beta,P}\epsilon_{\alpha\beta}^{f,f',P}[\bar{\nu}_{\beta}\gamma^{\mu}P_{L}l_{\alpha}][\bar{f}\gamma_{\mu}Pf']+h.c.,\label{eq:CC-NSI}\\
\mathscr{L}_{\textrm{NC-NSI}} & =-2\sqrt{2}G_{F}\sum_{f,\alpha,\beta,P}\epsilon_{\alpha\beta}^{f,P}[\bar{\nu}_{\alpha}\gamma^{\mu}P_{L}\nu_{\beta}][\bar{f}\gamma_{\mu}Pf],
\label{eq:NC-NSI}
\end{align}
where the lepton flavor index $\alpha,\beta=e,\mu,\tau$, the fermions
$f\neq f'=u,d$ for CC-NSI and $f=e,u,d$ for NC-NSI. The chirality
projection operator $P$ can take on the values of either $P_{L}=(1-\gamma^{5})/2$
or $P_{R}=(1+\gamma^{5})/2$. The dimensionless parameters $\epsilon_{\alpha\beta}^{f,f',P}$
and $\epsilon_{\alpha\beta}^{f,P}$ quantify the relative strength
of the neutrino NSI with respect to the SM Fermi constant $G_{F}$.
In general, both the CC and NC NSI parameters $\epsilon_{\alpha\beta}^{f,f',P}$
and $\epsilon_{\alpha\beta}^{f,P}$ are complex parameters.
It is expected that the size of each NSI parameter is of order $\left|\epsilon\right|\sim g_{\textrm{NP}}^{2}M_W^{2}/M_{\textrm{NP}}^{2}$
\citep{Kopp:2007ne,Proceedings:2019qno} where $M_W$, $g_{\textrm{NP}}$ and $M_{\textrm{NP}}$
are the $W$ boson mass, the coupling constant and the mass of the new mediator, respectively.
The existence of non-vanishing CC-NSI parameters $\epsilon_{\alpha\beta}^{f,f',P}$
for $\alpha\neq\beta$ indicates violation of the lepton flavor number conservation, and $\epsilon_{\alpha\alpha}^{f,f',P}\neq\epsilon_{\beta\beta}^{f,f',P}$ violation of lepton flavor universality. In the case that $\epsilon_{\alpha\beta}^{f,f',P}=0$, SM CC weak interactions are recovered. Note that the total lepton number is conserved in both the NSI described by eqs. (\ref{eq:CC-NSI}) and (\ref{eq:NC-NSI}) and SM at classical level. In the presence of CC-NSI, the production and detection processes of neutrinos would be modified. The NC-NSI could also affect neutrino propagation in matter. Both CC-NSI and NC-NSI can thus be probed in experiments involving the measurement of the Fermi constant $G_{F}$, the unitarity of the Cabibbo-Kobayash-Maskawa (CKM) matrix, and pion-related decay rates, among many others \citep{Davidson:2003ha,Biggio:2009nt,Falkowski:2019xoe}. These precision experiments could constrain $\left|\epsilon\right|$ or Re($\epsilon$) to $\mathcal{O}(10^{-6})$ under different assumptions. Of course, both CC-NSI and NC-NSI may also manifest themselves in
neutrino oscillation experiments and give rise to effective mixing angles and mass squared differences \citep{Fornengo:2001pm,Li:2014mlo,Liao:2016orc,Liao:2017awz,ANTARES:2021crm,IceCube:2022ubv}. In this paper, we use the full data set of the Daya Bay experiment to probe the effects of CC-NSI with two different approaches.\footnote{The effect of NC-NSI on neutrino propagation in matter can be ignored and only CC-NSI are relevant for short baseline reactor neutrino oscillation experiments \citep{Leitner:2011aa,Girardi:2014gna}.}  We assume that the effects of NSI are subdominant and the shifts between the standard and the effective oscillation parameters except $\theta_{13}$ are small.

The rest of the paper is organized as follows: in section \ref{sec:Two-approaches},
the two approaches to formulate CC-NSI in neutrino oscillation
experiments and their corresponding CC-NSI parameters are introduced.
Section \ref{sec:DayaBay-experiment} gives a brief description of
the Daya Bay reactor neutrino experiment. The constraints on CC-NSI
parameters extracted from the Daya Bay experiment are shown in section
\ref{sec:Constraints-on-NSI}. We summarize and conclude in section
\ref{sec:Summary}.

\section{Two approaches to CC-NSI \label{sec:Two-approaches}}

There are two approaches to describe CC-NSI in neutrino oscillation experiments. One approach is based on the ordinary quantum mechanics (QM), and referred to as QM-NSI \citep{Du:2020dwr, Falkowski:2019kfn}. The second approach deals with CC-NSI under the framework of the weak effective field theory (WEFT) \citep{Falkowski:2019xoe}, and is denoted as WEFT-NSI in this paper. 

\subsection{Neutrino transition probability in the standard case}

In the standard three-flavor neutrino oscillation framework, the survival probability of the electron antineutrinos with energy $E_\nu$ propagating in vacuum over a distance $L_\nu$ is 
\begin{align}
P_{\bar{\nu}_{e}\rightarrow\bar{\nu}_{e}}^{\textrm{std}} & =\sum^3_{j,k}\left|U_{ej}\right|^{2}\left|U_{ek}\right|^{2}\exp\left(-\frac{\Delta m_{jk}^{2}L_\nu}{2E_\nu}\right)\nonumber \\
 & =1-\sin^{2}(2\theta_{13})\left[\cos^{2}\theta_{12}\sin^{2}\left(\frac{\Delta m_{31}^{2}L_\nu}{4E_\nu}\right)\right.+\left.\sin^{2}\theta_{12}\sin^{2}\left(\frac{\Delta m_{32}^{2}L_\nu}{4E_\nu}\right)\right]\nonumber \\
 &\quad -\cos^{4}\theta_{13}\sin^{2}(2\theta_{12})\sin^{2}\left(\frac{\Delta m_{21}^{2}L_\nu}{4E_\nu}\right),\label{eq:Pee_std}
\end{align}
under the plane-wave approximation. The Pontecorvo-Maki-Nakagawa-Sakata (PMNS) lepton mixing matrix $U$
\citep{Pontecorvo:1957cp,Pontecorvo:1957qd,Maki:1960ut,Maki:1962mu,Pontecorvo:1967fh} relates the neutrino fields in the flavor basis to the mass basis and $UU^{\dagger}=I$ is assumed. The neutrino mixing parameters $\theta_{12}$, $\theta_{13}$ and the mass squared differences $\Delta m_{21}^{2}$ and $\Delta m_{32}^{2}$ are involved in eq.\,(\ref{eq:Pee_std}), while the mixing parameter $\theta_{23}$ and the Dirac CP-violating phase $\delta_{\text{CP}}$ are not relevant. With NSI being present, however, the dependence on $\theta_{23}$ and $\delta_{\text{CP}}$ emerges in general, as can be seen below. 

We note that the survival probability of eq.\,(\ref{eq:Pee_std}) is insensitive to the mass ordering for Daya Bay experiment, since the difference in the survival probability
of the two orderings is small (of order $\sin^{2}(2\theta_{13})\cos^{2}\theta_{12}$ $\sin(\Delta m_{32}^{2}L_{\nu}/2E_{\nu})$) in this case. When the effects of CC-NSI are included, the difference depends on the CC-NSI parameters also. The survival probability remains insensitive to the mass ordering, if the CC-NSI parameters are smaller than unity. In the following, we probe the constraints of the Daya Bay experiment on CC-NSI assuming the normal mass ordering. We have checked that the results are similar for the case of the inverted mass ordering. 

We also note that eq.\,(\ref{eq:Pee_std}) is dominated by the first two terms with the third term, depending on $\Delta m_{21}^2 L_\nu/4E_\nu$, negaligible for Daya Bay experiment. This leads to an approximate symmetry of the survival probability, i.e., $P_{\bar{\nu}_{e}\rightarrow\bar{\nu}_{e}}^{\textrm{std}}$ is invariant under the exchange of $\theta_{13} \leftrightarrow \pi/2-\theta_{13}$, which may still be a good symmetry when CC-NSI are present.

\subsection{QM-NSI with parameters $\epsilon^{s}$ and $\epsilon^{d}$ at production and detection
\label{subsec:QM-NSI-formulas}}

Under the framework of QM-NSI, the interaction eigenstate $\left|\nu_{\alpha}^{s/d}\right\rangle $
(where $s/d$ represents source or detection) with the presence of NSI is assumed
to be in a superposition of the SM weak eigenstates $\left|\nu_{\alpha}\right\rangle $
with $\alpha=e,\mu,\tau$ \citep{Grossman:1995wx,Gonzalez-Garcia:2001snt,Ohlsson:2008gx,Meloni:2009cg,Leitner:2011aa,Ohlsson:2013nna,Agarwalla:2014bsa},
i.e., 
\begin{equation}
\left|\nu_{\alpha}^{s}\right\rangle =\frac{1}{N_{\alpha}^{s}}\left(\left|\nu_{\alpha}\right\rangle +\sum_{\gamma}\epsilon_{\alpha\gamma}^{s}\left|\nu_{\gamma}\right\rangle \right),
\label{eq:QM-NSI-epsilon-s}
\end{equation}
and 
\begin{equation}
\left\langle \nu_{\beta}^{d}\right|=\frac{1}{N_{\beta}^{d}}\left(\left\langle \nu_{\beta}\right|+\sum_{\gamma}\epsilon_{\gamma\beta}^{d}\left\langle \nu_{\gamma}\right|\right),
\label{eq:QM-NSI-epsilon-d}
\end{equation}
such that $\left|\nu_{\beta}^{d}\right\rangle =\left(\left|\nu_{\beta}\right\rangle +\sum_{\gamma}\epsilon_{\beta\gamma}^{d\dagger}\left|\nu_{\gamma}\right\rangle \right)/N_{\beta}^{d}$, where $N_{\alpha}^{s}=\sqrt{[(I+\epsilon^{s})(I+\epsilon^{s\dagger})]_{\alpha\alpha}}$
and $N_{\beta}^{d}=\sqrt{[(I+\epsilon^{d\dagger})(I+\epsilon^{d})]_{\beta\beta}}$
are the normalization factors. Note these states are not orthogonal \citep{Langacker:1988up}, similar to the case of the
non-unitary mixing matrix \citep{Antusch:2006vwa}. The NSI parameters $\epsilon^{s/d}$ defined here are the effective coefficients which are different from those defined at the Lagrangian level in eq. (\ref{eq:CC-NSI}). We distinguish the coefficients $\epsilon^{s}$ and $\epsilon^{d}$ since the effect of NSI at the source and detector may be different. In matrix form, we can write
\begin{align}
[\left|\nu_{\alpha}^{s}\right\rangle ] & =(N^{s})^{-1}(I+\epsilon^{s})[\left|\nu_{\gamma}\right\rangle ],\\{}
[\left|\nu_{\beta}^{d}\right\rangle ] & =(N^{d})^{-1}(I+\epsilon^{d\dagger})[\left|\nu_{\gamma}\right\rangle ],
\end{align}
where $[\left|\nu_{\alpha}^{s/d}\right\rangle ]=(\left|\nu_{e}^{s/d}\right\rangle ,\left|\nu_{\mu}^{s/d}\right\rangle ,\left|\nu_{\tau}^{s/d}\right\rangle )^{T}$,
\begin{equation}
N^{s/d}=\begin{pmatrix}N_{e}^{s/d} & 0 & 0\\
0 & N_{\mu}^{s/d} & 0\\
0 & 0 & N_{\tau}^{s/d}
\end{pmatrix},
\end{equation}
and 
\begin{equation}
\epsilon^{s/d}=\begin{pmatrix}\epsilon_{ee}^{s/d} & \epsilon_{e\mu}^{s/d} & \epsilon_{e\tau}^{s/d}\\
\epsilon_{\mu e}^{s/d} & \epsilon_{\mu\mu}^{s/d} & \epsilon_{\mu\tau}^{s/d}\\
\epsilon_{\tau e}^{s/d} & \epsilon_{\tau\mu}^{s/d} & \epsilon_{\tau\tau}^{s/d}
\end{pmatrix}.
\label{eq:epsilon_s_d}
\end{equation}
The matrix of the normalization factors is factored out for convenience. Connecting to mass basis, we can define
\begin{equation}
U^{s}\equiv(I+\epsilon^{s*})U,\ \textrm{and}\ U^{d}\equiv(I+{\epsilon^d}^T)U.
\end{equation}
We note that the transformation matrix $(N^{s})^{-1}U^{s}$ or $(N^{d})^{-1}U^{d}$
becomes non-unitary, in contrast to the standard PMNS matrix $U$. With NSI, the survival probability of the electron antineutrinos becomes
\begin{align}
P_{\bar{\nu}_{e}^{s}\rightarrow\bar{\nu}_{e}^{d}}^{\textrm{QM-NSI}} & =\frac{1}{\left|N_{e}^{s}\right|^{2}\left|N_{e}^{d}\right|^{2}}\sum_{j,k}U_{ej}^{s}U_{ek}^{s*}U_{ej}^{d*}U_{ek}^{d}\exp\left(-i\frac{\Delta m_{jk}^{2}L_\nu}{2E_\nu}\right),\label{eq:QM-NSI-s_d}
\end{align}
where $U_{ej}^{s}=\sum_{\alpha}(\delta_{e\alpha}+\epsilon_{e\alpha}^{s*})U_{\alpha j}$
and $U_{ej}^{d}=\sum_{\alpha}(\delta_{e\alpha}+\epsilon_{\alpha e}^{d})U_{\alpha j}$.
Among the eighteen complex parameters $\epsilon_{\alpha\beta}^{s}$ and $\epsilon_{\alpha\beta}^{d}$ of eq.\,(\ref{eq:epsilon_s_d}), only the six associated
with electrons, i.e., $\epsilon_{e\alpha}^{s}$ and $\epsilon_{\alpha e}^{d}$,
are involved in this expression. In our analysis below, we decompose each complex NSI parameter into its absolute value and phase as
\begin{equation}
\epsilon_{\alpha\beta}^{s}=\left|\epsilon_{\alpha\beta}^{s}\right|e^{i\phi_{\alpha\beta}^{s}}\ \textrm{and}\ \epsilon_{\alpha\beta}^{d}=\left|\epsilon_{\alpha\beta}^{d}\right|e^{i\phi_{\alpha\beta}^{d}}.
\end{equation}
The neutrino fluxes and cross sections are needed to determine the rate of inverse beta-decay (IBD) events at the detector. With
the presence of CC-NSI, they are modified by ${\color{purple}{\color{green}{\normalcolor \Phi_{\bar{\nu}_{e}^{s}}(E,\epsilon^{s})}}}$$=$$\left|N_{e}^{s}\right|^{2}$
$\Phi_{\bar{\nu}_{e}}(E_\nu)$ and ${\color{purple}{\color{green}{\normalcolor \sigma_{\bar{\nu}_{e}^{d}}(E_\nu,\epsilon^{d})=\left|N_{e}^{d}\right|^{2}\sigma_{\bar{\nu}_{e}}(E_\nu)}}}$
\citep{Antusch:2006vwa} where $\Phi_{\bar{\nu}_{e}}(E_\nu)$ and $\sigma_{\bar{\nu}_{e}}(E_\nu)$
are the neutrino fluxes and cross sections in the SM, respectively, while $\Phi_{\bar{\nu}_{e}^{s}}(E_\nu,\epsilon^{s})$
and $\sigma_{\bar{\nu}_{e}^{d}}(E_\nu,\epsilon^{d})$ denote the corresponding
quantities with the presence of CC-NSI. We can define an effective survival probability through the detected number of IBD events in the detector: 
\begin{align}
N & \propto \int dE_\nu\frac{d\Phi_{\bar{\nu}_{e}^{s}}(E_\nu,\epsilon^{s})}{dE_\nu}{{\normalcolor P_{\bar{\nu}_{e}^{s}\rightarrow\bar{\nu}_{e}^{d}}^{\textrm{QM-NSI}}(E_\nu,L_\nu,\epsilon^{s},\epsilon^{d}})}\sigma_{\bar{\nu}_{e}^{d}}(E_\nu,\epsilon^{d})\nonumber\\
 & =\int dE_\nu\frac{d\Phi_{\bar{\nu}_{e}}(E_\nu)}{dE_\nu}{\color{magenta}{\normalcolor P_{\bar{\nu}_{e}^{s}\rightarrow\bar{\nu}_{e}^{d}}^{\textrm{QM-NSI-eff}}(E_\nu,L_\nu,\epsilon^{s},\epsilon^{d})}}\sigma_{\bar{\nu}_{e}}(E_\nu),
\end{align}
where
\begin{equation}
P_{\bar{\nu}_{e}^{s}\rightarrow\bar{\nu}_{e}^{d}}^{\textrm{QM-NSI-eff}}=\sum_{j,k}U_{ej}^{s}U_{ek}^{s*}U_{ej}^{d*}U_{ek}^{d}\exp\left(-i\frac{\Delta m_{jk}^{2}L_\nu}{2E_\nu}\right).\label{eq:QM-NSI-s_d_eff}
\end{equation}
We can see that the normalization factor $1/\left|N_{e}^{s}\right|^{2}\left|N_{e}^{d}\right|^{2}$
is cancelled out compared to eq.\,(\ref{eq:QM-NSI-s_d}). 

At reactor neutrino oscillation experiments, we can assume 
$\epsilon_{e\alpha}^{s}=\epsilon_{\alpha e}^{d*}$ since the primary
source of NSI is of the $V\pm A$ type \citep{Kopp:2007ne}. We consider
this special case first then extend our discussion to the general
case. With the assumption $\epsilon_{e\alpha}^{s}=\epsilon_{\alpha e}^{d*}\equiv\epsilon_{e\alpha}$
or $U_{ej}^{s}=U_{ej}^{d}\equiv U_{ej}^{sd}$, we have 
\begin{equation}
P_{\bar{\nu}_{e}^{s}\rightarrow\bar{\nu}_{e}^{d}}^{\textrm{QM-NSI-eff}}=\sum_{j,k}\left|U_{ej}^{sd}\right|^{2}\left|U_{ek}^{sd}\right|^{2}\exp\left(-i\frac{\Delta m_{jk}^{2}L_\nu}{2E_\nu}\right),
\label{eq:QM-NSI-s=d}
\end{equation}
where $U_{ej}^{sd}=\sum_{\alpha}(\delta_{e\alpha}+\epsilon_{e\alpha}^{*})U_{\alpha j}$.
The number of free complex parameters is reduced to three,
i.e., $\epsilon_{e\alpha}$ for $\alpha=e,\mu$ and $\tau$. We accordingly
use the decomposition $\epsilon_{e\alpha}=\left|\epsilon_{e\alpha}\right|e^{i\phi_{e\alpha}}$.
The analytical expressions eq.\,(\ref{eq:QM-NSI-s_d_eff}) and eq.\,(\ref{eq:QM-NSI-s=d}) will be used in the fit to experimental data. 

For the general case, $\epsilon_{e\alpha}^{s} \neq \epsilon_{\alpha e}^{d*}$ \citep{Leitner:2011aa}, we discuss the effects of $\epsilon_{e\alpha}^{s}$
and $\epsilon_{\alpha e}^{d}$ separately. The effective survival probability for these two cases $P_{\bar{\nu}_{e}^{s}\rightarrow\bar{\nu}_{e}^{d}}^{\textrm{QM-NSI-eff}}(\epsilon_{e\alpha}^{s},\epsilon_{\alpha e}^{d}=0)$ and $P_{\bar{\nu}_{e}^{s}\rightarrow\bar{\nu}_{e}^{d}}^{\textrm{QM-NSI-eff}}(\epsilon_{e\alpha}^{s}=0,\epsilon_{\alpha e}^{d})$ for CC-NSI present only in the antineutrino production and detection processes, respectively, are connected by 
\begin{equation}
P_{\bar{\nu}_{e}^{s}\rightarrow\bar{\nu}_{e}^{d}}^{\textrm{QM-NSI-eff}}(\epsilon_{e\alpha}^{s}=0,\epsilon_{\alpha e}^{d})\leftrightarrow P_{\bar{\nu}_{e}^{s}\rightarrow\bar{\nu}_{e}^{d}}^{\textrm{QM-NSI-eff}}(\epsilon_{e\alpha}^{s},\epsilon_{\alpha e}^{d}=0),
\end{equation}
under the transformation of $U_{ej}^{d}\leftrightarrow U_{ej}^{s*}$ and $U\leftrightarrow U^{*}$ or
\begin{equation}
\epsilon_{\alpha e}^{d}\leftrightarrow\epsilon_{e\alpha}^{s}\ \textrm{and}\ \delta_{\text{CP}}\leftrightarrow\pi-\delta_{\text{CP}}.\label{eq:QM-NSI-s-d-symmetry}
\end{equation}
We examine the effect of $\epsilon_{e\alpha}^{s}$ first. The constraints
on $\epsilon_{\alpha e}^{d}$ can be deduced from those on $\epsilon_{e\alpha}^{s}$ by this transformation. 

\begin{figure}
\begin{centering}
\includegraphics[width=8.5cm]{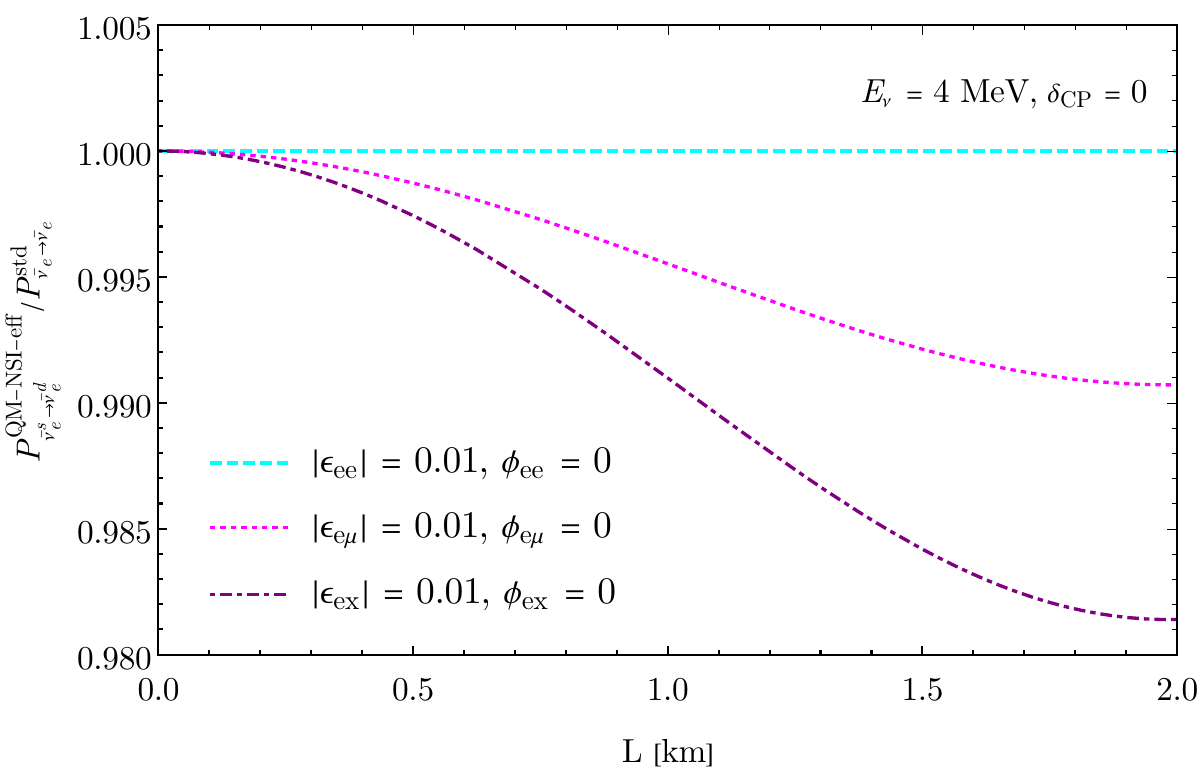}
\par\end{centering}
\caption{The ratio of the effective survival probability with NSI to the standard survival probability as a function of the distance shows the effect of QM-NSI on the shape of the survival probability. The different curves shown are for the effect of the respective QM-NSI parameter $\epsilon_{ee}$, $\epsilon_{e\mu}$ and $\epsilon_{ex}$, with their magnitudes equaling 0.01 and phases equaling zero. For the cases of $\epsilon_{ee}$ and $\epsilon_{e\mu}$, other NSI parameters are set to be zero. Each ratio curve is shifted by the amount $1-P_{\bar{\nu}_{e}^{s}\rightarrow\bar{\nu}_{e}^{d}}^{\textrm{QM-NSI-eff}}(L_\nu=0)$ to remove the zero-distance effect on the curve. $E_{\nu}=4$ MeV, $\sin^{2}\theta_{13}=0.022$ and $\delta_{\text{CP}}=0$ are used for this figure. Values of other oscillation parameters are listed in Table \ref{tab:Input-Parameters}. More details can be found in the text of section \ref{subsec:QM-NSI-formulas}.
\label{fig:QM-NSI-effect-shape}}
\end{figure}

For the presence of NSI, the so-called zero-distance effect $P_{\bar{\nu}_{e}^{s}\rightarrow\bar{\nu}_{e}^{d}}^{\textrm{QM-NSI-eff}}(L_\nu=0)\neq1$
\citep{Langacker:1988up,Kopp:2007ne} occurs. Explicitly, we have
\begin{gather}
P_{\bar{\nu}_{e}^{s}\rightarrow\bar{\nu}_{e}^{d}}^{\textrm{QM-NSI-eff}}(L_\nu=0)
=\begin{cases}
\left(1+2\left|\epsilon_{ee}\right|\cos\phi_{ee}+\left|\epsilon_{ee}\right|^{2}+\left|\epsilon_{e\mu}\right|^{2}+\left|\epsilon_{e\tau}\right|^{2}\right)^{2}, \ 
\textrm{when}\ \epsilon_{e\alpha}^{s}=\epsilon_{\alpha e}^{d*}\equiv\epsilon_{e\alpha};\\
1+2\left|\epsilon_{ee}^{s/d}\right|\cos\phi_{ee}^{s/d}+\left|\epsilon_{ee}^{s/d}\right|^{2}, \ 
\textrm{when}\ \epsilon_{e\alpha}^{s}\neq\epsilon_{\alpha e}^{d*}\ \textrm{and}\ \epsilon_{\alpha e}^{d}=0\ (\textrm{or}\ \epsilon_{e\alpha}^{s}=0).
\end{cases}
\label{eq:zero-distance-QM-NSI}
\end{gather}
To illustrate the effect of QM-NSI on the shape of the survival probability, we first calculate the ratio of the effective survival probability with NSI to the survival probability of the standard case as a function of the distance, i.e., $P_{\bar{\nu}_{e}^{s}\rightarrow\bar{\nu}_{e}^{d}}^{\textrm{QM-NSI-eff}}/P_{\bar{\nu}_{e}\rightarrow\bar{\nu}_{e}}^{\textrm{std}}$. The ratio is not unity at $L_\nu = 0$ because of the zero-distance effect. We then remove the zero-distance effect by shifting the ratio by the amount $1-P_{\bar{\nu}_{e}^{s}\rightarrow\bar{\nu}_{e}^{d}}^{\textrm{QM-NSI-eff}}(L_\nu=0)$. An illustration of the ratio curves are shown in figure \ref{fig:QM-NSI-effect-shape} for a typical choice of the parameter values of $E_{\nu}=4$ MeV, $\sin^{2}\theta_{13}=0.022$
and $\delta_{\text{CP}}=0$ with values of other oscillation parameters listed in Table \ref{tab:Input-Parameters}. The values of the QM-NSI parameters are chosen to be $\left|\epsilon_{e\alpha}\right|=0.01$ and $\phi_{e\alpha}=0$ for $\alpha=e,\mu,x$ (where $\epsilon_{ex}\equiv\epsilon_{ee}=\epsilon_{e\mu}=\epsilon_{e\tau}$). When the zero-distance effect is removed, the effective survival probability with $\epsilon_{ee}$ non-zero coincides with the standard survival probability and produces a ratio of unity. With the choice of the parameter values here, the presence of non-zero $\left|\epsilon_{e\mu}\right|$ or $\left|\epsilon_{ex}\right|$
reduces the survival probability, a role similar to an increased $\sin\theta_{13}$
in the standard case. We thus expect an anti-correlation between these QM-NSI parameters and $\sin\theta_{13}$ in these cases and
indeed these relationships are manifest in our results below.
\begin{table}
\centering{}
\begin{tabular}{|c|c|c|} 
\hline 
Parameters & Central value$\pm1\sigma$ & Origin\tabularnewline
\hline 
$\sin^{2}2\theta_{12}$ & 0.851$\pm$0.020 & PDG\cite{Workman:2022ynf}\tabularnewline
$\sin^{2}\theta_{23}$ & 0.546$\pm$0.021 & PDG\tabularnewline
$\Delta m_{21}^{2}$ {[}$10^{-5}$ eV$^{2}${]} & 7.53$\pm$0.18 & PDG\tabularnewline
$\Delta m_{32}^{2}$ {[}$10^{-3}$ eV$^{2}${]} & 2.45$\pm$0.07 & T2K\citep{T2K:2019bcf}\tabularnewline
\hline 
\end{tabular}
\caption{Values of standard oscillation parameters in the case of the normal
mass ordering. 
\label{tab:Input-Parameters}}
\end{table}

\subsection{WEFT-NSI with parameters $\varepsilon_X$
\label{subsec:WEFT-NSI-formulas} }

From the perspective of the effective field theory (EFT), the new physics at a high
scale $\Lambda_{\textrm{NP}}$ demonstrates their effects at a low scale by adding a series of higher dimensional operators $O_{i}^{(d)}$ (with dimension $d$), which are suppressed by powers of the scale $\Lambda_{\textrm{NP}}$, to the SM Lagrangian. An example of the EFT is the Standard Model effective field theory (SMEFT) which reads 
\begin{equation}
\mathscr{L}_{\textrm{SMEFT}}=\mathscr{L}_{\textrm{SM}}+\frac{1}{\Lambda_{\textrm{NP}}}\sum_{i=1}^{n_{5}}c_{i}^{(5)}O_{i}^{(5)}+\frac{1}{{\Lambda}_{\textrm{NP}}^{2}}\sum_{i=1}^{n_{6}}c_{i}^{(6)} O_{i}^{(6)}+\mathcal{O}(\frac{1}{\Lambda_{\textrm{NP}}^{3}})
\end{equation}
for the scale being above the weak scale. The higher dimensional operators $O_{i}^{(d)}$ consist of SM fields only and the Lagrangian respects the SM gauge symmetries and/or baryon/lepton number conservation \citep{Buchmuller:1985jz,Grzadkowski:2010es}. The dimensionless Wilson coefficients ${\color{olive}{\color{purple}{\normalcolor c}}}_{i}^{(d)}$ \citep{Grzadkowski:2010es} can be experimentally determined. The dimension-5 operators are responsible for the neutrino mass generation and mixing. Their effects on neutrino production and detection amplitudes can be ignored. Among the dimension-6 operators, there are four-fermion
operators involving neutrinos which correspond to the neutrino NSI.
The effect of the higher dimensional operators are suppressed by higher
powers of $\Lambda_{\textrm{NP}}$ and are ignored here. Analysis on CC-NSI based on the SMEFT and the combination
of the reactor neutrino experiments can be found in Refs.\, \citep{Falkowski:2019xoe,Du:2020dwr}. Global analysis including solar neutrino experiment can also be found, see e.g. ref.\,\citep{Chaves:2021kxe}. Since the reactor neutrino oscillation experiments are carried out
at much lower scales, new physics with scales lower than the weak
scale may also affect such experiments. The neutrino NSI in this case are better
defined in the so called weak effective field theory (WEFT) which
is an EFT with the heavy particles $W^{\pm}$, $Z^{0}$, the Higgs
boson, the top quark and the possible new heavy particles at a scale less than $M_{W}$ integrated out. The effective Lagrangian then takes the form \citep{Falkowski:2019xoe} 
\begin{align}
\mathscr{L}_{\textrm{WEFT}} & \supset-\frac{2V_{ud}}{v^{2}}
 \left\{(1+\epsilon_{L})_{\alpha\beta}(\overline{u}\gamma^{\mu}P_{L}d)(\overline{l}_{\alpha}\gamma_{\mu}P_{L}\nu_{\beta})+[\epsilon_{R}]_{\alpha\beta}(\overline{u}\gamma^{\mu}P_{R}d)(\overline{l}_{\alpha}\gamma_{\mu}P_{L}\nu_{\beta})\right.\nonumber \\
 & \quad+\frac{1}{2}[\epsilon_{S}]_{\alpha\beta}(\overline{u}d)(\overline{l}_{\alpha}P_{L}\nu_{\beta})-\frac{1}{2}[\epsilon_{P}]_{\alpha\beta}(\overline{u}\gamma_{5}d)(\overline{l}_{\alpha}P_{L}\nu_{\beta})\nonumber \\
 &\quad+\left.\frac{1}{4}[\epsilon_{T}]_{\alpha\beta}(\overline{u}\sigma^{\mu\nu}P_{L}d)(\overline{l}_{\alpha}\sigma_{\mu\nu}P_{L}\nu_{\beta})+h.c.\right\}.
\label{eq:Lagrangian_WEFT}
\end{align}
The fields $u$, $d$ and $l_{\alpha}$ are in their mass basis, while
the left-handed neutrino fields $\nu_{\beta}$ are in the flavor basis.
The quantities $V_{ud}$ and $v$ are the CKM matrix element and the
vacuum expectation value of the Higgs field, respectively. In addition to the SM-like
V-A type interactions $(1+\varepsilon_{L})$, the right-handed $(\varepsilon_{R})$,
scalar $(\varepsilon_{S})$, pseudoscalar $(\varepsilon_{P})$, and tensor
$(\varepsilon_{T})$ type CC interactions between leptons and quarks
are all present. This Lagrangian can thus be seen as a generalization
of eq.\,(\ref{eq:CC-NSI}). Note the NSI parameters
$\varepsilon_{L}$, $\varepsilon_{R}$, $\varepsilon_{S}$, $\varepsilon_{P}$,
and $\varepsilon_{T}$ are $3\times3$ matrices in the lepton flavor space. The analytical expression for the
transition probability $P_{\nu_{\alpha}\rightarrow\nu_{\beta}}^{\textrm{WEFT-NSI}}$ was derived in the framework of quantum field theory in ref.\,\citep{Falkowski:2019kfn}. The $\bar{\nu}_{e}\rightarrow\bar{\nu}_{e}$
survival probability can be written as 
\begin{align}
P_{\bar{\nu}_{e}\rightarrow\bar{\nu}_{e}}^{\textrm{WEFT-NSI}} & =N_{ee}^{-1}\sum_{k,l}\exp(-i\frac{\Delta m_{kl}^{2}L}{2E})\nonumber \\
 & \quad\times\left[U_{ek}U_{el}^{*}+\sum_{X}p_{XL}(\epsilon_{X}U)_{ek}U_{el}^{*}+\sum_{X}p_{XL}U_{ek}(\epsilon_{X}U)_{el}^{*}+\sum_{X,Y}p_{XY}(\epsilon_{X}U)_{ek}(\epsilon_{Y}U)_{el}^{*}\right]\nonumber \\
 & \quad\times\left[U_{ek}^{*}U_{el}+\sum_{X}d_{XL}(\epsilon_{X}U)_{ek}^{*}U_{el}+\sum_{X}d_{XL}U_{ek}^{*}(\epsilon_{X}U)_{el}+\sum_{X,Y}d_{XY}(\epsilon_{X}U)_{ek}^{*}(\epsilon_{Y}U)_{el}\right],
\label{eq:prob_EFT_complete}
\end{align}
where 
\begin{align}
N_{ee} & =\left[1+\sum_{X}p_{XL}\varepsilon_{X}+\sum_{X}p_{XL}\varepsilon_{X}^{*}+\sum_{X,Y}p_{XY}\epsilon_{Y}^{*}\varepsilon_{X}^{T}\right]_{ee}\nonumber \\
 & \quad\left[1+\sum_{X}d_{XL}\varepsilon_{X}^{*}+\sum_{X}d_{XL}\varepsilon_{X}+\sum_{X,Y}d_{XY}\varepsilon_{X}^{*}\epsilon_{Y}^{T}\right]_{ee},
\label{eq:prob_EFT_complete_norm}
\end{align}
and $X,Y=L,R,S,T$ with the dependence on $\varepsilon_{P}$ suppressed. The production (detection) coefficient $p_{XY}$ ($d_{XY}$) depends on the neutrino production (detection) amplitude and their values can be found in ref.\,\citep{Falkowski:2019kfn} for nuclear beta decay and inverse beta decay. The flavor diagonal Wilson coefficients $[\varepsilon_{X}]_{ee}$ have no effect on the survival probability, i.e., 
\begin{align}
P_{\bar{\nu}_{e}\rightarrow\bar{\nu}_{e}}^{\textrm{WEFT-NSI}}([\varepsilon_{X}]_{ee}\ \textrm{only}) & =\sum_{k,l}\left|U_{ek}\right|^{2}\left|U_{el}\right|^{2}\exp(-i\frac{\Delta m_{kl}^{2}L_\nu}{2E_\nu}),
\end{align}
which is just the standard expression of eq.\,(\ref{eq:Pee_std}). As to their effects on neutrino production and detection in reactor oscillation experiments, the effect of the coefficients $[\varepsilon_{L}]_{ee}$ and $[\varepsilon_{R}]_{ee}$ is completely absorbed
into the phenomenological values of $V_{ud}$ and $g_{A}$ which are
used to determine the event rate. The effects of the scalar and
tensor coefficients $[\varepsilon_{S}]_{ee}$ and $[\varepsilon_{T}]_{ee}$
are highly suppressed since these couplings are stringently bounded
by nuclear beta decays and their effects can be ignored in reactor
oscillation experiments. The flavor nondiagonal coefficients $[\varepsilon_{X}]_{e\alpha}$
with $\alpha\neq e$ have no effect on the neutrino production rate and detection cross section \citep{Gonzalez-Alonso:2018omy,Falkowski:2019xoe} and only manifest their effects through the survival probability. We thus
use $P_{\bar{\nu}_{e}\rightarrow\bar{\nu}_{e}}^{\textrm{WEFT-NSI}}$ as the effective survival probability.

\begin{figure}
\begin{centering}
\includegraphics[width=8.5cm]{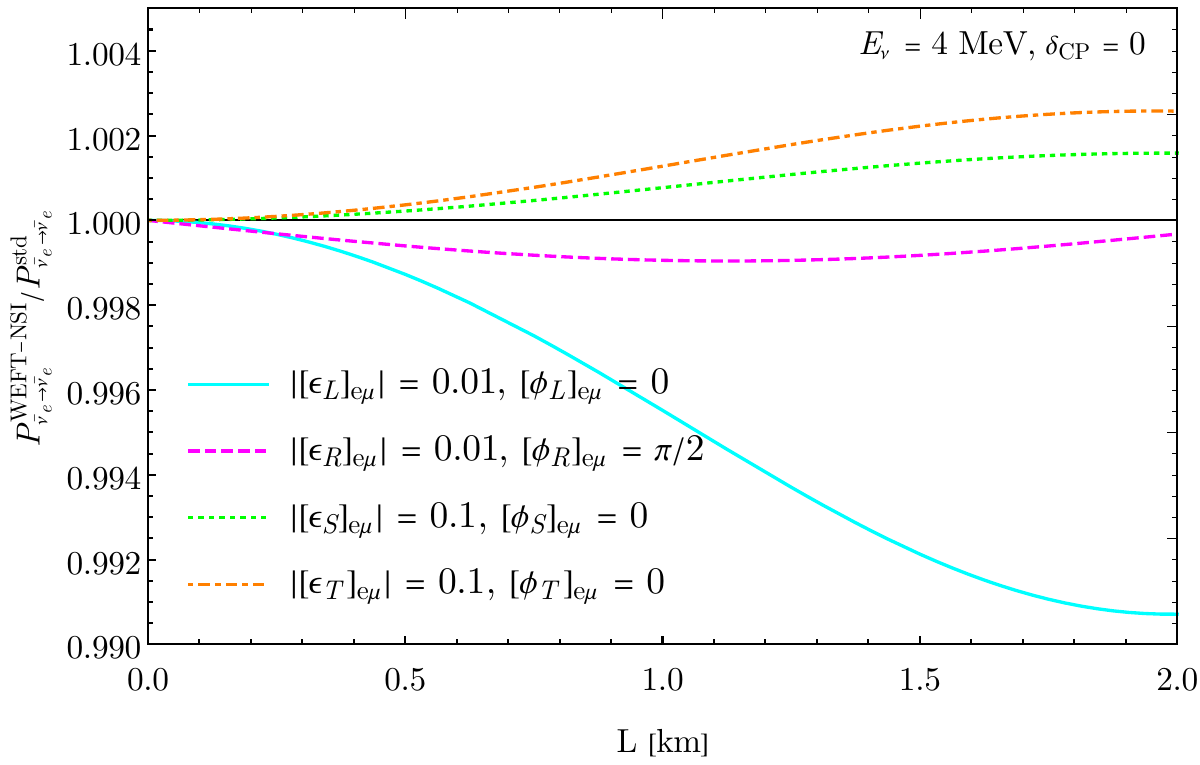}
\par\end{centering}
\caption{Similarly to figure \ref{fig:QM-NSI-effect-shape}, these ratios
as a function of the distance indicate the effect of WEFT-NSI on the
shape of the survival probability. The magnitude of $[\varepsilon_{S}]_{e\mu}$
and $[\varepsilon_{T}]_{e\mu}$ is taken to be ten times larger than
that of $[\varepsilon_{L}]_{e\mu}$ and $[\varepsilon_{R}]_{e\mu}$ to show
their effect clearly. More details can be found in the text of section \ref{subsec:WEFT-NSI-formulas}.
\textcolor{blue}{{} \label{fig:WEFT-NSI-effect-shape}}}
\end{figure}

As for the case of QM-NSI, we examine the effect of the WEFT-NSI on the shape of the survival probability through the ratio $P_{\bar{\nu}_{e}\rightarrow\bar{\nu}_{e}}^{\textrm{WEFT-NSI}}/P_{\bar{\nu}_{e}\rightarrow\bar{\nu}_{e}}^{\textrm{std}}$. The zero-distance effect in the WEFT-NSI framework can be simplified as 
\begin{equation}
P_{\bar{\nu}_{e}\rightarrow\bar{\nu}_{e}}^{\textrm{WEFT-NSI}}(L_\nu=0)=\frac{1+2p_{XL}d_{XL}\left|[\epsilon_{X}]_{e\alpha}\right|^{2}+p_{XX}d_{XX}\left|[\epsilon_{X}]_{e\alpha}\right|^{4}}{1+(p_{XX}+d_{XX})\left|[\epsilon_{X}]_{e\alpha}\right|^{2}+p_{XX}d_{XX}\left|[\epsilon_{X}]_{e\alpha}\right|^{4}}\label{eq:zero-distance-WEFT-NSI}
\end{equation}
if only one NSI parameter $[\varepsilon_{X}]_{e\alpha}$ ($\alpha\neq e$) is considered at a time. The quantity $P_{\bar{\nu}_{e}\rightarrow\bar{\nu}_{e}}^{\textrm{WEFT-NSI}}($ $L_\nu=0)$ is always less than unity for each nonvanishing parameter $[\varepsilon_{X}]_{e\alpha}$ except for $[\varepsilon_{L}]_{e\alpha}$ for which $P_{\bar{\nu}_{e}\rightarrow\bar{\nu}_{e}}^{\textrm{WEFT-NSI}}(L_\nu=0)=1$. With the zero-distance effect removed,  figure \ref{fig:WEFT-NSI-effect-shape} shows the ratios for the NSI parameters $[\varepsilon_{X}]_{e\mu}$ for $X=L,R,S$ and $T$, respectively. As can be seen from the figure,
the effect of $[\varepsilon_{L}]_{e\mu}$ or $[\varepsilon_{R}]_{e\mu}$ is similar to that of $\epsilon{}_{e\alpha}$ of QM-NSI. We thus expect a similar anti-correlation between these parameters and $\sin\theta_{13}$. For the cases of $[\varepsilon_{S}]_{e\mu}$
and $[\varepsilon_{T}]_{e\mu}$, $\left|[\varepsilon_{S}]_{e\mu}\right|=\left|[\varepsilon_{T}]_{e\mu}\right|=0.1$
is taken to make the plot to show their effect on the shape of the
survival probability more clearly. The corresponding ratio curves
deviate from the unity line in just the opposite way as for the cases
of $[\varepsilon_{L}]_{e\mu}$ and $[\varepsilon_{R}]_{e\mu}$, and they
will be forced to increase with $\sin\theta_{13}$ to fit the data
appropriately.
As in the QM-NSI approach, each of the complex NSI parameters is decomposed as 
\begin{equation}
[\varepsilon_{X}]_{e\alpha}=\left|[\varepsilon_{X}]_{e\alpha}\right|e^{i[\phi_{X}]_{e\alpha}},
\end{equation}
where $[\phi_{X}]_{e\alpha}\in[0,2\pi)$ for $\alpha=\mu,\tau$.

\section{Daya Bay reactor neutrino experiment \label{sec:DayaBay-experiment}}

The main goal of the Daya Bay reactor neutrino experiment is detecting MeV-scale
electron antineutrinos produced in nuclear reactors to determine the
mixing angle $\theta_{13}$ via the study of $\overline{\nu}_{e}$
disappearance. The $\overline{\nu}_{e}$'s are detected through the
IBD reaction $\overline{\nu}_{e}+p\rightarrow e^{+}+n$ and are identified
with the combination of a prompt-energy signal due to the positron kinetic energy loss and annihiliation and a delayed-energy signal due to the subsequent neutron capture.

The electron antineutrinos are emitted from the three pairs of 2.9 GW$_{th}$
reactors at the Daya Bay-Ling Ao nuclear power facility in Shenzhen,
China, and are detected by up to eight antineutrino detectors (ADs)
which were installed in three underground experimental halls (EH1,
EH2 and EH3) with a flux-averaged baseline of about 500 m, 500 m,
and 1650 m from the reactors, respectively. Twenty tonnes of liquid scintillator doped with
0.1\% gadolinium by weight (GdLS) in each AD \citep{YEH2007329,DING2008238,BERIGUETE201482}
were used to detect the IBD events. More information
about the experiment can be found in Refs. \citep{DayaBay:2014cmr,DayaBay:2015kir}. 

There were three different configurations of ADs in the three EHs in the operation of the Daya Bay experiment (i.e., 6-AD, 8-AD and 7-AD operation periods). With a total of 3158 days of data acquisition, a final sample of $5.55\times10^{6}$
IBD candidates with the final-state neutron captured on gadolinium were obtained \cite{DayaBay:2022orm}. Here we also probe the CC-NSI effect with the same data sample. As mentioned in the Introduction, we only consider the NSI effects on the measurement of the oscillation parameter $\theta_{13}$. 

The $\chi^{2}$ is constructed based on the binned
maximum poisson likelihood method as 
\begin{align}
\chi^{2} & =2\sum_n^{N_{\text{period}}}\sum_{j}^{N_{\textrm{ADs}}}\sum_{i}^{N_{\textrm{Ebins}}}[N_{nji}^{\textrm{pred}}-N_{nji}^{\textrm{obs}}+N_{nji}^{\textrm{obs}}\ln\frac{N_{nji}^{\textrm{obs}}}{N_{nji}^{\textrm{pred}}}]\nonumber \\
 & \quad+\sum_{j}^{N_{\text{period}}\times N_{\textrm{EHs}}\times N_{\textrm{Ebins}}}\sum_{k}^{N_{\text{period}}\times N_{\textrm{EHs}}\times N_{\textrm{Ebins}}}f_{j}V_{jk}^{-1}f_{k}\nonumber \\
 & \quad+\sum_{l}^{E\textrm{scale}}\frac{\eta_{l}^{2}}{\sigma_{l}^{2}}+\sum_{m}^{N_{\textrm{ADs}}}\frac{\zeta_{m}^{2}}{\sigma_{m}^{2}}+\sum_{n}^{N_{\textrm{ADs}}\times\textrm{bkg}}\frac{b_{n}^{2}}{\sigma_{n}^{2}} +\left(\frac{\delta^{\text{corr}}}{\sigma^{\text{corr}}}\right)^{2}+\sum_{i}^{N_\textrm{osc}}\left(\frac{\delta_{i}^{\textrm{osc}}}{\sigma_{i}^{\textrm{osc}}}\right)^{2},
\end{align}
where the expected number of events $N_{nji}^{\textrm{pred}}\equiv N_{nji}^{\textrm{pred}}(\theta_{13},\vec{\epsilon}_{\textrm{NSI}}|\vec{f},\vec{\eta},\vec{\zeta},\vec{b},\delta^{\text{corr}},\vec{\delta}^{\textrm{osc}})$
in the $i$-th energy bin of the $j$-th AD of the $n$-th operation period is obtained from the prediction of a model with the standard oscillation parameter $\theta_{13}$, the NSI parameters $\vec{\epsilon}_{\textrm{NSI}}$ and the estimation of the background. The effect of NSI on the measurement of the standard neutrino oscillation parameters except $\theta_{13}$ are assumed to be negligible for the strong constraints from other experiments alluded to in section \ref{sec:Introduction}. $N_{nji}^{\textrm{obs}}$ is the corresponding observed number of IBD candidate events. There are 26 bins of the reconstructed energy spectrum with the first bin ranging from 0.7 MeV to 1.3 MeV, the last from 7.3 MeV to 12.0 MeV and the other 24 bins uniformly distributed from 1.3 to 7.3 MeV. The parameters $\vec{f}$, $\vec{\eta}$, $\vec{\zeta}$, $\vec{b}$ and $\vec{\delta}^{\textrm{osc}}$
are reactor related, energy nonlinearity response related, AD related,
background related and external oscillation parameter related systematic
nuisance parameters, respectively. The nuisance parameter $\delta^{\text{corr}}$ represents
the overall normalization which comes from the correlated detector
efficiency and the reactor flux model normalization.
These nuisance parameters are constrained by the corresponding uncertainties
$\sigma_{j}$ except for the parameter $\vec{f}$ for which the covariance
matrix $V$ is used to reduce the number of the nuisance parameters for
the reactor flux model. More details about the the nuisance parameters can be found in \citep{DayaBay:2016ggj}. Central values and uncertainties of oscillation parameters for the case of the normal mass ordering are listed in Table \ref{tab:Input-Parameters}. The neutrino flux is evaluated using the Huber-Mueller model \citep{PhysRevC.84.024617,PhysRevC.83.054615} where we have conservatively enlarged the overall uncertainty in the flux to $\sigma^{\text{corr}}=5\%$  
given the lack of detailed knowledge of the structure of the forbidden transitions \cite{Hayes:2013wra,Hayes:2015ctl} and uncertainties from other possible sources.

\section{Constraints on NSI parameters \label{sec:Constraints-on-NSI}}

Since there are multiple parameters, we initially consider variations in a single CC-NSI parameter at a time. We start with finding the allowed regions in the $(\sin^{2}\theta_{13},\left|\epsilon\right|)$ plane for the corresponding CC-NSI phase $\phi$ and/or the CP-violating phase $\delta_{\textrm{CP}}$ to be set to zero and vary freely, respectively. When necessary, we show the allowed regions when $\phi$ and/or $\delta_{\textrm{CP}}$ take certain values, i.e., $\pi/2,\pi$ and/or $3\pi/2$ to help understand the formation of the allowed regions when these phases vary freely. We also provide constraints in the $(\phi,\left|\epsilon\right|)$ plane with $\sin^{2}\theta_{13}$ set to vary freely and $\delta_{\text{CP}}=0$,
and in the $(\left|\epsilon_{1}\right|,\left|\epsilon_{2}\right|)$ plane with $\sin^{2}\theta_{13}$ set to vary freely and $\phi=\delta_{\text{CP}}=0$. 

\subsection{Constraints on QM-NSI parameters $\epsilon_{e\alpha}$ for $\epsilon_{e\alpha}^{s}=\epsilon_{\alpha e}^{d*}\equiv\epsilon_{e\alpha}$}

The results below are for the allowed regions and constraints of the non-universal NSI parameters $\epsilon_{ee}$, $\epsilon_{e\mu}$, $\epsilon_{e\tau}$ and the universal NSI parameter $\epsilon_{ex}\equiv\epsilon_{ee}=\epsilon_{e\mu}=\epsilon_{e\tau}$,
respectively. 

\subsubsection{Constraints on electron-NSI coupling $\epsilon_{ee}$
\label{subsec:Constraint-epsilon_ee}}

\begin{figure*}[t]
\begin{centering}
\subfigure[]{\includegraphics[width=7.0cm]{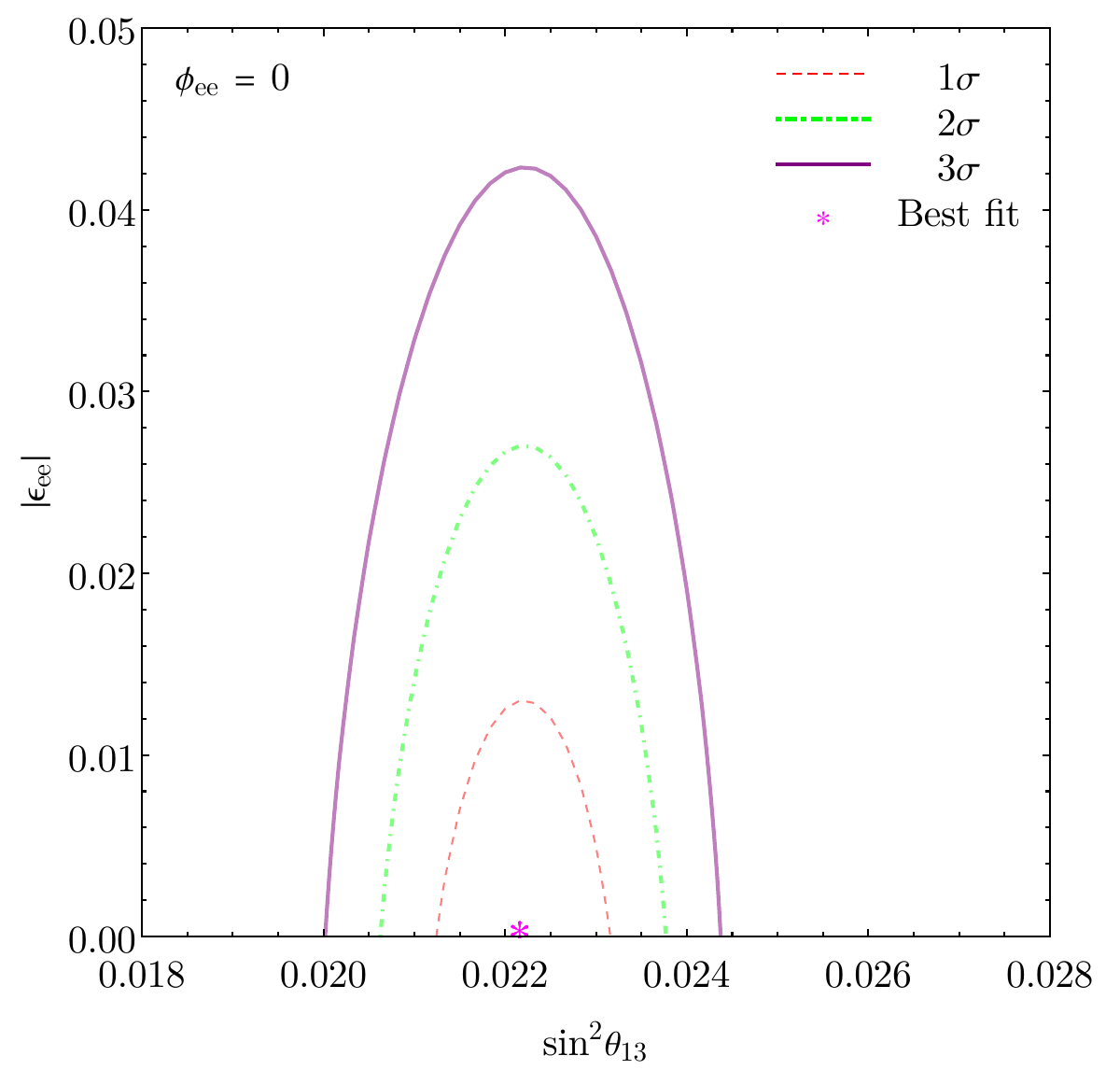}
\label{fig:s=d_ee_a}}
\subfigure[]{\includegraphics[width=7.0cm]{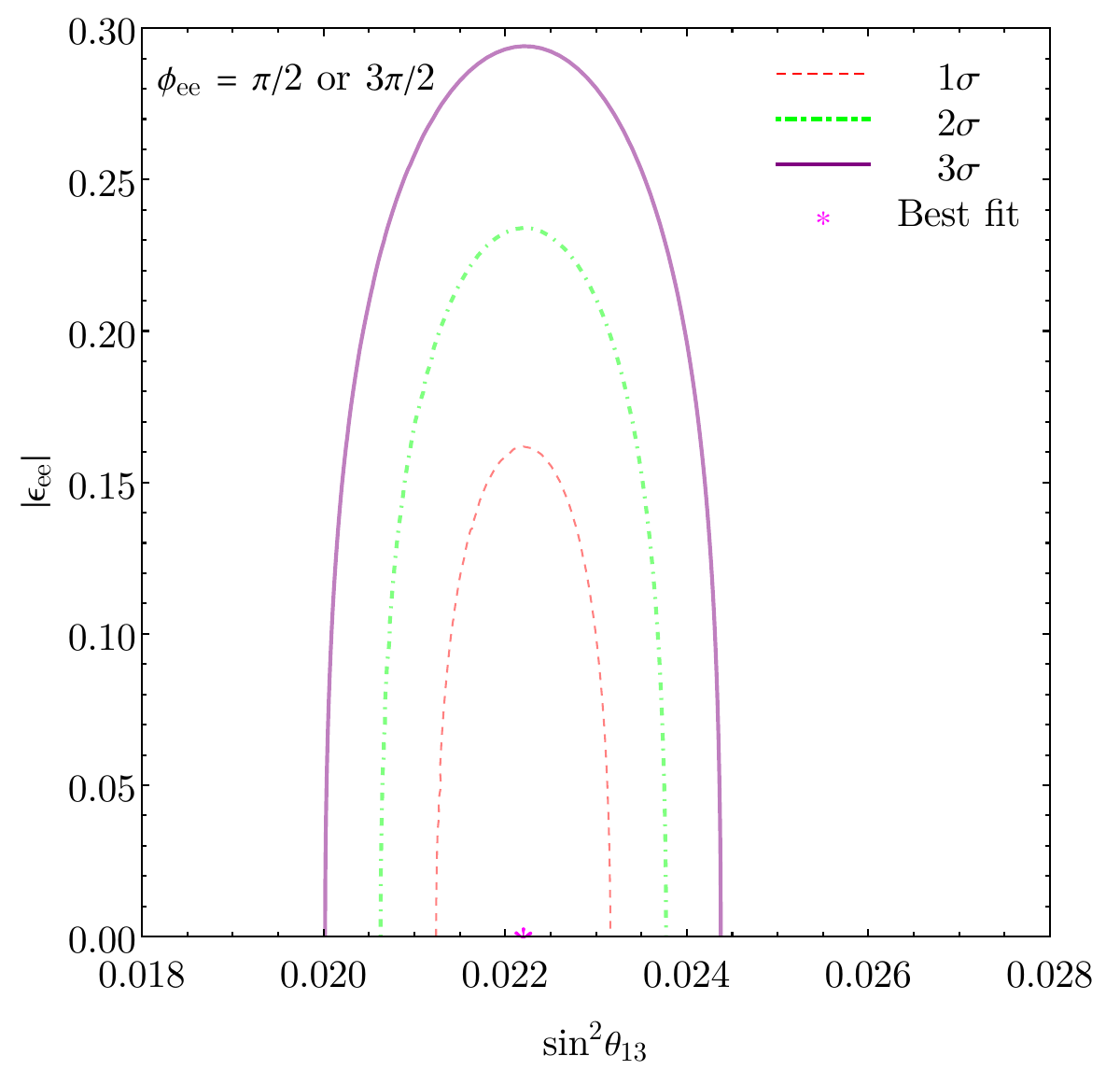}
\label{fig:s=d_ee_b}}
\par\end{centering}
\begin{centering}
\subfigure[]{\includegraphics[width=7.0cm]{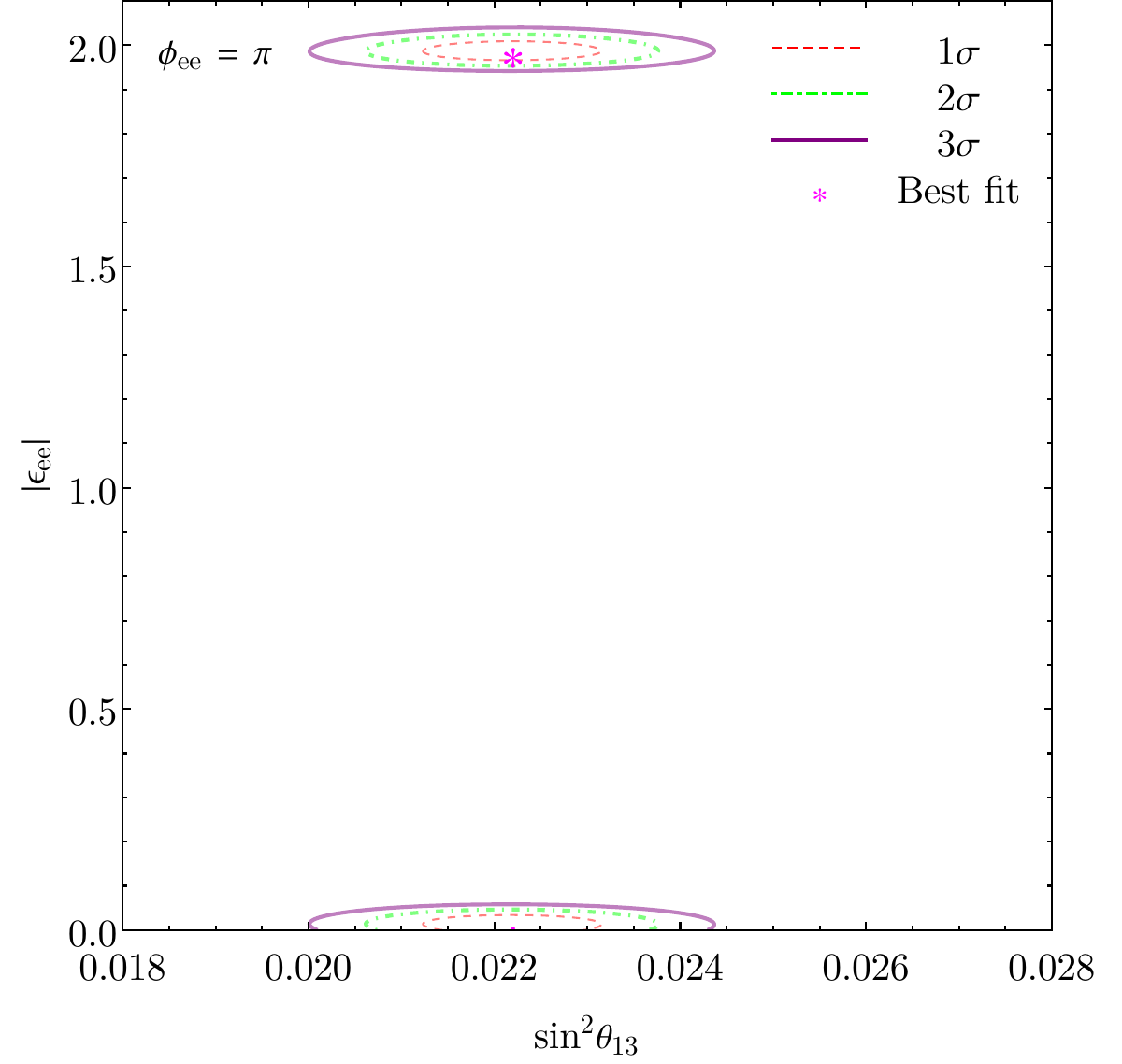}
\label{fig:s=d_ee_c}}
\subfigure[]{\includegraphics[width=7.0cm]{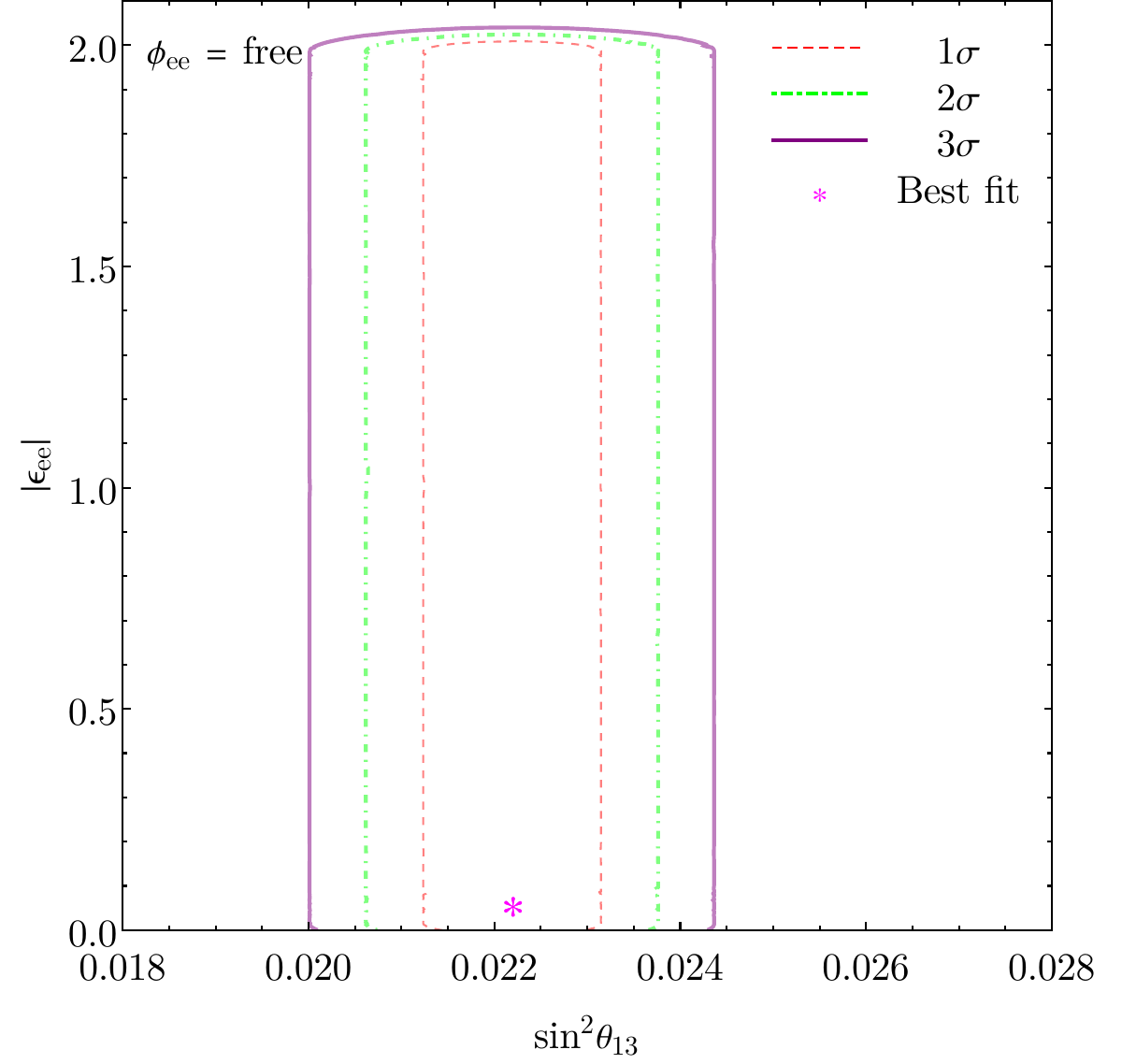}
\label{fig:s=d_ee_d}}
\par\end{centering}
\caption{The first three panels \ref{fig:s=d_ee_a}, \ref{fig:s=d_ee_b} and \ref{fig:s=d_ee_c} shows the dependence of the allowed regions
in the $(\sin^{2}\theta_{13},\left|\epsilon_{ee}\right|)$ plane on
the values of the CC-NSI phase $\phi_{ee}$ for $\phi_{ee}=0,\pi/2,\pi$
and $3\pi/2$, respectively. The allowed regions are the same for
$\phi_{ee}=\pi/2$ and $3\pi/2$. The lower right panel \ref{fig:s=d_ee_d} is for $\phi_{ee}$
being marginalized over ($\phi_{ee}=$free). Details of the analysis are provided in section \ref{subsec:Constraint-epsilon_ee}. 
\label{fig:s=d_ee}}
\end{figure*}
The parameter $\epsilon_{ee}$ represents a kind of flavor-conserving non-universal
NSI associated with $\bar{\nu}_{e}$ present in both production and
detection processes. We have $U_{ej}^{sd}=(1+\epsilon_{ee}^{*})U_{ej}$
. The effective survival probability is 
\begin{align}
P_{\bar{\nu}_{e}^{s}\rightarrow\bar{\nu}_{e}^{d}}^{\textrm{QM-NSI-eff}} & =(1+\left|\epsilon_{ee}\right|^{2}+2\left|\epsilon_{ee}\right|\cos\phi_{ee})^{2}P_{\bar{\nu}_{e}\rightarrow\bar{\nu}_{e}}^{\textrm{std}},\label{eq:s=d_ee}
\end{align}
which has no dependence on $\delta_{\text{CP}}$ and $\theta_{23}$ as in
the standard case of eq.\,(\ref{eq:Pee_std}). This type of NSI effectively changes the normalization of the number of events. And the approximate symmetry of the standard survival probability is inherited, i.e., $P_{\bar{\nu}_{e}^{s}\rightarrow\bar{\nu}_{e}^{d}}^{\textrm{QM-NSI-eff}}$ is approximately invariant under the exchange of $\theta_{13} \leftrightarrow \pi/2-\theta_{13}$. We thus provide the allowed regions in the $(\sin^{2}\theta_{13},\left|\epsilon_{ee}\right|)$ plane for $\theta_{13}$ being small only. Figures \ref{fig:s=d_ee_a}, \ref{fig:s=d_ee_b} and \ref{fig:s=d_ee_c} show the allowed regions in the $(\sin^{2}\theta_{13},\left|\epsilon_{ee}\right|)$ plane for $\phi_{ee}=0,\pi/2$ ($\text{or } 3\pi/2$)
and $\pi$, respectively. It is
easy to see from eq.\,(\ref{eq:s=d_ee}) that the allowed regions
for $\phi_{ee}=\pi/2$ and $3\pi/2$ are the same. This is a typical
feature for the case of $\epsilon_{e\alpha}^{s}=\epsilon_{\alpha e}^{d*}$
and we will see it again in the cases with $\epsilon_{e\mu}$, $\epsilon_{e\tau}$
and $\epsilon_{ex}$ below. For $\phi_{ee}=0$, $P_{\bar{\nu}_{e}^{s}\rightarrow\bar{\nu}_{e}^{d}}^{\textrm{eff}}=(1+\left|\epsilon_{ee}\right|)^{4}P_{\bar{\nu}_{e}\rightarrow\bar{\nu}_{e}}$, the most stringent constraint is found which reads $\left|\epsilon_{ee}\right|<0.0148$ at $90\%$ confidence level (C.L.) with one degree of freedom (d.o.f.). For $\phi_{ee}=\pi$, we have $P_{\bar{\nu}_{e}^{s}\rightarrow\bar{\nu}_{e}^{d}}^{\textrm{eff}}=(1-\left|\epsilon_{ee}\right|)^{4}P_{\bar{\nu}_{e}\rightarrow\bar{\nu}_{e}}$.
The allowed region is separated into two subregions. One is consistent
with $\left|\epsilon_{ee}\right|=0$, the other with $\left|\epsilon_{ee}\right|=2$.
The allowed region of $\left|\epsilon_{ee}\right|$ becomes large if we marginalize over $\phi_{ee}$ from $0$ to $2\pi$ which leads to the constraint $\left|\epsilon_{ee}\right|<2.01$. All the allowed region plots show that the Daya Bay experimental data is consistent with the standard oscillation framework ($\left|\epsilon_{ee}\right|=0$) within 1$\sigma$ C.L.. The numerical values of the $90\%$ C.L. constraints (1 d.o.f.) on $\left|\epsilon_{ee}\right|$ under different conditions are listed in Table \ref{tab:s=d_ee}. 
\begin{table}
\centering{}
\begin{tabular}{|c|c|}
\hline 
$\phi_{ee}$ & $\left|\epsilon_{ee}\right|$\tabularnewline
\hline 
$0$ & $\left|\epsilon_{ee}\right|<0.0148$\tabularnewline
$\pi/2,3\pi/2$ & $\left|\epsilon_{ee}\right|<0.172$\tabularnewline
$\pi$ & $\left|\epsilon_{ee}\right|<0.0371$ or $1.97<\left|\epsilon_{ee}\right|<2.01$\tabularnewline
free & $\left|\epsilon_{ee}\right|<2.01$\tabularnewline
\hline 
\end{tabular}
\caption{$90\%$ C.L. constraints (1 d.o.f) on the QM-NSI parameter $\left|\epsilon_{ee}\right|$
projected from the $(\sin^{2}\theta_{13},\left|\epsilon_{ee}\right|)$
plane for $\phi_{ee}$ taking on values of $0,\pi/2,\pi$, $3\pi/2$
and being marginalized over ($\phi_{ee}=$free), respectively. \label{tab:s=d_ee}}
\end{table}

The constraints on $\left|\epsilon_{ee}\right|$ depend primarily on the normalization uncertainty $\sigma^{\text{corr}}$ when the phase $\phi_{ee}$ is fixed at some special values, as discussed in ref. \cite{VanegasForero:2019mqo}. This dependence can be understood as shown in figure \ref{fig:QM-NSI-effect-shape} or eq.\,(\ref{eq:s=d_ee}). Both $|\epsilon_{ee}|$ and the neutrino flux have the same effect which is independent of $L_\nu$. In the future if the neutrino flux can be accurately predicted, the constraints on $\left|\epsilon_{ee}\right|$ can be further improved.

\subsubsection{Constraints on muon-NSI and tau-NSI couplings $\epsilon_{e\mu}$
and $\epsilon_{e\tau}$ 
\label{subsec:Constraint-epsilon_emu}} 

\begin{figure*}
\begin{centering}
\subfigure[]{\includegraphics[width=7.0cm]{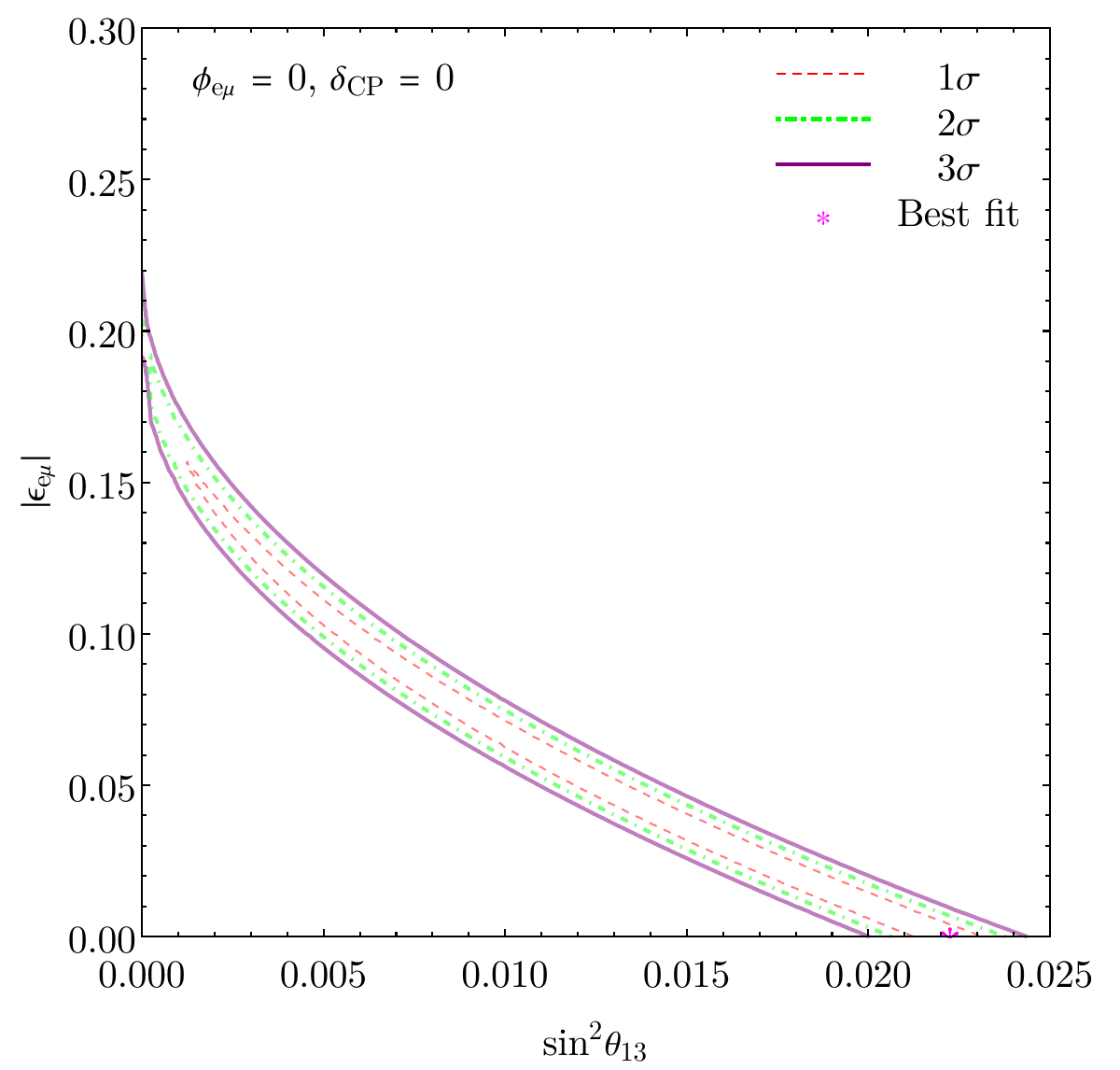}
\label{fig:s=d_emu_a}}
\subfigure[]{\includegraphics[width=7.0cm]{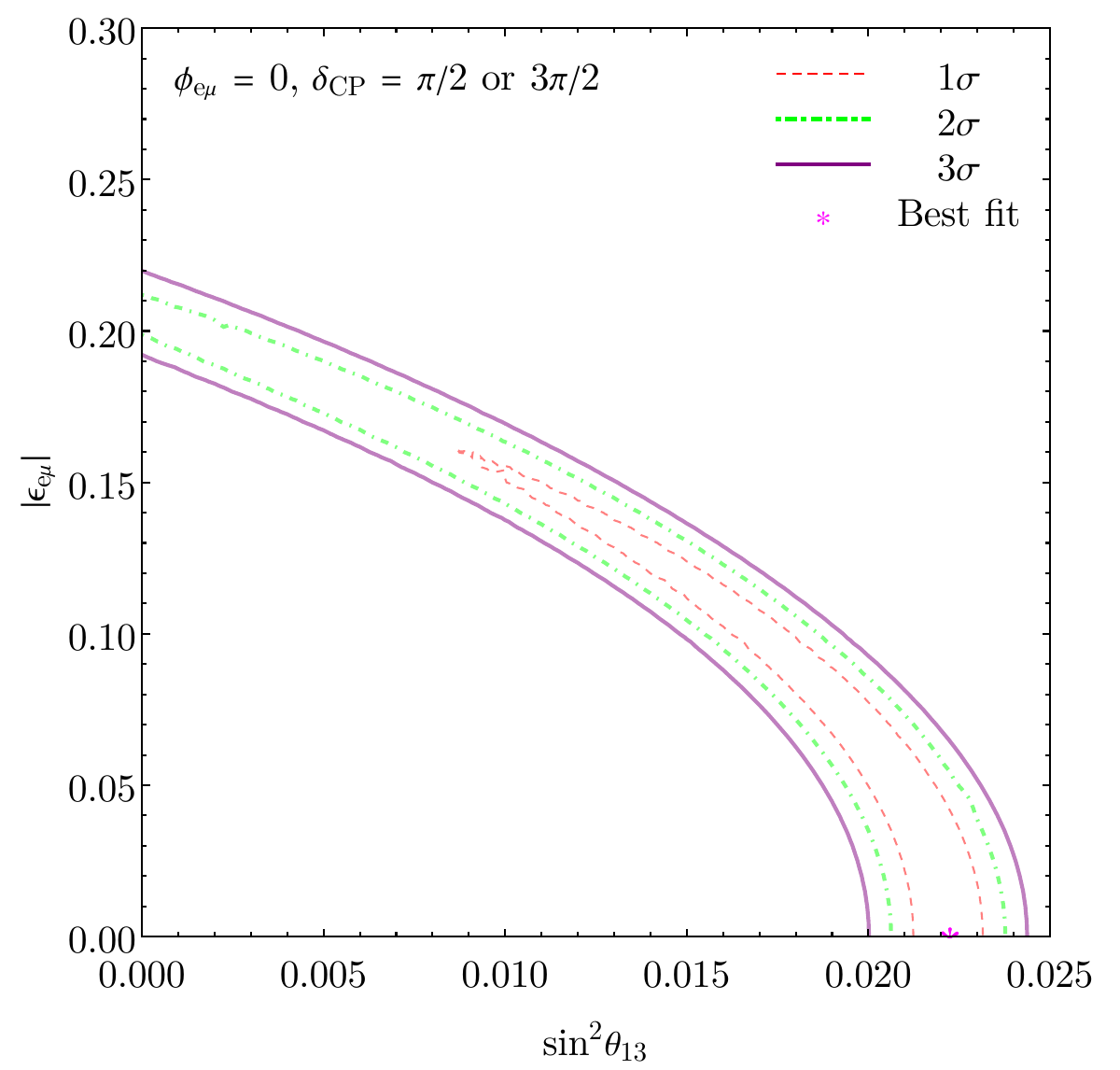}
\label{fig:s=d_emu_b}}
\par\end{centering}
\begin{centering}
\subfigure[]{\includegraphics[width=7.0cm]{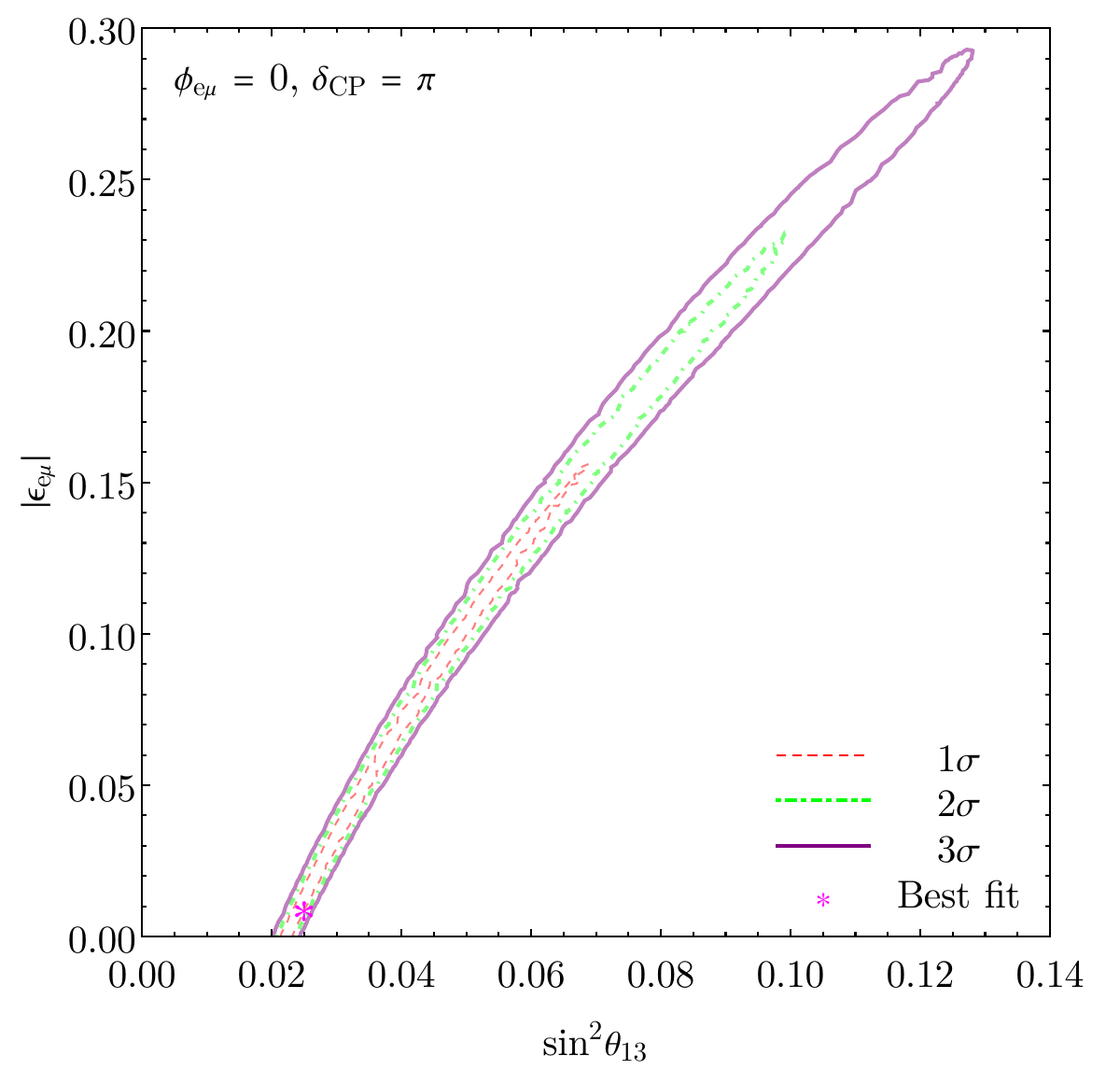}
\label{fig:s=d_emu_c}}
\subfigure[]{\includegraphics[width=7.0cm]{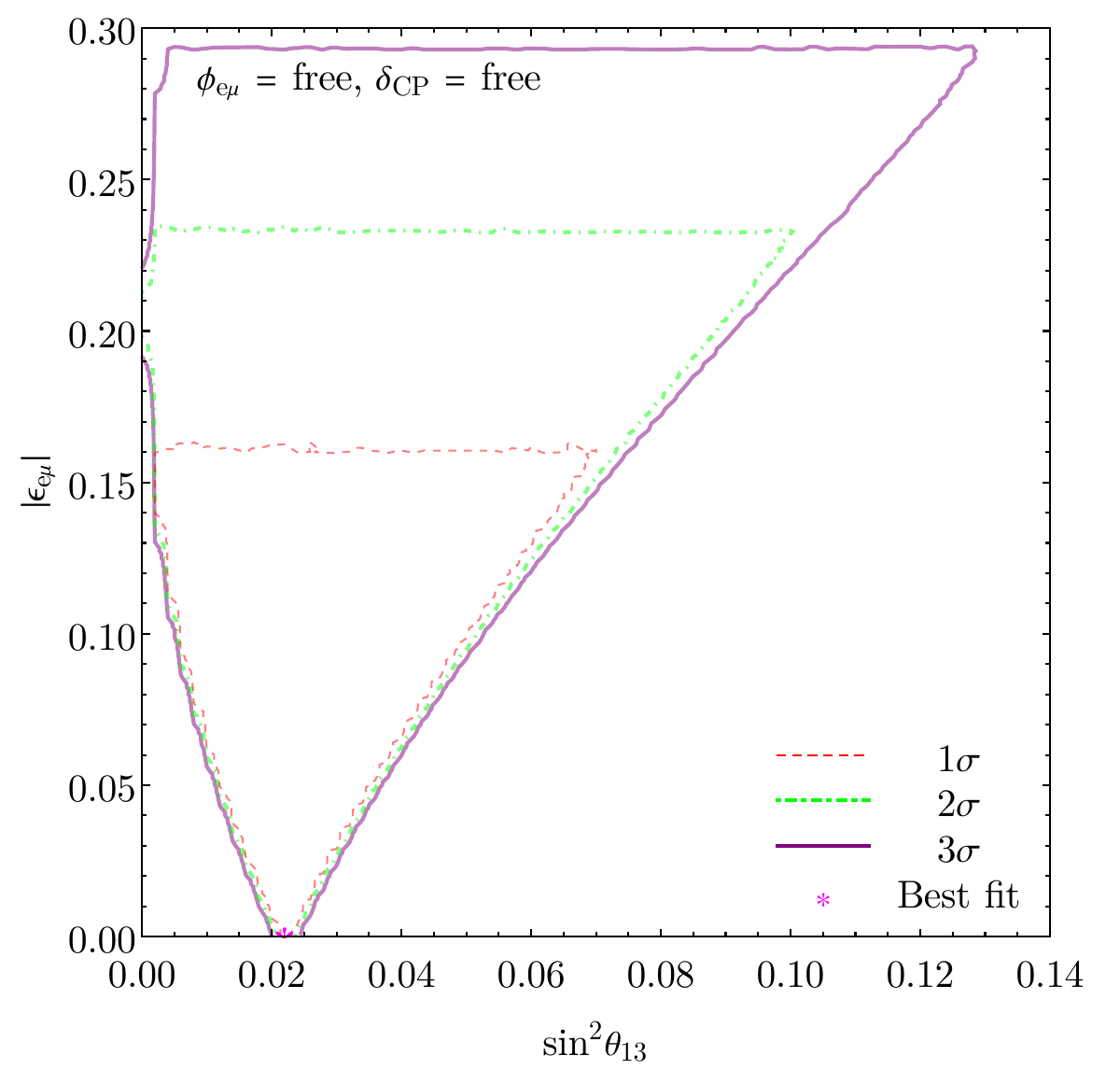}
\label{fig:s=d_emu_d}}
\par\end{centering}
\caption{The first three panels \ref{fig:s=d_emu_a}, \ref{fig:s=d_emu_b} and \ref{fig:s=d_emu_c} shows the dependence of the allowed regions
in the $(\sin^{2}\theta_{13},\left|\epsilon_{e\mu}\right|)$ plane on
the values of the CC-NSI phase $\phi_{e\mu}$ and the Dirac CP-violating
phase $\delta_{\textrm{CP}}$ for $\phi_{e\mu}=0$ and $\delta_{\textrm{CP}}=0,\pi/2,\pi$
and $3\pi/2$, respectively. The corresponding allowed regions for
$\delta_{\textrm{CP}}=0$ and $\phi_{e\mu}=\pi/2,\pi$ and $3\pi/2$,
respectively, are similar. The lower right panel \ref{fig:s=d_emu_d} is for both phases
being marginalized over ($\delta_{\text{CP}}=$free, $\phi_{e\mu}=$free). Details of the analysis can be found in section \ref{subsec:Constraint-epsilon_emu}.
\label{fig:s=d_emu}}
\end{figure*}

The flavor-violating non-universal NSI parameter $\epsilon_{e\mu}$
associates the electron (positron) with $\bar{\nu}_{\mu}$ in the
production (detection) processes. When $\epsilon_{e\mu}$ is non-zero, we have $U_{ej}^{sd}=U_{ej}+\epsilon_{e\mu}^{*}U_{\mu j}$
and 
\begin{equation}
\left|U_{ej}^{sd}\right|^{2}=\left|U_{ej}\right|^{2}+\left|\epsilon_{e\mu}\right|^{2}\left|U_{\mu j}\right|^{2}+2\textrm{Re}(\epsilon_{e\mu}U_{ej}U_{\mu j}^{*}).
\end{equation}
The 2nd term on the right hand side depends on $\delta_{\textrm{CP}}$
in the form of $\cos(\delta_{\textrm{CP}})$ for $j=1,2$. The 3rd
term is dependent on $\delta_{\textrm{CP}}$ and $\phi_{e\mu}$ in
the form of $\cos(\delta_{\textrm{CP}}-\phi_{e\mu})$ and/or $\cos(\phi_{e\mu})$. For this reason, the effective survival
probability is the same for $\delta_{\textrm{CP}}=\pi/2$ and $3\pi/2$
when $\phi_{e\mu}=0$ or $\phi_{e\mu}=\pi/2$ and $3\pi/2$ when $\delta_{\textrm{CP}}=0$. The roles played by $\delta_{\textrm{CP}}$ and $\phi_{e\mu}$ are similar. For the presence of NSI with the parameter $\epsilon_{e\mu}$, the effective mixing angle $\tilde{\theta}_{13}$ (what is measured in the reactor oscillation experiment) might be different from the true mixing angle  $\theta_{13}$. We find that the effective survival probability in this case is approximately invariant under the exchange of $\theta_{13} \leftrightarrow \pi/2-2\tilde{\theta}_{13}+\theta_{13}$ which reduces to $\theta_{13} \leftrightarrow \pi/2-\theta_{13}$ for the standard survival probability for which $\tilde{\theta}_{13} \rightarrow \theta_{13}$, or of $\theta_{13} \leftrightarrow \pi/2-\theta_{13}$, depending on the values of  $\phi_{e\mu}$ and $\delta_{\textrm{CP}}$. We thus provide allowed regions in the $(\sin^{2}\theta_{13},\left|\epsilon_{e\mu}\right|)$
plane around small $\theta_{13}$ only. Figures \ref{fig:s=d_emu_a}, \ref{fig:s=d_emu_b} and \ref{fig:s=d_emu_c} show such allowed regions for $\delta_{\textrm{CP}}=0,\pi/2,\pi$ and $3\pi/2$, respectively,
when setting $\phi_{e\mu}=0$. The approximate expressions
of the effective survival probability is useful in explaining the
behavior of the allowed regions. The reactor data can be fitted with an approximation to the standard case using 
\begin{equation}
\sin^{2}\tilde{\theta}_{13}\approx\sin^{2}\theta_{13}+2\sin\theta_{13}\sin\theta_{23}\left|\epsilon_{e\mu}\right|\cos(\delta_{\textrm{CP}}-\phi_{e\mu}),\label{eq:s=d_emu_1stOrder}
\end{equation}
for $\theta_{13}$ and $\left|\epsilon_{e\mu}\right|$
being small \citep{Agarwalla:2014bsa}. For the case that $\phi_{e\mu}=\delta_{\textrm{CP}}=0$,
$\left|\epsilon_{e\mu}\right|$ must decrease with $\sin^{2}\theta_{13}$
to maintain the good agreement with the experimental data. For $\phi_{e\mu}=0$
and $\delta_{\textrm{CP}}=\pi$, $\left|\epsilon_{e\mu}\right|$ increases with $\sin^{2}\theta_{13}$. The case that $\phi_{e\mu}=0$
and $\delta_{\textrm{CP}}=\pi/2$ or $3\pi/2$ indicates that $\left|\epsilon_{e\mu}\right|$
is independent of $\sin^{2}\theta_{13}$ for a vanishing $\left|\epsilon_{e\mu}\right|$.
The cases that setting $\delta_{\textrm{CP}}=0$ and $\phi_{e\mu}=\pi/2,\pi$
and $3\pi/2$ are almost the same and thus are not shown. The allowed
region for marginalizing over $\delta_{\textrm{CP}}$ with $\phi_{e\mu}=0$ is the combination of the allowed regions with 
$\delta_{\textrm{CP}}$ taking any special value in the range $[0,2\pi)$ when $\phi_{e\mu}=0$.
The situation for $\delta_{\textrm{CP}}=0$ and $\phi_{e\mu}$ to
vary freely is the same, and so is the allowed region for both $\delta_{\textrm{CP}}$
and $\phi_{e\mu}$ to vary freely as shown in figure \ref{fig:s=d_emu_d}. As in the case of $\epsilon_{ee}$,
the data is consistent with the standard oscillation
framework ($\left|\epsilon_{e\mu}\right|=0$) less than 1$\sigma$ C.L.
For the NSI parameter $\epsilon_{e\tau}$ being non-zero, we have $U_{ej}^{sd}=U_{ej}+\epsilon_{e\tau}^{*}U_{\tau j}$.
And given that $\left|U_{\mu i}\right|\approx\left|U_{\tau i}\right|$
from measurements for $i=1,2$ and $3$ \citep{Esteban:2020cvm} ,
we see the role the parameter $\epsilon_{e\tau}$ plays is similar
to that of the parameter $\epsilon_{e\mu}$. Thus the allowed regions
on $\left|\epsilon_{e\mu}\right|$ and $\left|\epsilon_{e\tau}\right|$
are close to one another. The constraints on $\left|\epsilon_{e\mu}\right|$
and $\left|\epsilon_{e\tau}\right|$ are listed in Table \ref{tab:s=d_emu__etau_ex}. 

Unlike the case for $\left|\epsilon_{ee}\right|$ which is mostly affected by the reactor flux uncertainty, the constraints on $\left|\epsilon_{e\mu}\right|$ or $\left|\epsilon_{e\tau}\right|$ depend on both the systematical and statistical uncertainties. As shown in figure \ref{fig:QM-NSI-effect-shape}, the parameter $\left|\epsilon_{e\mu}\right|$ or $\left|\epsilon_{e\tau}\right|$ could be determined through  the far/near relative  measurement at different baselines, which is quite similar to the $\theta_{13}$ oscillation measurement. Thus, the parameter $\left|\epsilon_{e\mu}\right|$ or $\left|\epsilon_{e\tau}\right|$ is not sensitive to the neutrino flux uncertainty.
\begin{figure*}
\begin{centering}
\subfigure[]{\includegraphics[width=7.0cm]{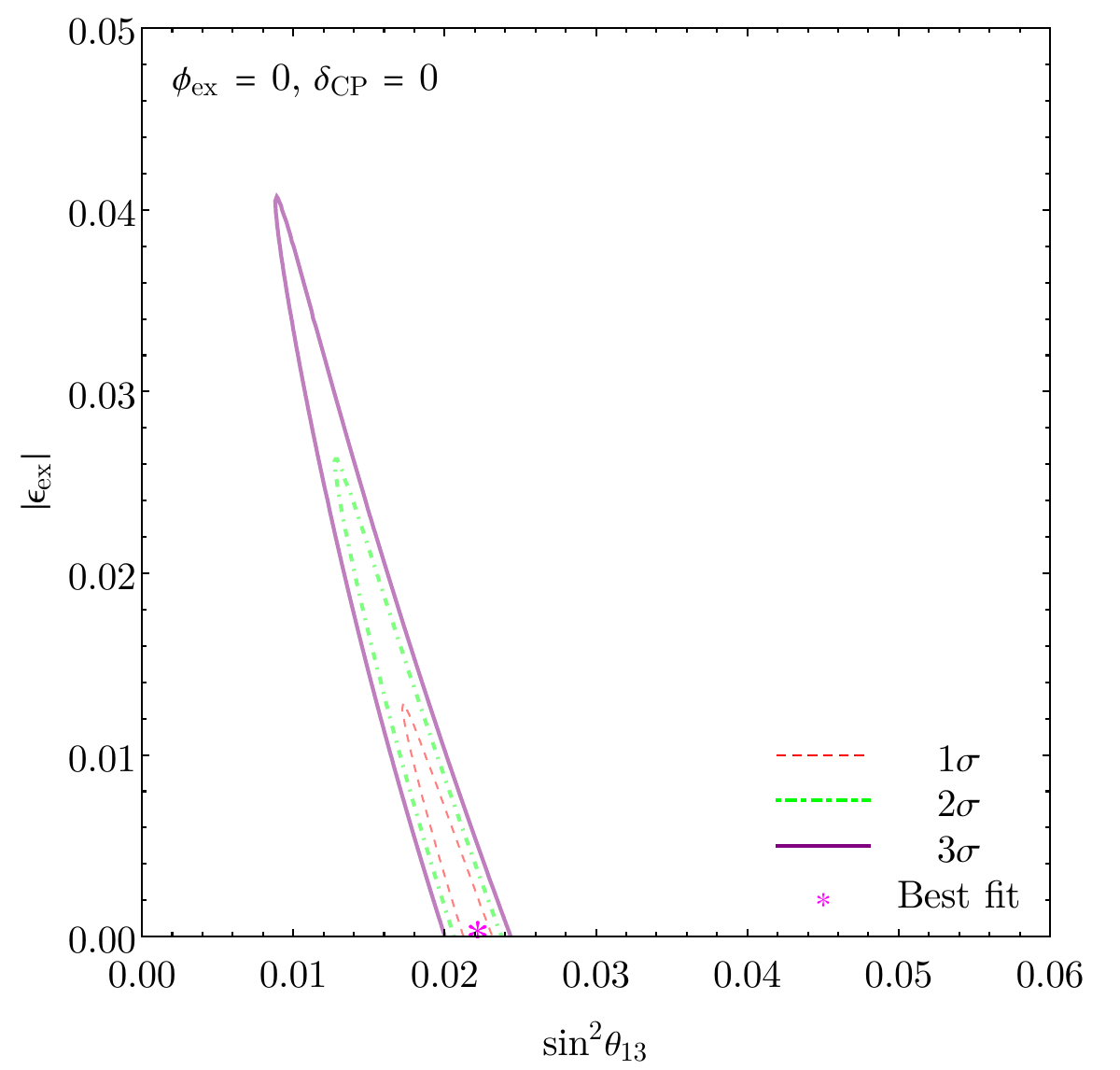}
\label{fig:s=d_ex_a}}
\subfigure[]{\includegraphics[width=7.0cm]{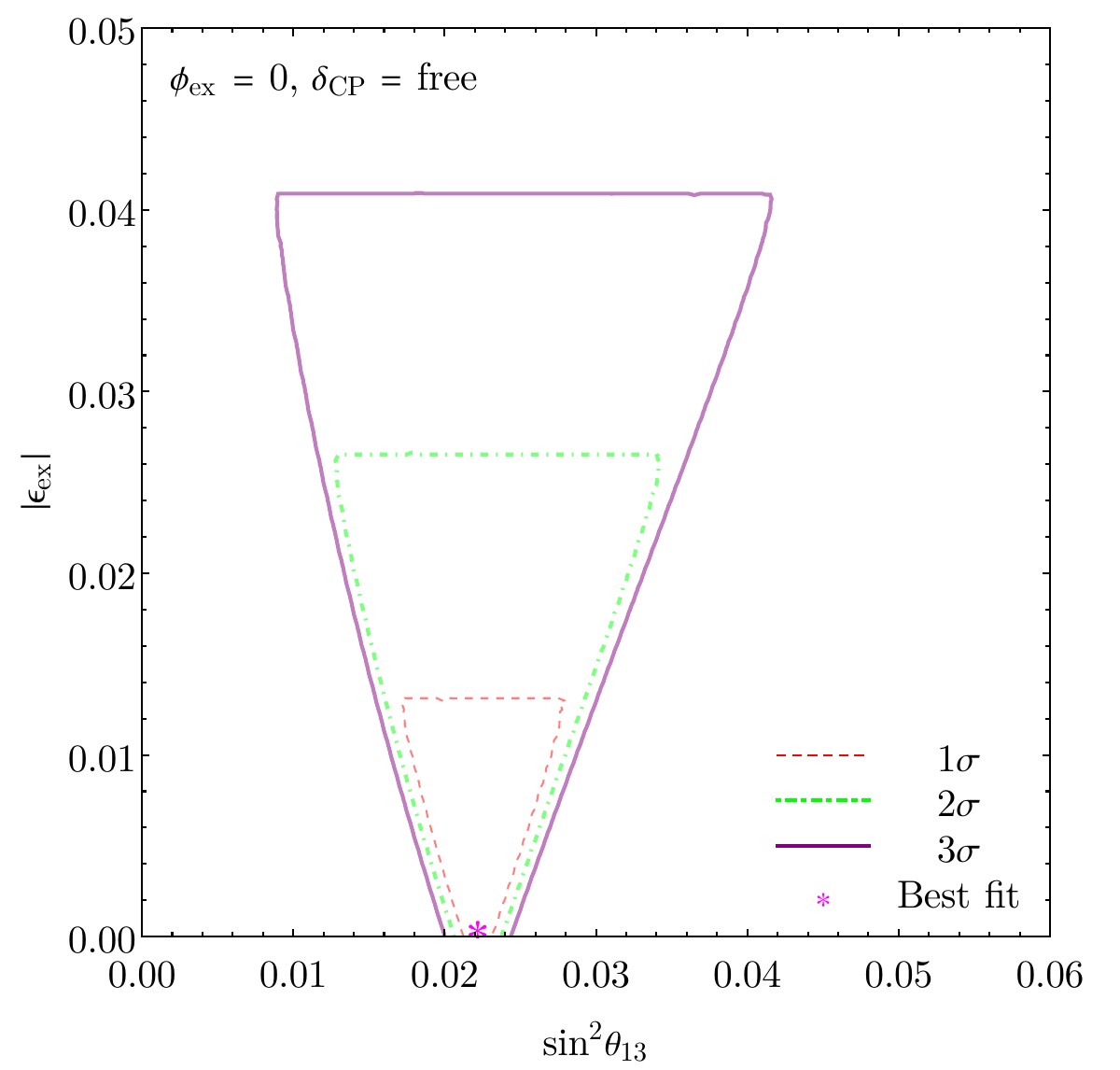}
\label{fig:s=d_ex_b}}
\par\end{centering}
\begin{centering}
\subfigure[]{\includegraphics[width=7.0cm]{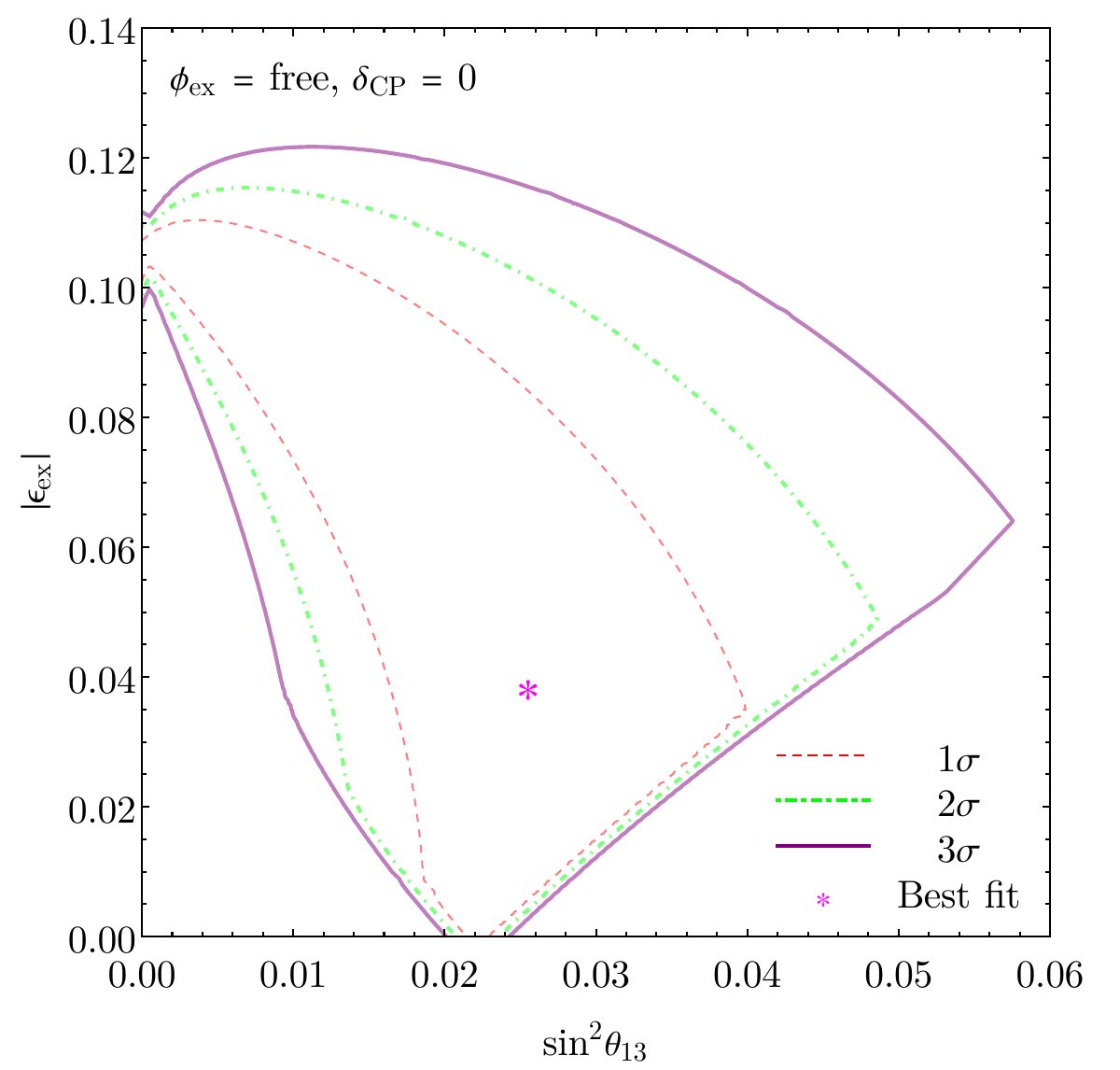}
\label{fig:s=d_ex_c}}
\subfigure[]{\includegraphics[width=6.9cm]{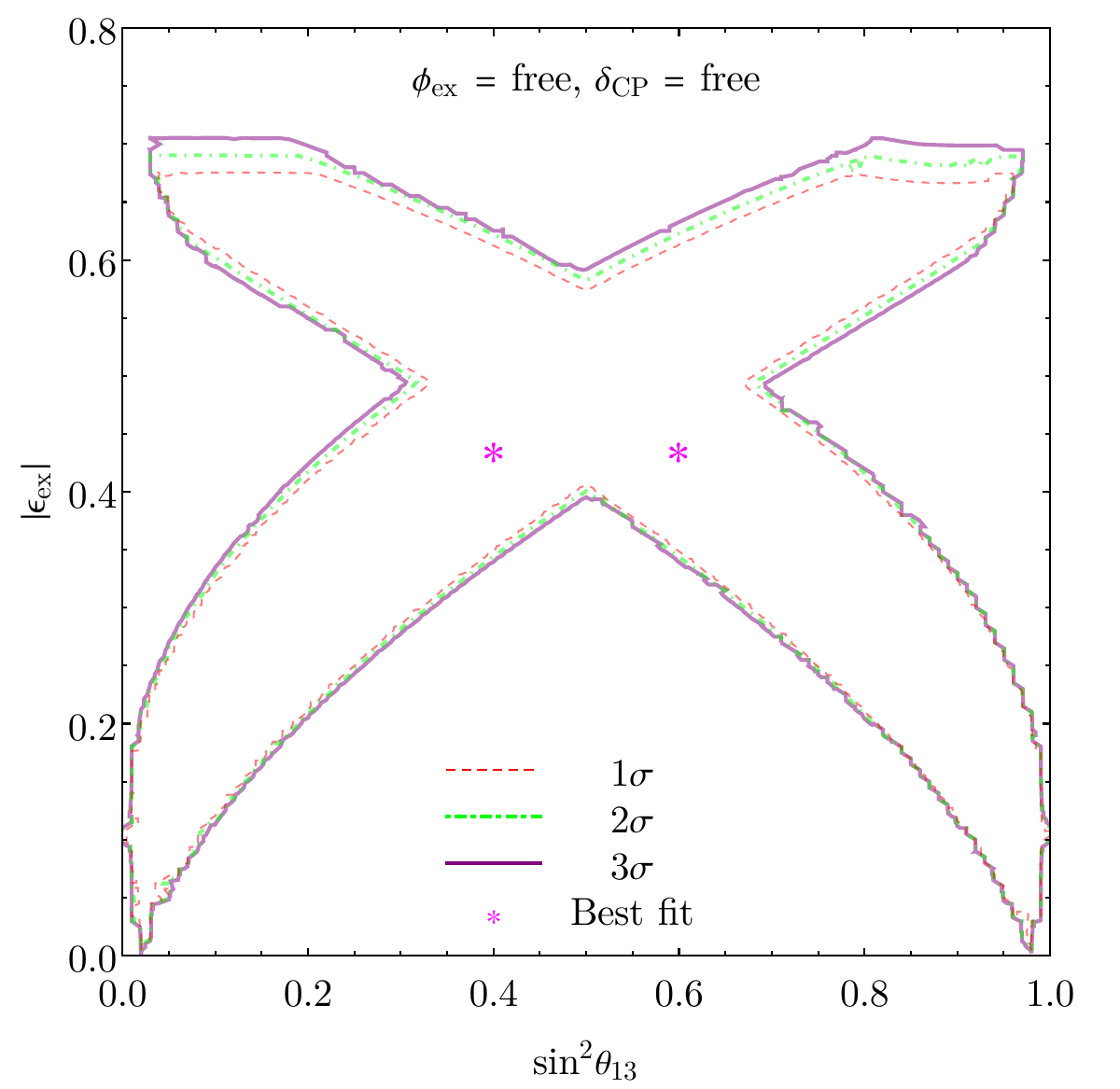}
\label{fig:s=d_ex_d}}
\par\end{centering}
\caption{Allowed regions in the $(\sin^{2}\theta_{13},\left|\epsilon_{ex}\right|)$
plane for $\phi_{ex}=\delta_{\textrm{CP}}=0$, marginalizing over
$\delta_{\text{CP}}$ ($\delta_{\text{CP}}=$free) while $\phi_{ex}=0$, over $\phi_{ex}$
($\phi_{ex}=$free) while $\delta_{\text{CP}}=0$ and over both phases ($\delta_{\text{CP}}=$free,
$\phi_{ex}=$free) as indicated in the plots. Details of the analysis can be found in section \ref{subsec:Constraint-epsilon_ex}.
\label{fig:s=d_ex}}
\end{figure*}

\subsubsection{Constraints on flavor-universal NSI coupling $\epsilon_{ex}$
\label{subsec:Constraint-epsilon_ex}}

The universal NSI parameter $\epsilon_{ex}\equiv\epsilon_{ee}=\epsilon_{e\mu}=\epsilon_{e\tau}$
associates the electron (or positron) with all three flavors of neutrinos
with the same strength in both production and detection processes.
We have $U_{ej}^{sd}=U_{ej}+\epsilon_{ex}^{*}\sum_{\alpha}U_{\alpha j}$
which can be seen as a combination of the three cases considered above
for $\epsilon_{ee}$, $\epsilon_{e\mu}$ and $\epsilon_{e\tau}$.
Similar to the case with $\epsilon_{e\mu},$ the effective survival
probability depends on $\delta_{\textrm{CP}}$ and $\phi_{ex}$ in
the form of $\cos(\phi_{ex})$, $\cos(\delta_{\textrm{CP}})$ and
$\cos(\delta_{\textrm{CP}}-\phi_{ex})$ with degeneracy when either $\phi_{ex}$ or $\delta_{\textrm{CP}}$ is $\pi/2$ and $3\pi/2$  and the other phase is zero. 
However, the roles $\delta_{\textrm{CP}}$ and $\phi_{e\mu}$ play are different as seen explicitly in the expression up to the first order in $\left|\epsilon_{ex}\right|$ \citep{Agarwalla:2014bsa}
\begin{equation}
\sin^{2}\tilde{\theta}_{13} \approx\sin^{2}\theta_{13}+2\sin\theta_{13}(\sin\theta_{23}+\cos\theta_{23})\left|\epsilon_{ex}\right|\cos(\delta_{\textrm{CP}}-\phi_{ex})-\frac{\left|\epsilon_{ex}\right|\cos\phi_{ex}}{\sin^{2}(\Delta m_{31}^{2}L_\nu/(4E_\nu))}.\label{eq:sin13eff_ex}
\end{equation}
The allowed regions in the $(\sin^{2}\theta_{13},\left|\epsilon_{ex}\right|)$
plane are similar to those in the $(\sin^{2}\theta_{13},\left|\epsilon_{e\mu}\right|)$
plane, but with much stronger constraints on $\left|\epsilon_{ex}\right|$ when either $\delta_{\textrm{CP}}$ or $\phi_{ex}$ is zero. The effective survival probability in this case is also approximately invariant under the exchange of $\theta_{13} \leftrightarrow \pi/2-2\tilde{\theta}_{13}+\theta_{13}$ or $\theta_{13} \leftrightarrow \pi/2-\theta_{13}$, depending on the values of  $\phi_{ex}$ and $\delta_{\textrm{CP}}$. We provide allowed regions in the $(\sin^{2}\theta_{13},\left|\epsilon_{ex}\right|)$
plane around small $\theta_{13}$ when possible. Figure \ref{fig:s=d_ex_a} shows the allowed region when both $\phi_{ex}$ and $\delta_{\textrm{CP}}$ equal zero. The allowed region when $\delta_{\textrm{CP}}$ ($\phi_{ex}$)
varies freely with $\phi_{ex}$ ($\delta_{\textrm{CP}}$) set to zero is the combination of the allowed regions
of the corresponding phase being in the range $[0,2\pi)$.
The plots are shown in figures \ref{fig:s=d_ex_b} and \ref{fig:s=d_ex_c}, respectively. For the different dependence on the
two phases $\delta_{\text{CP}}$ and $\phi_{ex}$, the two allowed regions
appear very different, in contrast to the case of $\epsilon_{e\mu}$.
The constraint on $\left|\epsilon_{ex}\right|$ is much relaxed when
both $\delta_{\textrm{CP}}$ and $\phi_{ex}$ are marginalized over as can be seen in figure \ref{fig:s=d_ex_d} where the allowed regions in the small and large $\theta_{13}$ merge to a single one and appears symmetric under $\theta_{13} \leftrightarrow \pi/2-\theta_{13}$. The numerical
values of the constraints on $\left|\epsilon_{ex}\right|$ are listed in Table \ref{tab:s=d_emu__etau_ex}. 
\begin{figure*}
\begin{centering}
\subfigure[]{\includegraphics[width=7.0cm]{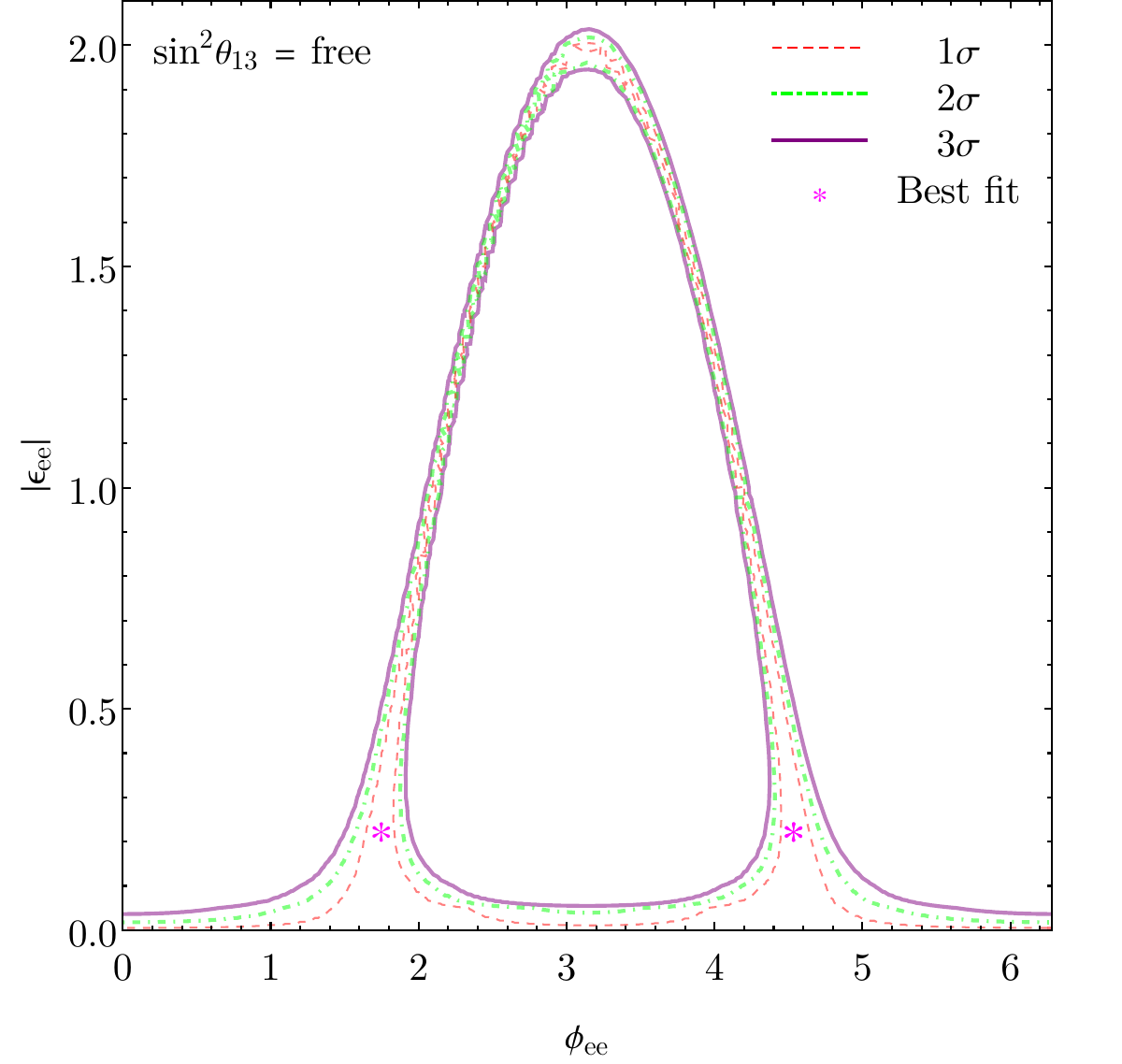}
\label{fig:QM-NSI-phi-epsilon_a}}
\subfigure[]{\includegraphics[width=7.0cm]{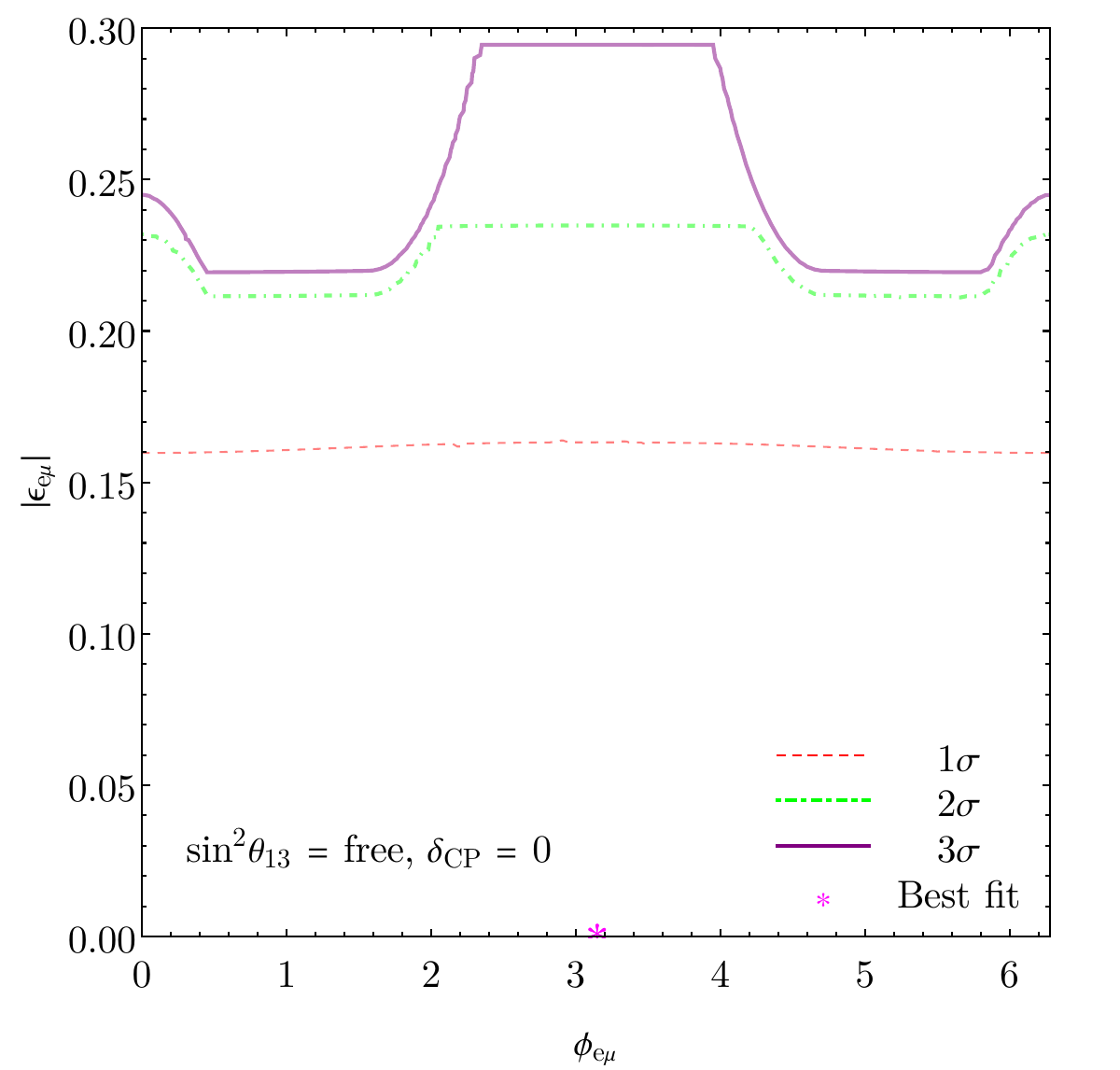}
\label{fig:QM-NSI-phi-epsilon_b}}
\par\end{centering}
\begin{centering}
\subfigure[]{\includegraphics[width=7.0cm]{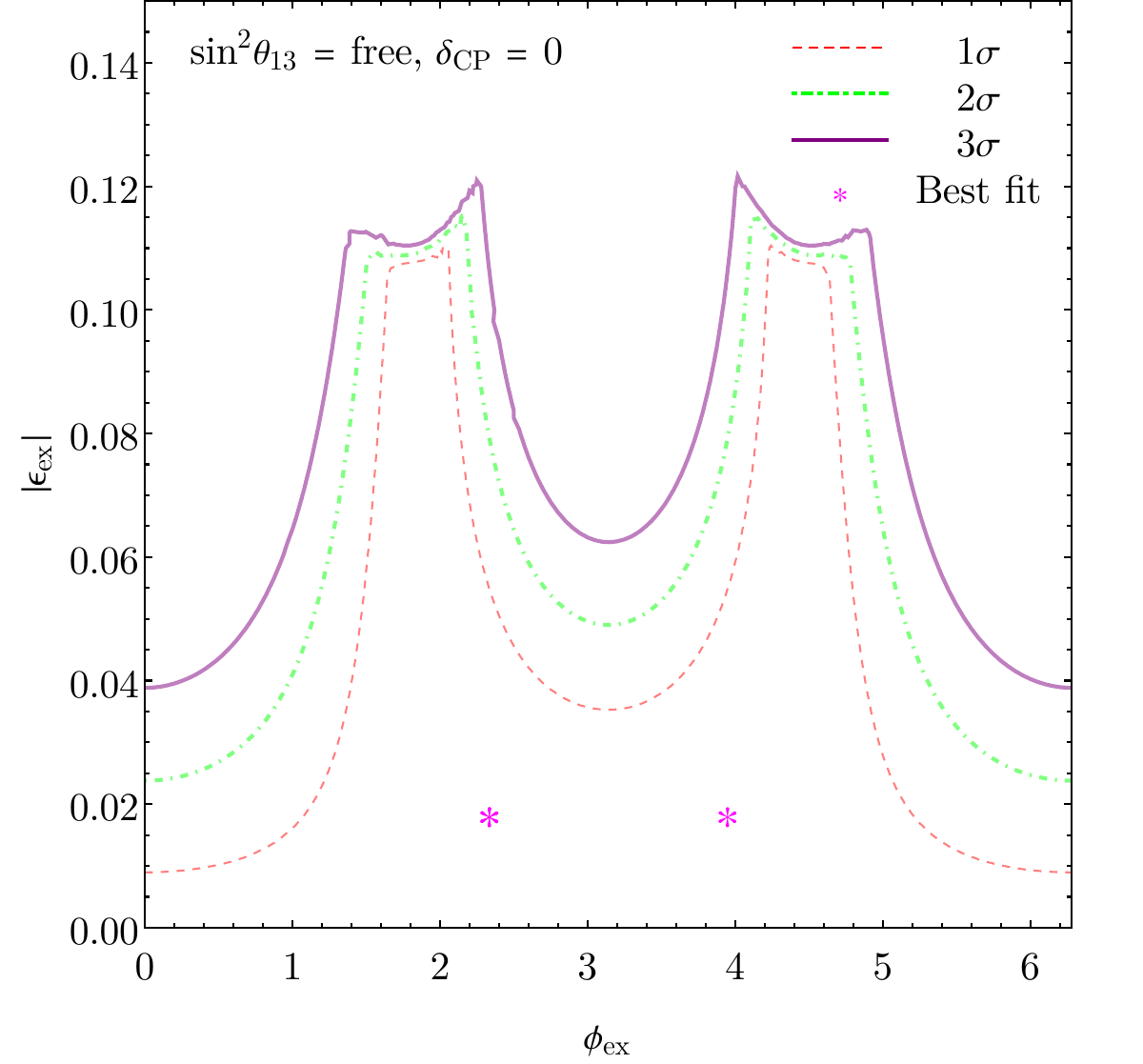}
\label{fig:QM-NSI-phi-epsilon_c}}
\par\end{centering}
\caption{Allowed region in the $(\phi_{e\alpha},\left|\epsilon_{e\alpha}\right|)$
plane marginalizing over $\sin^{2}\theta_{13}$ ($\sin^{2}\theta_{13}=$free)
for $\epsilon_{ee}$ (\ref{fig:QM-NSI-phi-epsilon_a}) and for $\delta_{\text{CP}}=0$ for
$\epsilon_{e\mu}$ (\ref{fig:QM-NSI-phi-epsilon_b}) and $\epsilon_{ex}$ (\ref{fig:QM-NSI-phi-epsilon_c}), respectively. The allowed region for $\epsilon_{e\tau}$
is similar to that of $\epsilon_{e\mu}$. Details of the analysis can be found in section \ref{subsec:Constraint-phi_epsilon}.
\label{fig:QM-NSI-phi-epsilon}}
\end{figure*}

\subsubsection{Allowed regions in $(\phi_{e\alpha},\left|\epsilon_{e\alpha}\right|)$ plane
\label{subsec:Constraint-phi_epsilon}}

We also determine the allowed regions in the $(\phi_{e\alpha},\left|\epsilon_{e\alpha}\right|)$
plane for $\delta_{\text{CP}}=0$ with $\sin^{2}\theta_{13}$ left to
vary freely. The plots are shown in figures \ref{fig:QM-NSI-phi-epsilon_a}, \ref{fig:QM-NSI-phi-epsilon_b} and \ref{fig:QM-NSI-phi-epsilon_c} for $\epsilon_{ee}$, $\epsilon_{e\mu}$ and $\epsilon_{ex}$, respectively. The shapes of the allowed regions can be understood by referring to the corresponding plots in
the $(\sin^{2}\theta_{13},\left|\epsilon_{e\alpha}\right|)$ plane.
For example, consider the allowed regions for $\epsilon_{ee}$ in figure \ref{fig:s=d_ee}. At $\phi_{ee}=0$,
the upper limit at $3\sigma$ on $\left|\epsilon_{ee}\right|$ is less than about $0.043$. At $\phi_{ee}=\pi/2$
or $3\pi/2$, the upper limit is no larger than about $0.3$. While for $\phi_{ee}=\pi$,
it reaches its peak and is just less than about $2.1$. All these
features can be read off directly from figures \ref{fig:s=d_ee_a}, \ref{fig:s=d_ee_b} and \ref{fig:s=d_ee_c}. For $\epsilon_{e\mu}$, the plots for $\delta_{\textrm{CP}}=0$
and $\phi_{e\mu}=\pi/2,\pi$ or $3\pi/2$ are almost the same as those
for $\phi_{e\mu}=0$ and $\delta_{\textrm{CP}}=\pi/2,\pi$ or $3\pi/2$
which are shown in figure \ref{fig:s=d_emu}. The constraint on
$\left|\epsilon_{e\mu}\right|$ at $\delta_{\textrm{CP}}=0$ and $\phi_{e\mu}=0$
is a little weaker than those at around $\delta_{\textrm{CP}}=0$
and $\phi_{e\mu}=\pi/2$ or $3\pi/2$. This is so because at $\delta_{\textrm{CP}}=0$
and $\phi_{e\mu}=0$ the constraint on $\left|\epsilon_{e\mu}\right|$
is relaxed a little bit when $\sin^{2}\theta_{13}\sim1$. The features for $\epsilon_{ex}$ is understood in a similar way. We note that the allowed regions in the $(\phi_{e\alpha},\left|\epsilon_{e\alpha}\right|)$ plane in figure \ref{fig:QM-NSI-phi-epsilon} is symmetric under the exchange $\phi_{e\alpha} \leftrightarrow 2\pi-\phi_{e\alpha}$ which arises from the effective survival probability depending on the phases in the form of $\cos(\phi_{e\alpha})$ when $\delta_{\text{CP}}=0$. The constraints on the magnitude of the NSI parameter $\left|\epsilon_{e\alpha}\right|$ obtained from the plots in the $(\phi_{e\alpha},\left|\epsilon_{e\alpha}\right|)$ plane are the same as those from the plots in the $(\sin^{2}\theta_{13},\left|\epsilon_{e\alpha}\right|)$ plane with $\delta_{\text{CP}}=0$ and the corresponding phase marginalized over. As to the NSI phases $\phi_{ee}$, $\phi_{e\mu}$ and $\phi_{ex}$, we see from figure \ref{fig:QM-NSI-phi-epsilon} that they are unconstrained for $\delta_{\text{CP}}=0$ and $\sin^{2}\theta_{13}$ varying freely. The allowed regions related to $\epsilon_{e\tau}$ are similar to those of $\epsilon_{e\mu}$.
\begin{table}
\centering{}
\begin{tabular}{|c|c|c|c|}
\hline 
$(\phi_{e\alpha},\delta_{\text{CP}})$ & $\left|\epsilon_{e\mu}\right|$ & $\left|\epsilon_{e\tau}\right|$ & $\left|\epsilon_{ex}\right|$\tabularnewline
\hline 
$(0,0)$ & $\left|\epsilon_{e\mu}\right|<0.165$ & $\left|\epsilon_{e\tau}\right|<0.171$ & $\left|\epsilon_{ex}\right|<0.0145$\tabularnewline
$(0,\textrm{free})$ & $\left|\epsilon_{e\mu}\right|<0.171$ & $\left|\epsilon_{e\tau}\right|<0.174$ & $\left|\epsilon_{ex}\right|<0.0146$\tabularnewline
$(\textrm{free},0)$ & $\left|\epsilon_{e\mu}\right|<0.174$ & $\left|\epsilon_{e\tau}\right|<0.174$ & $\left|\epsilon_{ex}\right|<0.110$\tabularnewline
$(\textrm{free},\textrm{free})$ & $\left|\epsilon_{e\mu}\right|<0.174$ & $\left|\epsilon_{e\tau}\right|<0.174$ & $\left|\epsilon_{ex}\right|<0.678$\tabularnewline
\hline 
\end{tabular}
\caption{$90\%$ C.L. constraints (1 d.o.f) on the QM-NSI parameters $\left|\epsilon_{e\mu}\right|$,
$\left|\epsilon_{e\tau}\right|$ and $\left|\epsilon_{ex}\right|$
projected from the $(\sin^{2}\theta_{13},\left|\epsilon_{e\alpha}\right|)$
plane for the phases $\phi_{e\alpha}$ and $\delta_{\text{CP}}$ taking on
different values and being marginalized over ($(\phi_{e\alpha},\delta_{\text{CP}})$=$(\textrm{free},\textrm{free})$),
respectively. Constraints on these NSI parameters for $\phi_{e\alpha}=0$
and $\delta_{\text{CP}}=\pi/2,3\pi/2$ and $\pi$ or the other way around
are close to those for $\phi_{e\alpha}=0$ and $\delta_{\text{CP}}=0$. \label{tab:s=d_emu__etau_ex}}
\end{table}

\subsection{Constraints on QM-NSI parameter $\epsilon_{e\alpha}^{s}$ for $\epsilon_{e\alpha}^{s}\protect\neq\epsilon_{\alpha e}^{d*}$}

In the general case, $\epsilon_{e\alpha}^{s} \neq \epsilon_{\alpha e}^{d*}$.
We assume they are independent and discuss the effect of $\epsilon_{e\alpha}^{s}$.
The constraints on $\epsilon_{\alpha e}^{d}$ can be obtained from eq.\,(\ref{eq:QM-NSI-s-d-symmetry}). The effective survival probability is still approximately invariant under the exchange of $\theta_{13} \leftrightarrow \pi/2-2\tilde{\theta}_{13}+\theta_{13}$ or $\theta_{13} \leftrightarrow \pi/2-\theta_{13}$, depending on the values of $\phi_{e\alpha}^s$ and/or $\delta_{\textrm{CP}}$. We focus on the allowed regions in the $(\sin^{2}\theta_{13},\left|\epsilon^s_{e\alpha}\right|)$ plane around small $\theta_{13}$ when possible. The dependence of the constraints on the systematical and statistical uncertainties is similar to the case of $\epsilon_{e\alpha}^{s}=\epsilon_{\alpha e}^{d*}$.

\subsubsection{Constraints on electron-NSI coupling $\epsilon_{ee}^{s}$
\label{subsec:Constraint-epsilon_s_ee}}

\begin{figure*}
\begin{centering}
\subfigure[]{\includegraphics[width=7.0cm]{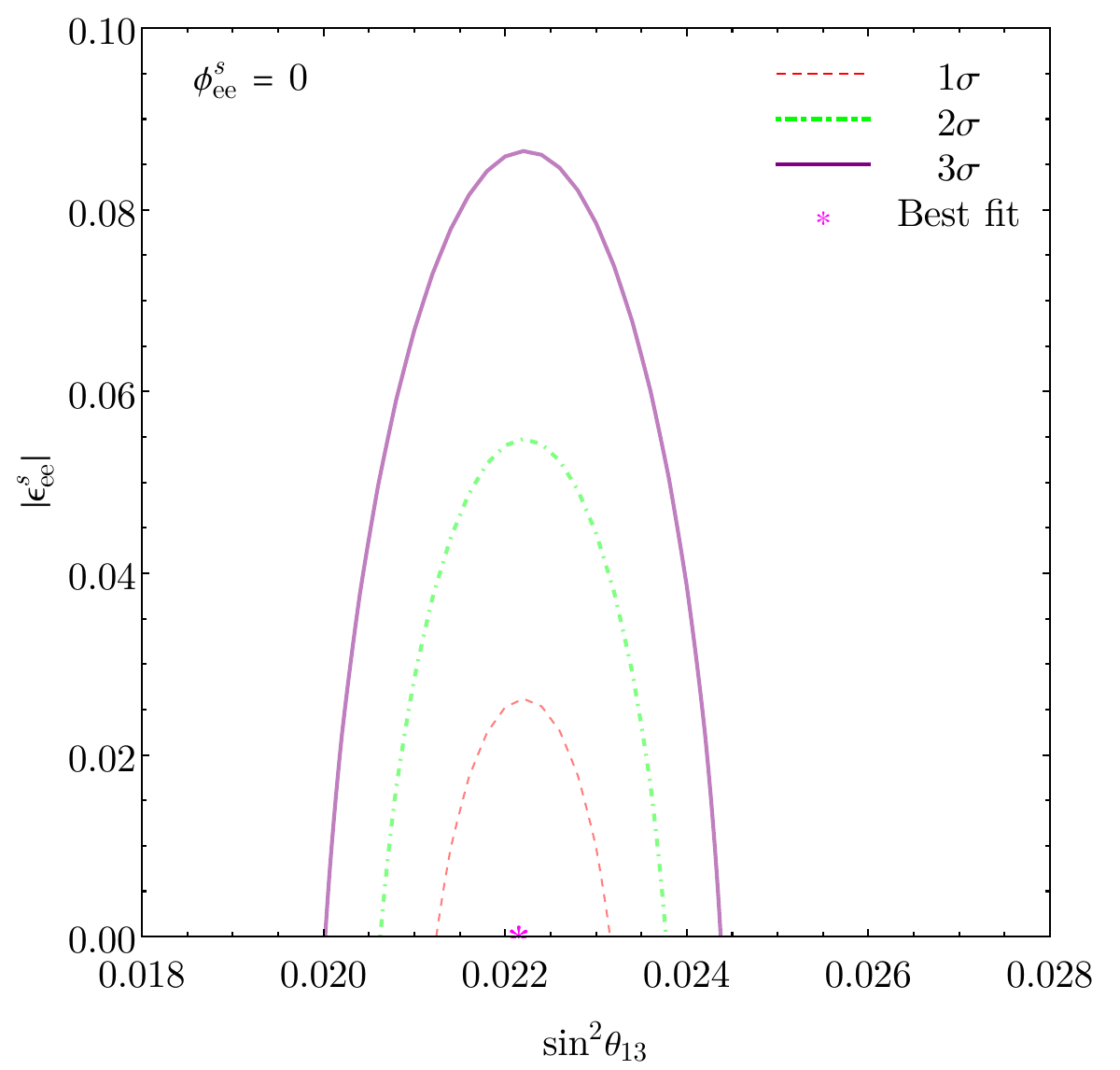}
\label{fig:s!=d_s_ee_a}}
\subfigure[]{\includegraphics[width=6.8cm]{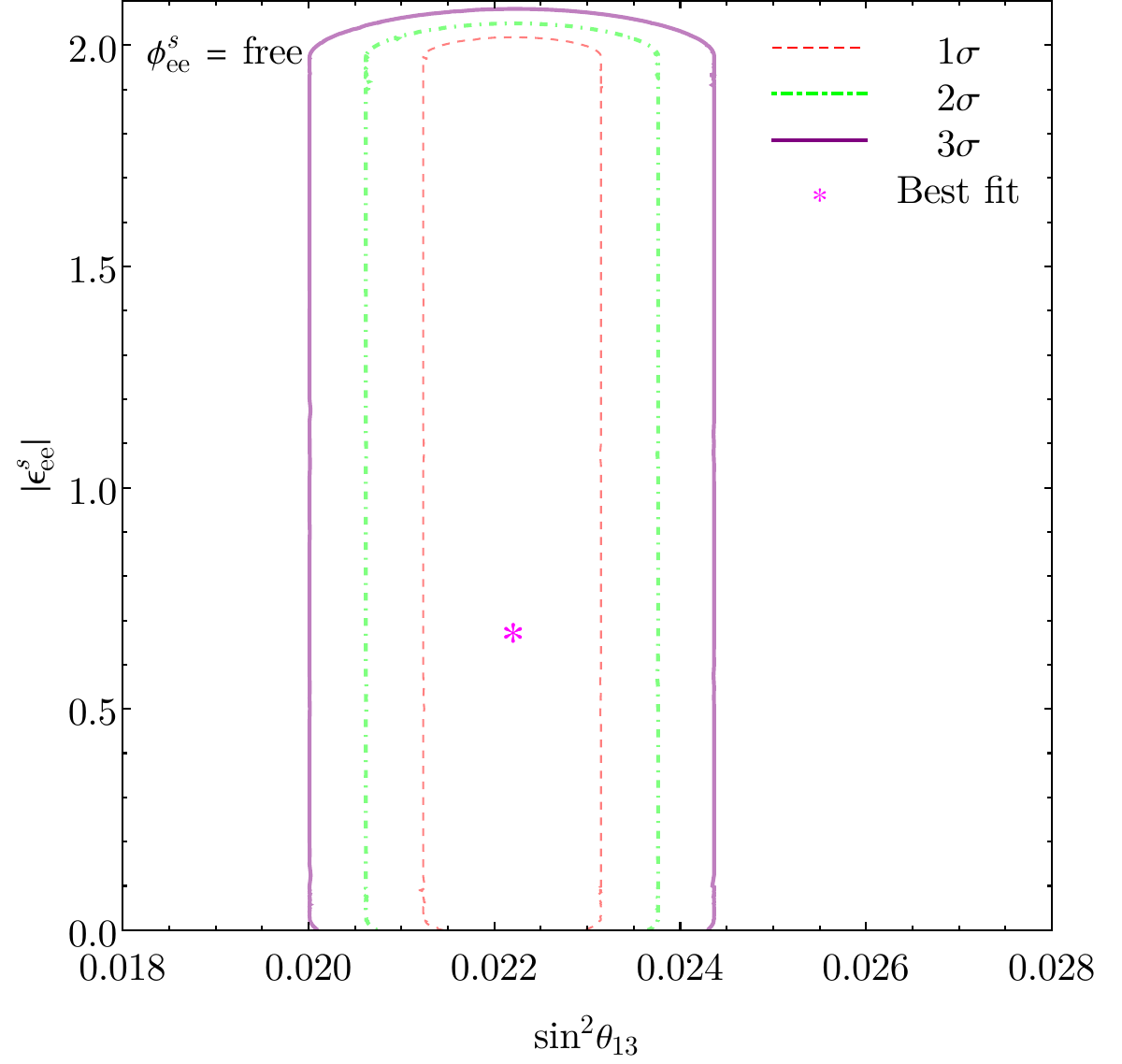}
\label{fig:s!=d_s_ee_b}}
\par\end{centering}
\caption{Allowed regions in the $(\sin^{2}\theta_{13},\left|\epsilon_{ee}^{s}\right|)$
plane for $\phi_{ee}^{s}=0$ (left panel) and for it being marginalized
over ($\phi_{ee}^{s}=$free, right panel). The situation is similar
to the case with $\epsilon_{ee}$ but with the constraints less stringent. Details of the analysis can be found in section \ref{subsec:Constraint-epsilon_s_ee}.
\label{fig:s!=d_s_ee}}
\end{figure*}

The non-universal NSI parameter $\epsilon_{ee}^{s}$ associates the
electron with $\bar{\nu}_{e}$ in the production processes and thus
conserves lepton flavor. We find
\begin{equation}
P_{\bar{\nu}_{e}^{s}\rightarrow\bar{\nu}_{e}^{d}}^{\textrm{QM-NSI-eff}}(\epsilon_{ee}^{s},\epsilon_{e\alpha}^{d}=0)
=(1+\left|\epsilon_{ee}^{s}\right|^{2}+2\left|\epsilon_{ee}^{s}\right|\cos\phi_{ee}^{s})P_{\bar{\nu}_{e}\rightarrow\bar{\nu}_{e}}^{\textrm{std}}.
\end{equation}
It can be seen that this effective survival probability is the same in form to that with $\epsilon_{ee}$ except that the power of the factor $(1+\left|\epsilon_{ee}^{s}\right|^{2}+2\left|\epsilon_{ee}^{s}\right|\cos\phi_{ee}^{s})$
is one, while it is two for $\epsilon_{ee}$ as can be seen from eq.\,(\ref{eq:s=d_ee}). Two consequences follow.
Firstly, the pattern of the allowed regions is similar to that with
$\epsilon_{ee}$. Secondly, the allowed ranges on $\left|\epsilon_{ee}^{s}\right|$
must be larger than those with $\left|\epsilon_{ee}\right|$. These results can be seen from comparing figures \ref{fig:s!=d_s_ee_a} and \ref{fig:s!=d_s_ee_b} with  figures \ref{fig:s=d_ee_a} and  \ref{fig:s=d_ee_d}, respectively, or from comparing the numerical values in Tables \ref{tab:s=d_ee} and \ref{tab:s!=d_see}. As for the case of $\epsilon_{ee}$, the Daya Bay experimental data is consistent with the standard oscillation framework ($\left|\epsilon^s_{ee}\right|=0$) within 1$\sigma$ C.L..
\begin{table}
\centering{}
\begin{tabular}{|c|c|}
\hline  
$\phi_{ee}^{s}$ & $\left|\epsilon_{ee}^{s}\right|$\tabularnewline
\hline 
$0$ & $\left|\epsilon_{ee}^{s}\right|<0.0298$\tabularnewline
$\pi/2,3\pi/2$ & $\left|\epsilon_{ee}^{s}\right|<0.246$\tabularnewline
$\pi$ & $\left|\epsilon_{ee}^{s}\right|<0.0702$ or $1.93<\left|\epsilon_{ee}\right|<2.02$\tabularnewline
free & $\left|\epsilon_{ee}^{s}\right|<2.02$\tabularnewline
\hline 
\end{tabular}
\caption{$90\%$ C.L. constraints (1 d.o.f) on the QM-NSI parameter $\left|\epsilon_{ee}^{s}\right|$
projected from the $(\sin^{2}\theta_{13},\left|\epsilon_{ee}^{s}\right|)$
plane for $\phi_{ee}^{s}$ taking on values of $0,\pi/2,\pi$, $3\pi/2$
and being marginalized over ($\phi_{ee}^{s}=$ free), respectively.
\label{tab:s!=d_see}}
\end{table}

\subsubsection{Constraints on muon-NSI and tau-NSI couplings $\epsilon_{e\mu}^{s}$
and $\epsilon_{e\tau}^{s}$ 
\label{subsec:Constraint-epsilon_s_emu}}

\begin{figure*}
\begin{centering}
\subfigure[]{\includegraphics[width=7.0cm]{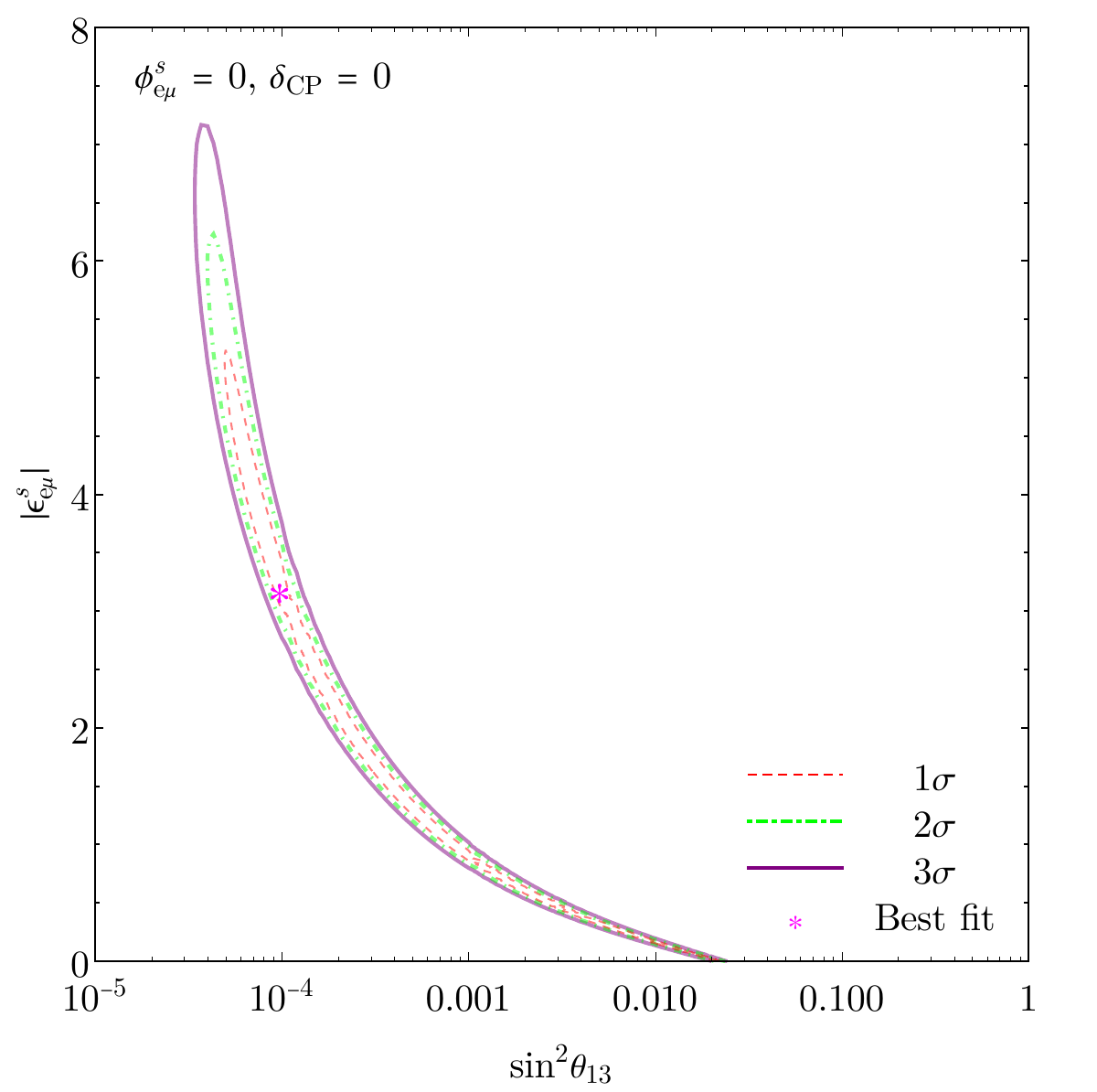}
\label{fig:s!=d_semu_a}}
\par\end{centering}
\begin{centering}
\subfigure[]{\includegraphics[width=7.0cm]{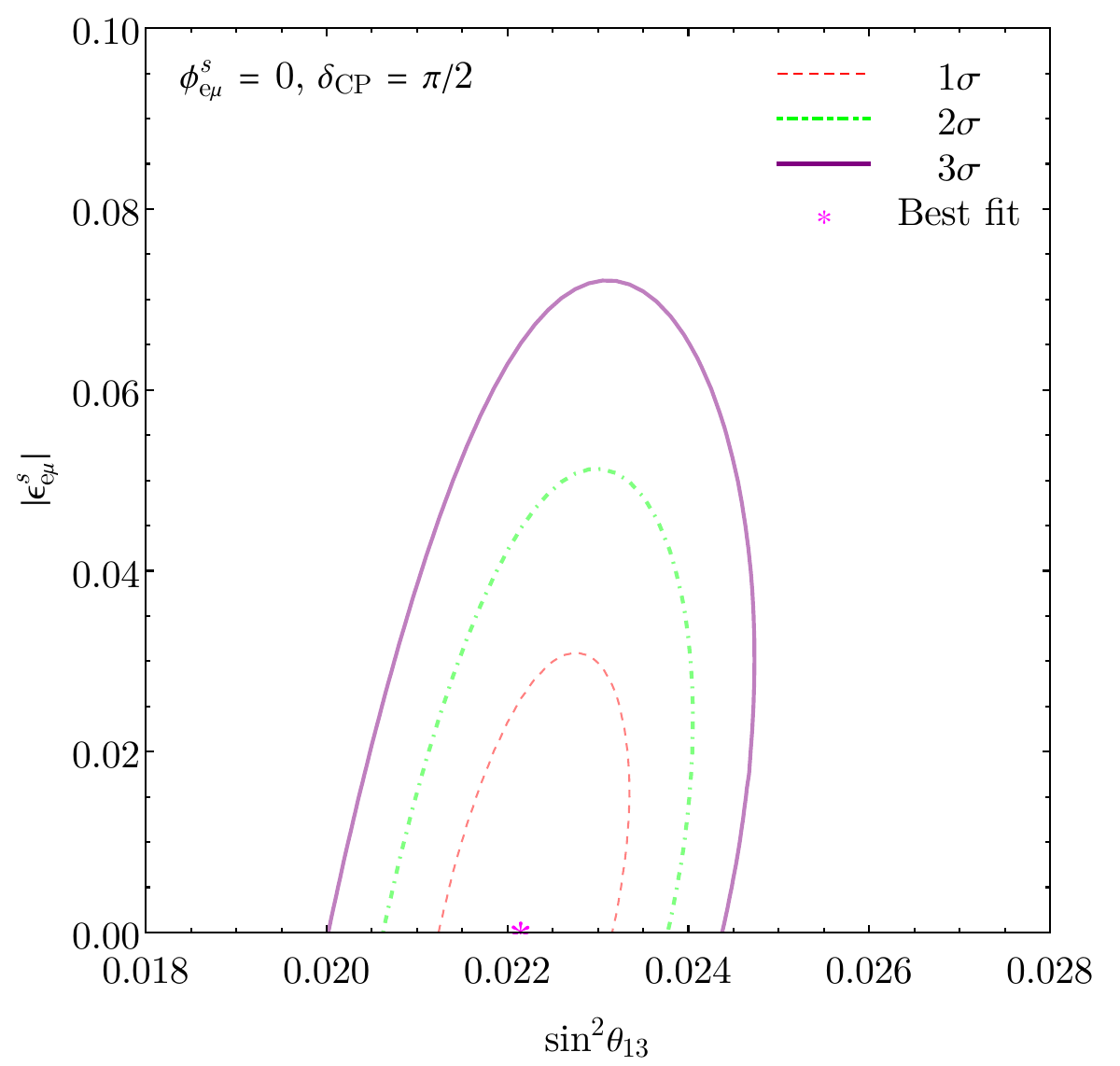}
\label{fig:s!=d_semu_b}}
\subfigure[]{\includegraphics[width=7.0cm]{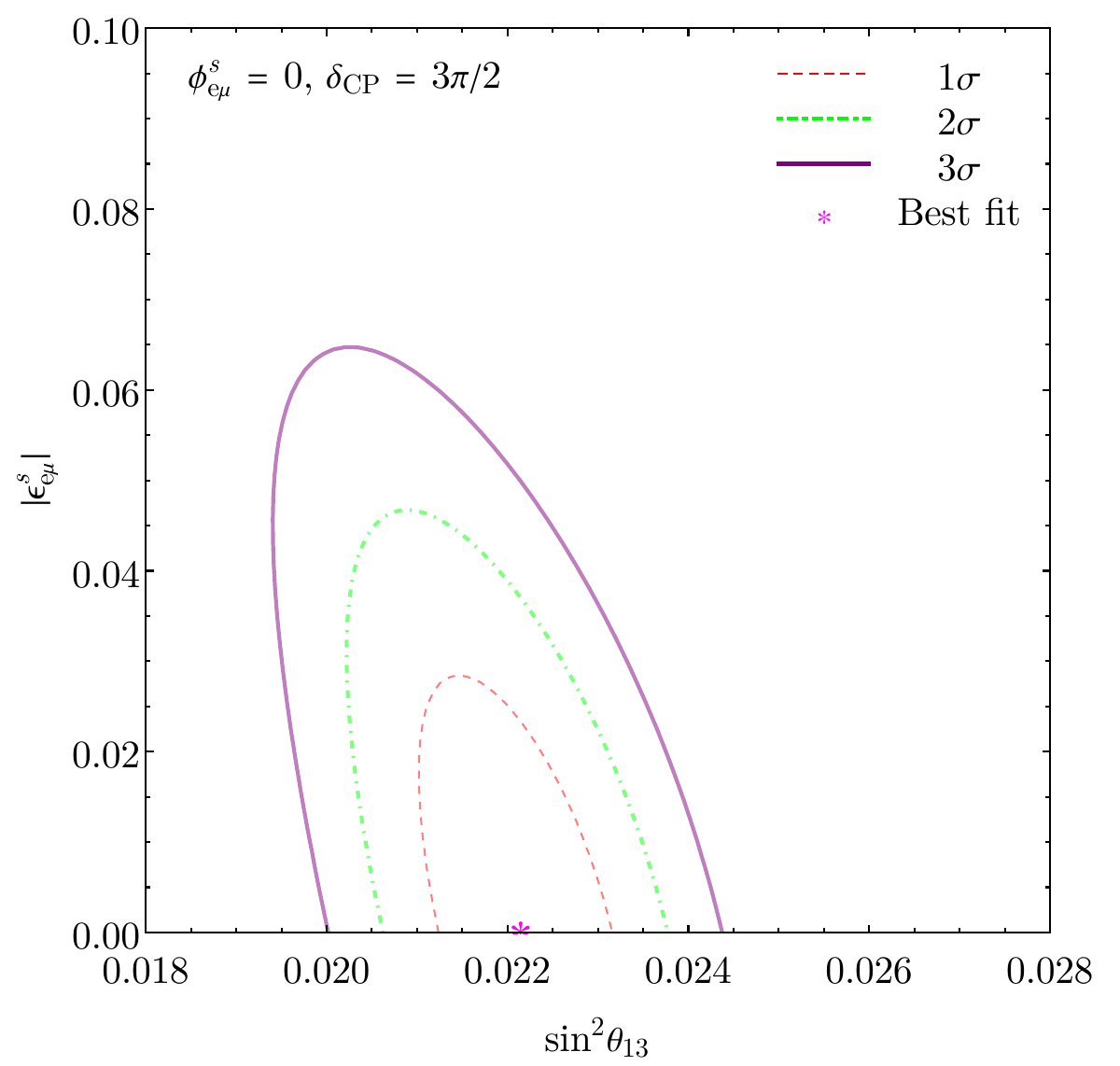}
\label{fig:s!=d_semu_c}}
\par\end{centering}
\caption{The dependence of the allowed regions in the $(\sin^{2}\theta_{13},\left|\epsilon_{e\mu}^{s}\right|)$
plane on the values of the phases $\phi_{e\mu}^{s}$ and $\delta_{\textrm{CP}}$
for $\phi_{e\mu}^{s}=0$ and $\delta_{\text{CP}}=0,\pi/2$ and $3\pi/2$,
respectively. The degeneracy for $\delta_{\text{CP}}=\pi/2$ and $3\pi/2$
is broken. The corresponding allowed regions for $\delta_{\text{CP}}=0$
and $\phi_{e\mu}^{s}=\pi/2$ and $3\pi/2$, respectively, are similar.
The parameter $\left|\epsilon_{e\mu}^{s}\right|$ is not constrained
for the case of $\phi_{e\mu}^{s}=0$ and $\delta_{\text{CP}}=\pi$ or $\phi_{e\mu}^{s}=\pi$
and $\delta_{\textrm{CP}}=0$ and thus not constrained when either
or both phases are marginalized over. Details of the analysis can be found in section \ref{subsec:Constraint-epsilon_s_emu}.
\label{fig:s!=d_semu}}
\end{figure*}

The neutrino NSI with parameter $\epsilon_{e\mu}^{s}$ associates
the electron with $\bar{\nu}_{\mu}$ in the production processes and
thus is non-universal and violates the lepton family number conservation. The effective survival probability valid to first order in $\left|\epsilon_{e\mu}^{s}\right|$ is helpful in interpreting the behavior of the allowed regions. We have \citep{Agarwalla:2014bsa}
\begin{align}
P_{\bar{\nu}_{e}^{s}\rightarrow\bar{\nu}_{e}^{d}}^{\textrm{QM-NSI-eff}}(\epsilon_{e\mu}^{s},\epsilon_{e\alpha}^{d}=0) & \approx P_{\bar{\nu}_{e}\rightarrow\bar{\nu}_{e}}^{\textrm{std}}+2\sin\theta_{13}\sin\theta_{23}\left|\epsilon_{e\mu}^{s}\right|\sin(\delta_{CP}-\phi_{e\mu}^{s})\sin(\Delta m_{31}^{2}L_\nu/(2E_\nu))\nonumber \\
 & \quad-4\sin\theta_{13}\sin\theta_{23}\left|\epsilon_{e\mu}^{s}\right|\cos(\delta_{CP}-\phi_{e\mu}^{s})\sin^{2}(\Delta m_{31}^{2}L_\nu/(4E_\nu)).
\label{eq:s!=d_epsilon_s_emu_1st} 
\end{align}
For $\phi_{e\mu}^{s}=0$ and $\delta_{\textrm{CP}}=0$ or $\pi$,
we can write 
\begin{equation}
\sin^{2}\tilde{\theta}_{13}\approx\sin^{2}\theta_{13}\pm\sin\theta_{13}\sin\theta_{23}\left|\epsilon_{e\mu}^{s}\right|,
\end{equation}
where the +($-$) sign corresponds to $\delta_{\textrm{CP}}=0$ ($\pi$). Comparing to eq.\,(\ref{eq:s=d_emu_1stOrder}) for
the corresponding cases, we see $\left|\epsilon_{e\mu}^{s}\right|$
plays the same role as $2\left|\epsilon_{e\mu}\right|$ if both are
small. It turns out that the upper limit on $\left|\epsilon_{e\mu}^{s}\right|$
is indeed much larger than that on $\left|\epsilon_{e\mu}\right|$
for $\delta_{\textrm{CP}}=0$. For $\delta_{\textrm{CP}}=\pi$, the
upper limit on $\left|\epsilon_{e\mu}^{s}\right|$ increases with
$\sin^{2}\theta_{13}$ and reaches infinity at $\sin^{2}\theta_{13}=1$.
Thus no bound can be set in this case. The upper limits exist for $\phi_{e\mu}^{s}=0$
and $\delta_{\textrm{CP}}=\pi/2$ or $3\pi/2$. But compared to the case of $\epsilon_{e\mu}$, the degeneracy of the effective survival probability for either phase to take the values of $\pi/2$ and $3\pi/2$ when the other is set to zero is broken due to the dependence on the phases in the forms of $\sin\phi_{e\mu}^{s}$ and $\sin\delta_{\textrm{CP}}$ as well as $\cos\phi_{e\mu}^{s}$ and $\cos\delta_{\textrm{CP}}$. This can also be seen from eq. \,(\ref{eq:s!=d_epsilon_s_emu_1st}) which reduces to  
\begin{equation}
\sin^{2}\tilde{\theta}_{13}\approx\sin^{2}\theta_{13}\mp\sin\theta_{13}\sin\theta_{23}\left|\epsilon_{e\mu}^{s}\right|\cot(\Delta m_{31}^{2}L_\nu/(4E_\nu)),
\end{equation}
with the sign $-$(+) corresponding to $\delta_{\textrm{CP}}=\pi/2$ ($3\pi/2$). Whether or not the NSI parameter $\left|\epsilon_{e\mu}^{s}\right|$
increases (or decreases) with $\sin^{2}\theta_{13}$ for $\delta_{\textrm{CP}}=\pi/2$
(or $3\pi/2$) depends on the value of $L_\nu/E_\nu$ through $\cot(\Delta m_{31}^{2}L_\nu/(4E_\nu))$.
It turns out that $\left|\epsilon_{e\mu}^{s}\right|$ increases with
$\sin^{2}\theta_{13}$ for $\delta_{\textrm{CP}}$ or $\phi_{e\mu}^{s}=\pi/2$,
and decreases with it for $\delta_{\textrm{CP}}$ or $\phi_{e\mu}^{s}=3\pi/2$,
with the other phase set to zero. See figure \ref{fig:s!=d_semu}
for the allowed regions for these three cases. $\left|\epsilon_{e\mu}^{s}\right|$
is unconstrained when either or both phases are left
to vary freely. The situation for the NSI parameter $\epsilon_{e\tau}^{s}$
is similar. The constraints are given in Table \ref{tab:s!=d_emu__etau_ex}. 

\subsubsection{Constraints on flavor-universal NSI coupling $\epsilon_{ex}^{s}$
\label{subsec:Constraint-epsilon_s_ex}}

\begin{figure*}
\begin{centering}
\subfigure[]{\includegraphics[width=7.0cm]{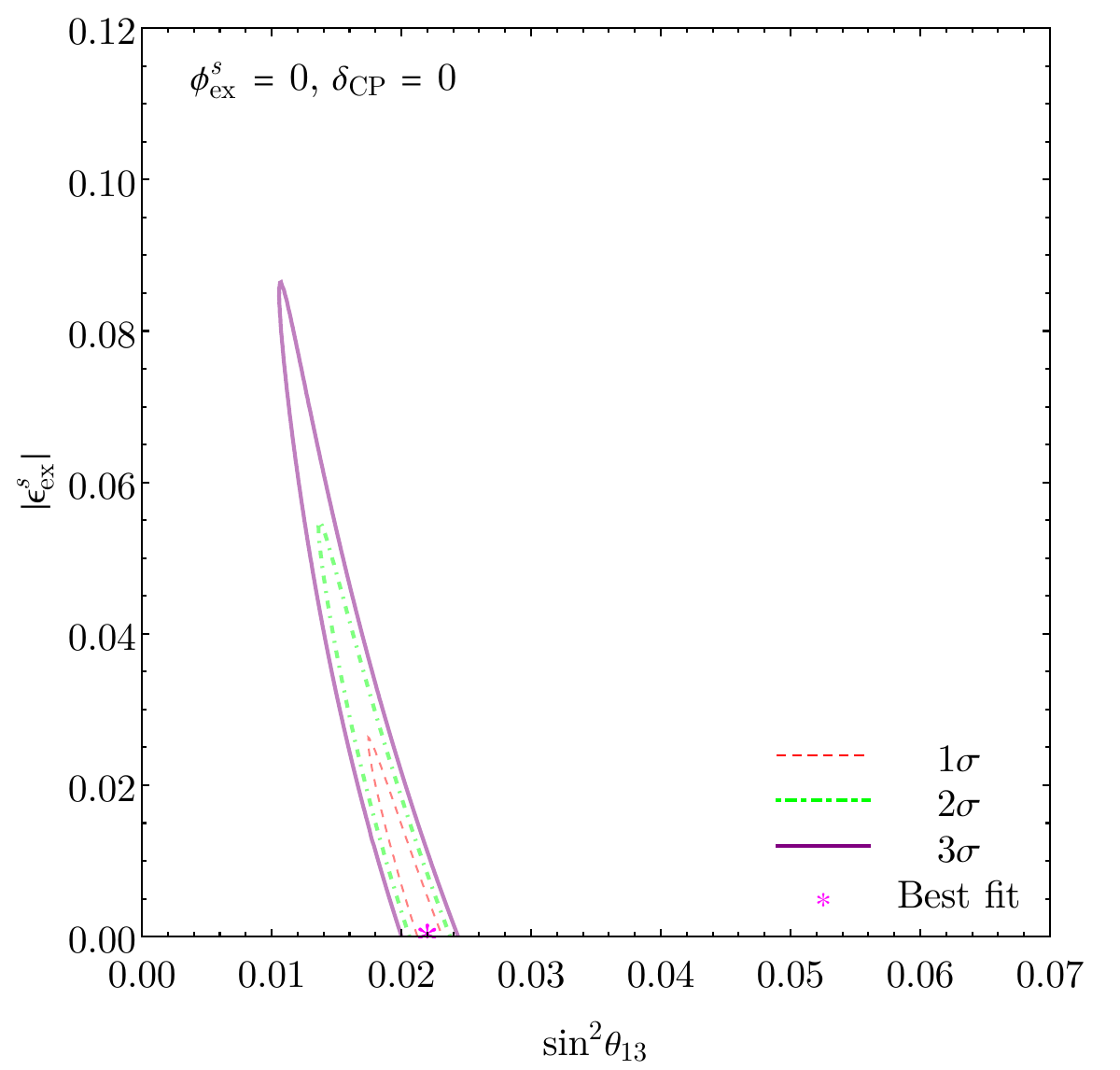}
\label{fig:s!=d_sex_a}}
\subfigure[]{\includegraphics[width=7.0cm]{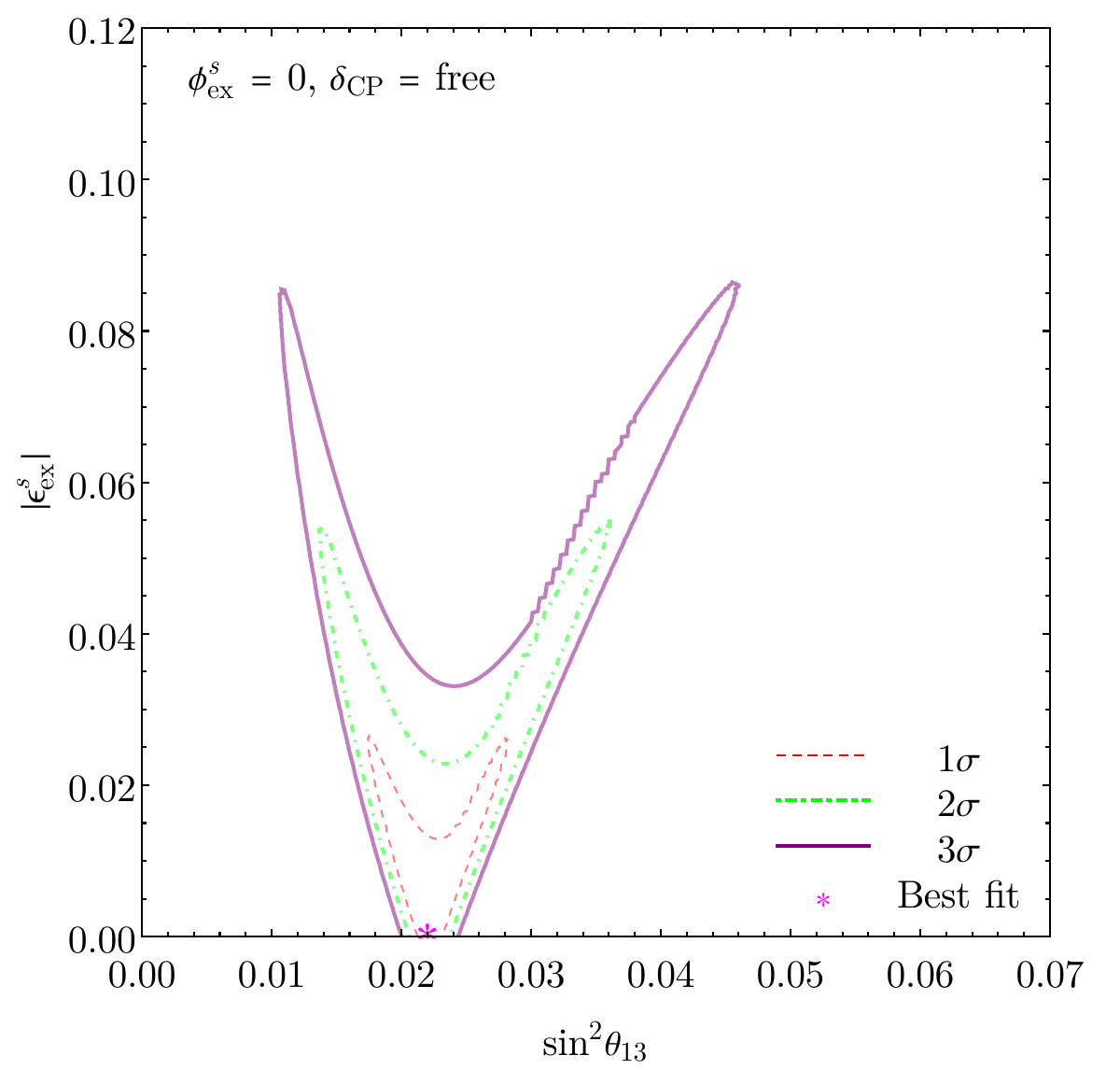}
\label{fig:s!=d_sex_b}}
\par\end{centering}
\begin{centering}
\subfigure[]{\includegraphics[width=7.0cm]{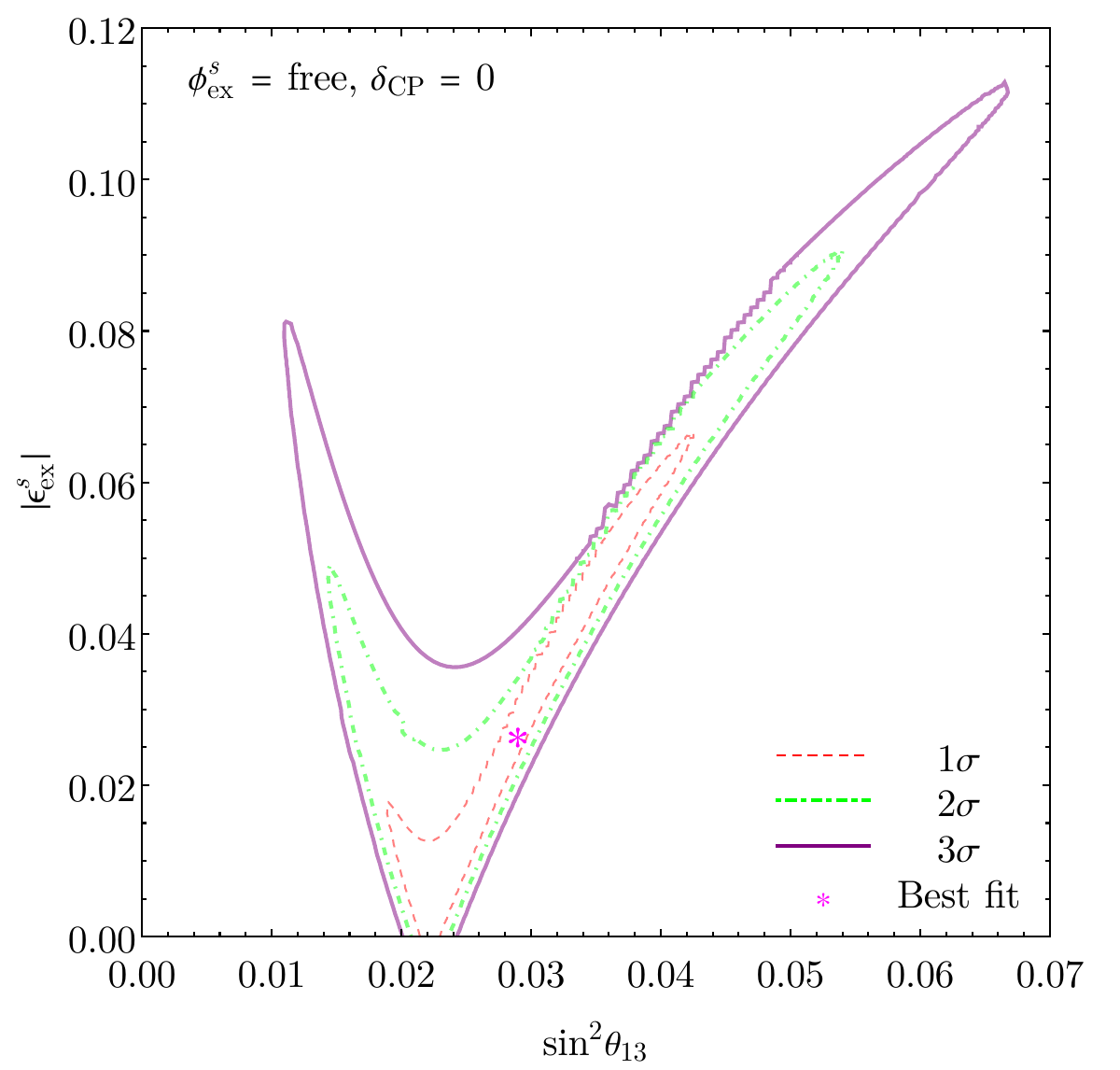}
\label{fig:s!=d_sex_c}}
\subfigure[]{\includegraphics[width=6.8cm]{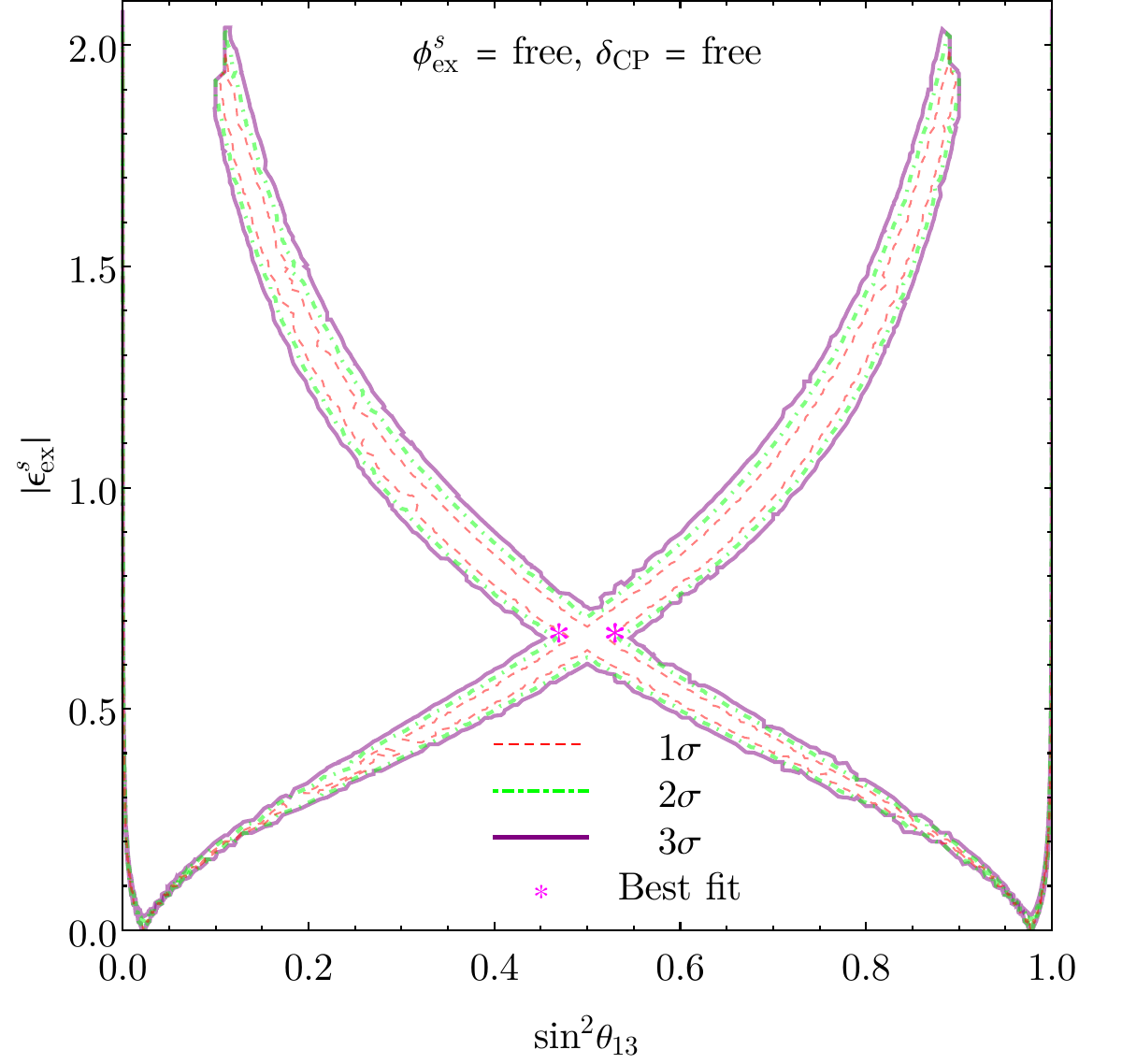}
\label{fig:s!=d_sex_d}}
\par\end{centering}
\caption{Allowed regions in the $(\sin^{2}\theta_{13},\left|\epsilon_{ex}^{s}\right|)$
plane for $\phi_{ex}^{s}=\delta_{\textrm{CP}}=0$, marginalizing over
$\delta_{\text{CP}}$ ($\delta_{\text{CP}}=$free) while $\phi_{ex}^{s}=0$, over
$\phi_{ex}^{s}$ ($\phi_{ex}^{s}=$free) while $\delta_{\text{CP}}=0$ and
over both phases ($\delta_{\text{CP}}=$ free, $\phi_{ex}^{s}=$ free) as
indicated in the plots. Details of the analysis can be found in section \ref{subsec:Constraint-epsilon_s_ex}.
\label{fig:s!=d_sex}}
\end{figure*}

The same reasoning above for $\epsilon_{e\mu}^{s}$ applies to the
universal NSI parameter $\epsilon_{ex}^{s}\equiv\epsilon_{ee}^{s}=\epsilon_{e\mu}^{s}=\epsilon_{e\tau}^{s}$.
Thus the allowed regions in the $(\sin^{2}\theta_{13},\left|\epsilon_{ex}^{s}\right|)$
plane look similar to those in the $(\sin^{2}\theta_{13},\left|\epsilon_{ex}\right|)$
plane for both $\left|\epsilon_{ex}\right|$ and $\left|\epsilon_{ex}^{s}\right|$
being small. And degeneracy of one of the phase equaling $\pi/2$
and $3\pi/2$ with the other one set to zero is broken also. The effective
survival probability to first order in $\left|\epsilon_{ex}^{s}\right|$
can be found in ref.\,\citep{Agarwalla:2014bsa}:
\begin{align}
P_{\bar{\nu}_{e}^{s}\rightarrow\bar{\nu}_{e}^{d}}^{\textrm{QM-NSI-eff}}(\epsilon_{e\mu}^{s},\epsilon_{e\alpha}^{d}=0) & \approx P_{\bar{\nu}_{e}\rightarrow\bar{\nu}_{e}}+2\left|\epsilon_{ex}^{s}\right|\cos(\phi_{ex}^{s})\nonumber \\
 & \quad+2\sin\theta_{13}(\sin\theta_{23}+\cos\theta_{23})\left|\epsilon_{ex}^{s}\right|\sin(\delta_{CP}-\phi_{ex}^{s})\sin(\Delta m_{31}^{2}L_\nu/(2E_\nu))\nonumber \\
 & \quad-4\sin\theta_{13}(\sin\theta_{23}+\cos\theta_{23})\left|\epsilon_{ex}^{s}\right|\cos(\delta_{CP}-\phi_{ex}^{s})\sin^{2}(\Delta m_{31}^{2}L_\nu/(4E_\nu)),
\end{align}
which leads to 
\begin{equation}
\sin^{2}\tilde{\theta}_{13}\approx\sin^{2}\theta_{13}\pm\left|\epsilon_{ex}^{s}\right|\left[\sin\theta_{13}(\sin\theta_{23}+\cos\theta_{23})\mp\frac{1}{2\sin^{2}(\Delta m_{31}^{2}L_\nu/(4E_\nu))}\right],
\end{equation}
for the cases $\phi_{ex}^{s}=0$ and $\delta_{\textrm{CP}}=0$ and
$\pi$, respectively. Similarly to the case of $\epsilon_{e\mu}^{s}$,
$\left|\epsilon_{ex}^{s}\right|$ plays the role of $2\left|\epsilon_{ex}\right|$,
if we compare this condition to the condition of eq.\,(\ref{eq:sin13eff_ex})
for the corresponding cases. Thus the upper limits on $\left|\epsilon_{ex}^{s}\right|$
are expected to be larger than those on $\left|\epsilon_{ex}\right|$
also. For the case of $\phi_{ex}^{s}$ or $\delta_{\textrm{CP}}$
taking on the values of $\pi/2$ or $3\pi/2$ while the other phase
set to zero, the situation depends on the value of $L_\nu/E_\nu$ through
$\cot(\Delta m_{31}^{2}L_\nu/(4E_\nu))$, as in the case of $\epsilon_{e\mu}^{s}$.
The results show that the bounds get stronger than those for the corresponding
cases of $\left|\epsilon_{ex}\right|$. These strong bounds are present
as the dips in the allowed regions for $\phi_{ex}^{s}=0$ and $\delta_{\textrm{CP}}$
varying freely or $\delta_{\textrm{CP}}=0$ and $\phi_{ex}^{s}$ varying
freely, as shown in figures \ref{fig:s!=d_sex_b} and \ref{fig:s!=d_sex_c} with the difference arises from the different
dependence on the two phases $\delta_{\textrm{CP}}$ and $\phi_{ex}^{s}$
as before. If both phases are marginalized over, the allowed region
is enormously enlarged, as can be seen in figure \ref{fig:s!=d_sex_d}. Although not very clear in figure \ref{fig:s!=d_sex_d}, 
the data is consistent less than $1\sigma$ C.L. with the standard oscillation framework ($\left|\epsilon^s_{ex}\right|=0$) in all the cases considered here.
\begin{table}
\centering{}
\begin{tabular}{|c|c|c|c|}
\hline 
$(\phi_{e\alpha}^{s},\delta_{\text{CP}})$ & $\left|\epsilon_{e\mu}^{s}\right|$ & $\left|\epsilon_{e\tau}^{s}\right|$ & $\left|\epsilon_{ex}^{s}\right|$\tabularnewline
\hline 
$(0,0)$ & $\left|\epsilon_{e\mu}^{s}\right|<5.38$ & $\left|\epsilon_{e\tau}^{s}\right|<2.14$ & $\left|\epsilon_{ex}^{s}\right|<0.0296$\tabularnewline
$(0,\pi/2)$ & $\left|\epsilon_{e\mu}^{s}\right|<0.0337$ & $\left|\epsilon_{e\tau}^{s}\right|<0.0363$ & $\left|\epsilon_{ex}^{s}\right|<0.0142$\tabularnewline
$(0,\pi)$ & no limit & no limit & $\left|\epsilon_{ex}^{s}\right|<0.0296$\tabularnewline
$(0,3\pi/2)$ & $\left|\epsilon_{e\mu}^{s}\right|<0.0309$ & $\left|\epsilon_{e\tau}^{s}\right|<0.0345$ & $\left|\epsilon_{ex}^{s}\right|<0.0130$\tabularnewline
(0,free) & no limit & no limit & $\left|\epsilon_{ex}^{s}\right|<0.0299$\tabularnewline
(free,0) & no limit & no limit & $\left|\epsilon_{ex}^{s}\right|<0.0696$\tabularnewline
(free,free) & no limit & no limit & $\left|\epsilon_{ex}^{s}\right|<2.02$\tabularnewline
\hline 
\end{tabular}
\caption{$90\%$ C.L. constraints (1 d.o.f) on the QM-NSI parameters $\left|\epsilon_{e\mu}^{s}\right|$,
$\left|\epsilon_{e\tau}^{s}\right|$ and $\left|\epsilon_{ex}^{s}\right|$
projected from the $(\sin^{2}\theta_{13},\left|\epsilon_{e\alpha}^{s}\right|)$
plane for the phases $\phi_{e\alpha}^{s}$ and $\delta_{\text{CP}}$ taking
on different values and being marginalized over ($(\phi_{e\alpha}^{s},\delta_{\text{CP}})$=$(\textrm{free},\textrm{free})$),
respectively. \label{tab:s!=d_emu__etau_ex}}
\end{table}

\subsubsection{Allowed regions in $(\phi_{e\alpha}^{s},\left|\epsilon_{e\alpha}^{s}\right|)$ and $(\left|\epsilon_{e\alpha}^{s}\right|,\left|\epsilon_{\alpha e}^{d}\right|)$ planes
\label{subsec:Constraint-phi_s_epsilon_s_d}}

\begin{figure*}
\begin{centering}
\subfigure[]{\includegraphics[width=6.9cm]{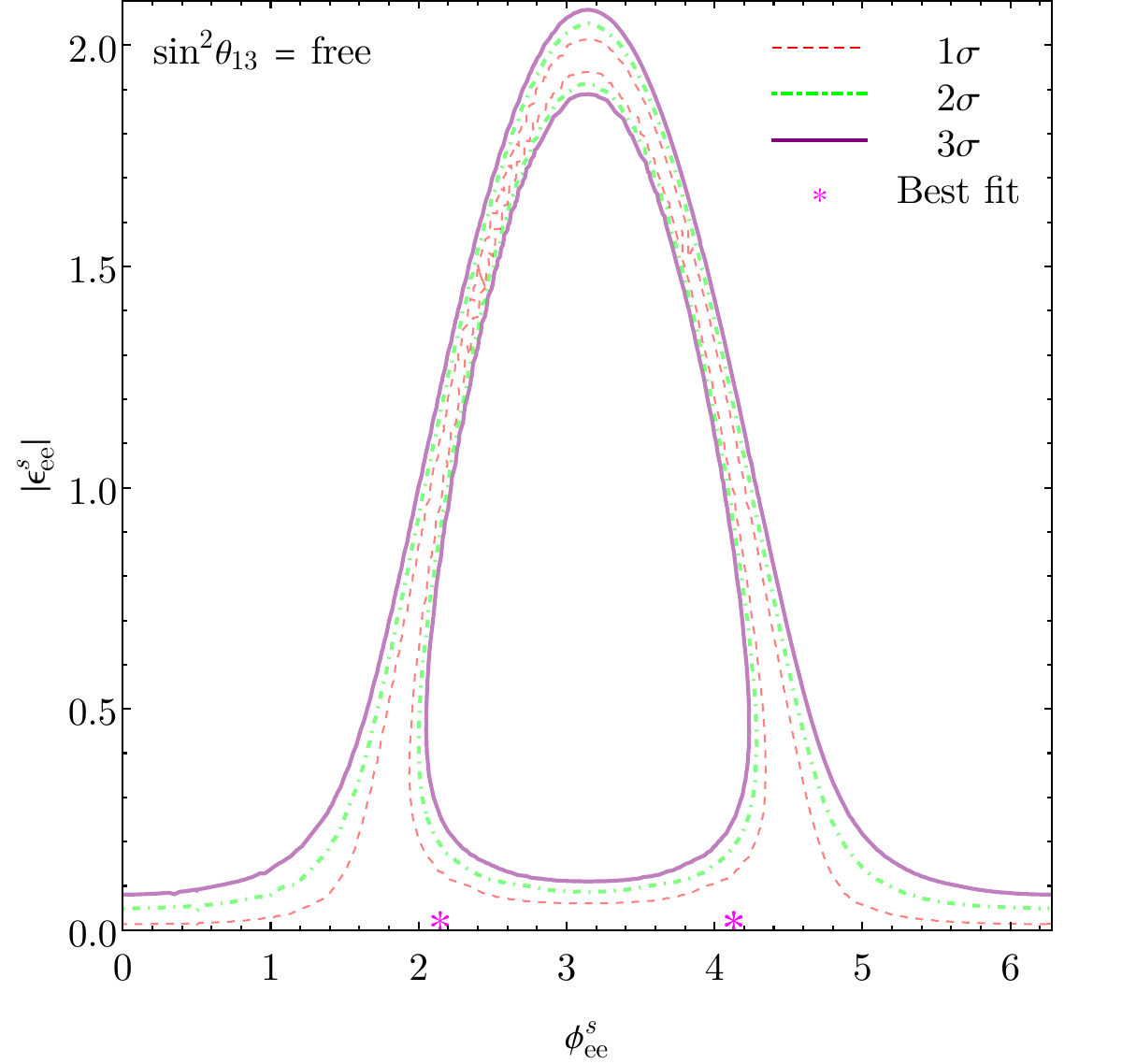}
\label{fig:QM-NSI-phi-epsilon-s_a}}
\subfigure[]{\includegraphics[width=7.0cm]{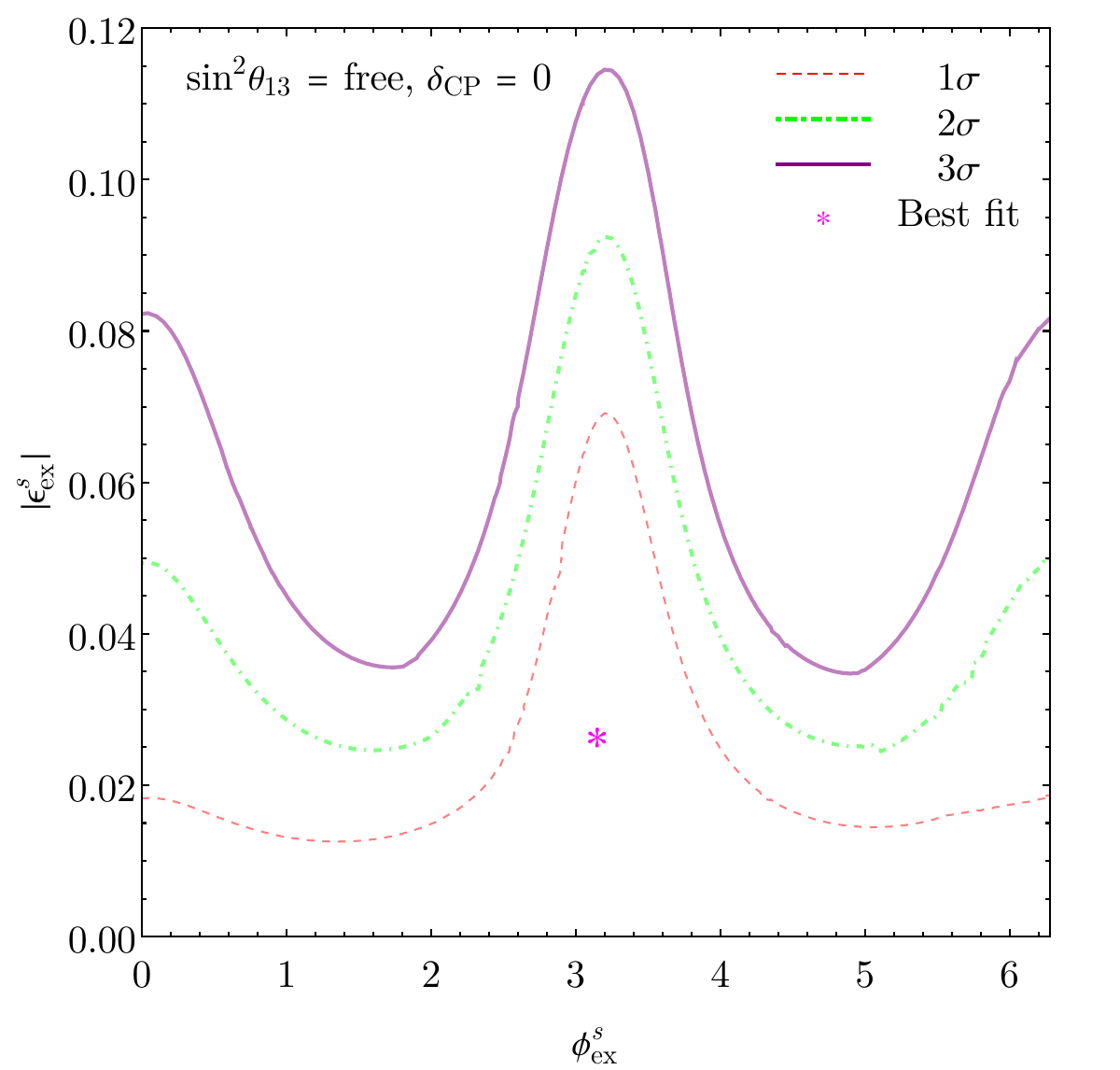}
\label{fig:QM-NSI-phi-epsilon-s_b}}
\par\end{centering}
\caption{Allowed region in the $(\phi_{e\alpha}^{s},\left|\epsilon_{e\alpha}^{s}\right|)$
plane marginalizing over $\sin^{2}\theta_{13}$ for $\delta_{\text{CP}}=0$.
The left panel is for $\epsilon_{ee}^{s}$, and the right for $\epsilon_{ex}^{s}$.
The corresponding allowed regions for the magnitude of $\epsilon_{e\mu}^{s}$
and $\epsilon_{e\tau}^{s}$ are not bound. Details of the analysis can be found in section \ref{subsec:Constraint-phi_s_epsilon_s_d}.
\label{fig:QM-NSI-phi-epsilon-s}}
\end{figure*}

\begin{figure*}
\begin{centering}
\subfigure[]{\includegraphics[width=7.0cm]{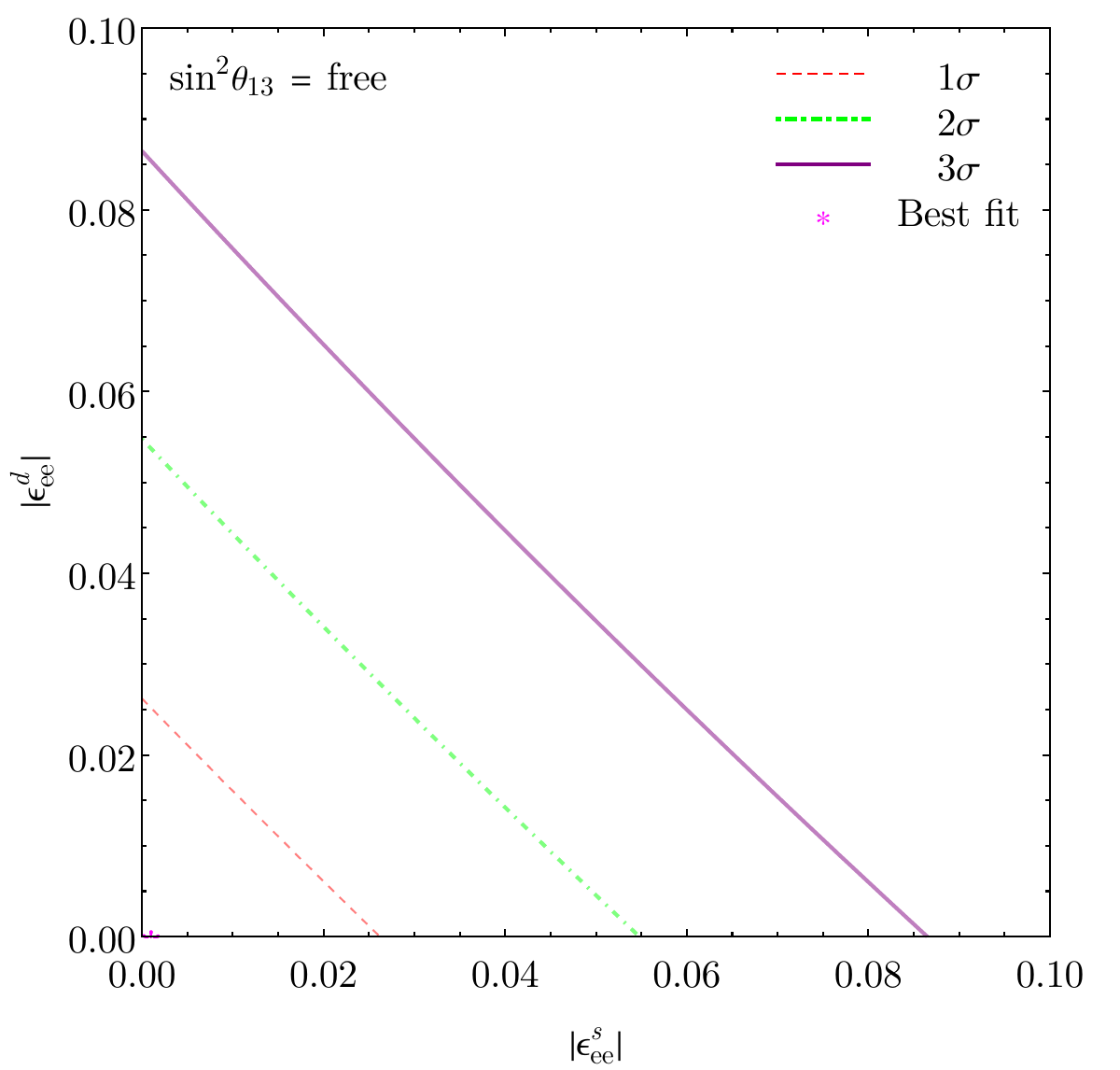}
\label{fig:QM-NSI-epsilons-epsilond_a}}
\subfigure[]{\includegraphics[width=7.0cm]{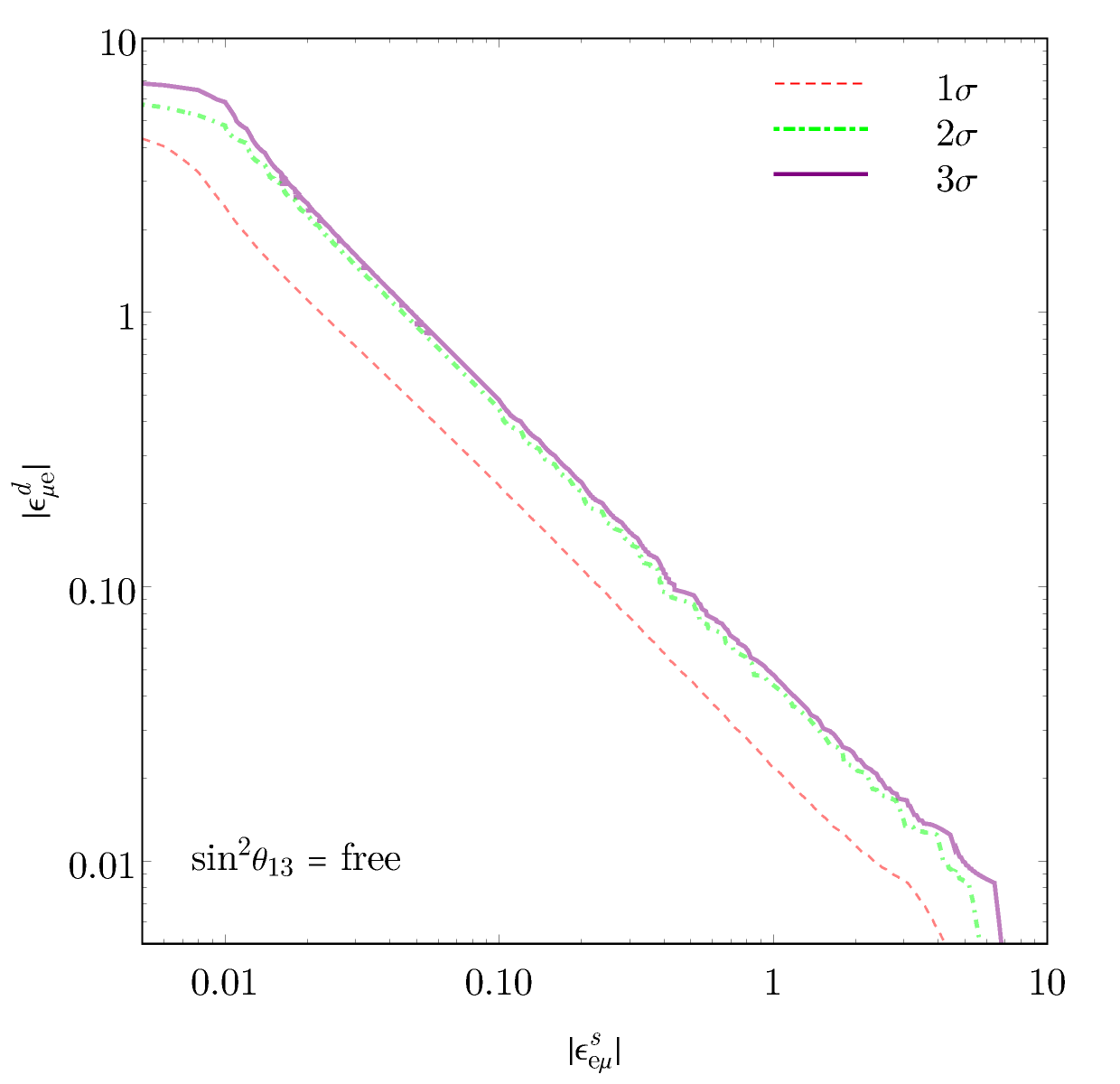}
\label{fig:QM-NSI-epsilons-epsilond_b}}
\par\end{centering}
\caption{Allowed region in the $(\left|\epsilon_{e\alpha}^{s}\right|,\left|\epsilon_{\alpha e}^{d}\right|)$
plane marginalizing over $\sin^{2}\theta_{13}$ with all phases fixed to zero. The left panel is for $\alpha=e$ , and the right
for $\alpha=\mu$. The plot for $\alpha=\tau$ is similar to that
of $\alpha=\mu$. Details of the analysis can be found in section \ref{subsec:Constraint-phi_s_epsilon_s_d}.
\label{fig:QM-NSI-epsilons-epsilond}}
\end{figure*}

We similarly determine the allowed regions in the $(\phi_{e\alpha}^{s},\left|\epsilon_{e\alpha}^{s}\right|)$
plane for $\delta_{\text{CP}}=0$ with $\sin^{2}\theta_{13}$ left to vary freely for $\alpha=e$ and $x$ in figures \ref{fig:QM-NSI-phi-epsilon-s_a} and \ref{fig:QM-NSI-phi-epsilon-s_b}, respectively. These allowed regions can be understood in the same way as for the case of $\epsilon_{e\alpha}^{s}=\epsilon_{\alpha e}^{d*}$. The bound on $\left|\epsilon_{e\mu}^{s}\right|$ can not be set at $\delta_{\text{CP}}=0$ and $\phi_{e\mu}^{s}=\pi$ as described above. The NSI phases $\phi^s_{ee}$ and $\phi^s_{ex}$ are not constrained either as can be seen in figure \ref{fig:QM-NSI-phi-epsilon-s}. We show in figure \ref{fig:QM-NSI-epsilons-epsilond}
the allowed regions in the $(\left|\epsilon_{e\alpha}^{s}\right|,\left|\epsilon_{\alpha e}^{d}\right|)$
plane for $\sin^{2}\theta_{13}$ varying freely and all phases fixed to zero. Again, the behavior can be understood in a similar way as for those in the $(\phi_{e\alpha},\left|\epsilon_{e\alpha}\right|)$ plane and the
corresponding constraints (1 d.o.f) at $90\%$ C.L. are the same as those listed in Table \ref{tab:s!=d_emu__etau_ex} for the case of all phases set to zero. It can be seen from figure \ref{fig:QM-NSI-epsilons-epsilond} that the allowed regions are symmetric about the line $\left|\epsilon^{d}\right|=\left|\epsilon^{s}\right|$, i.e., $\left|\epsilon^{s}\right|$ and $\left|\epsilon^{d}\right|$ play the same role in affecting the effective probability which is implied by the transformation of eq.\,(\ref{eq:QM-NSI-s-d-symmetry}) when all phases are taken to be zero. 

\subsection{Constraints on WEFT-NSI parameters $[\varepsilon_{X}]_{e\alpha}$ \label{subsec:Constraints-on-WEFT-NSI}}

We consider in this section the NSI parameters $[\varepsilon_{X}]_{e\alpha}$
for $X=L,R,S,T$ and $\alpha=\mu,\tau$, and again, one parameter at a time. The effective survival probability under the WEFT framework of eq.\,(\ref{eq:prob_EFT_complete}) is still approximately invariant under the exchange of $\theta_{13} \leftrightarrow \pi/2-2\tilde{\theta}_{13}+\theta_{13}$ or $\theta_{13} \leftrightarrow \pi/2-\theta_{13}$ depending on the values of $[\phi_X]_{e\alpha}$ and $\delta_{\textrm{CP}}$ if the magnitude of the WEFT-NSI parameters $\left|[\varepsilon_X]_{e\alpha}\right|$ are small. We first focus on the allowed region in the $(\sin^{2}\theta_{13},\left|[\varepsilon_X]_{e\mu}\right|)$ plane around small $\theta_{13}$ for the corresponding WEFT-NSI phase $[\phi_X]_{e\mu}$ and $\delta_{\textrm{CP}}$ set to zero and vary freely, respectively. The allowed regions for $[\phi_X]_{e\mu}=\pi$ and $\delta_{\textrm{CP}}=0$ are also shown if necessary. We
also provide allowed regions in the $([\phi_X]_{e\mu},\left|[\epsilon_X]_{e\mu}\right|)$
plane with $\sin^{2}\theta_{13}$ set to vary freely and $\delta_{\text{CP}}=0$. The numerical values of the constraints on the parameters $\left|[\varepsilon_X]_{e\alpha}\right|$ under different conditions are listed in Table \ref{tab:Constraint-WEFT-NSI_L_R_S_T}. The difference between the constraints on $\left|[\varepsilon_{X}]_{e\tau}\right|$ and $\left|[\varepsilon_{X}]_{e\mu}\right|$ are expected to be small since the only difference between the two cases is from the lower two rows of the PMNS mixing matrix $U_{\mu k}$
and $U_{\tau k}$ which are close in numerical values \citep{Esteban:2020cvm}. For this reason, we will show our results for $[\varepsilon_{X}]_{e\mu}$ only. To help understand the behavior of the WEFT-NSI parameters $[\varepsilon_{X}]_{e\mu}$, we refer to the survival probability valid to first order in $[\varepsilon_{X}]_{e\mu}$ \citep{Falkowski:2019xoe} in the discussion below.

\subsubsection{Constraints on left-handed NSI coupling $[\varepsilon_{L}]_{e\mu}$
\label{subsec:Constraints-on-EFT-epsilon_L}}

\begin{figure}
\begin{centering}
\includegraphics[width=7.0cm]{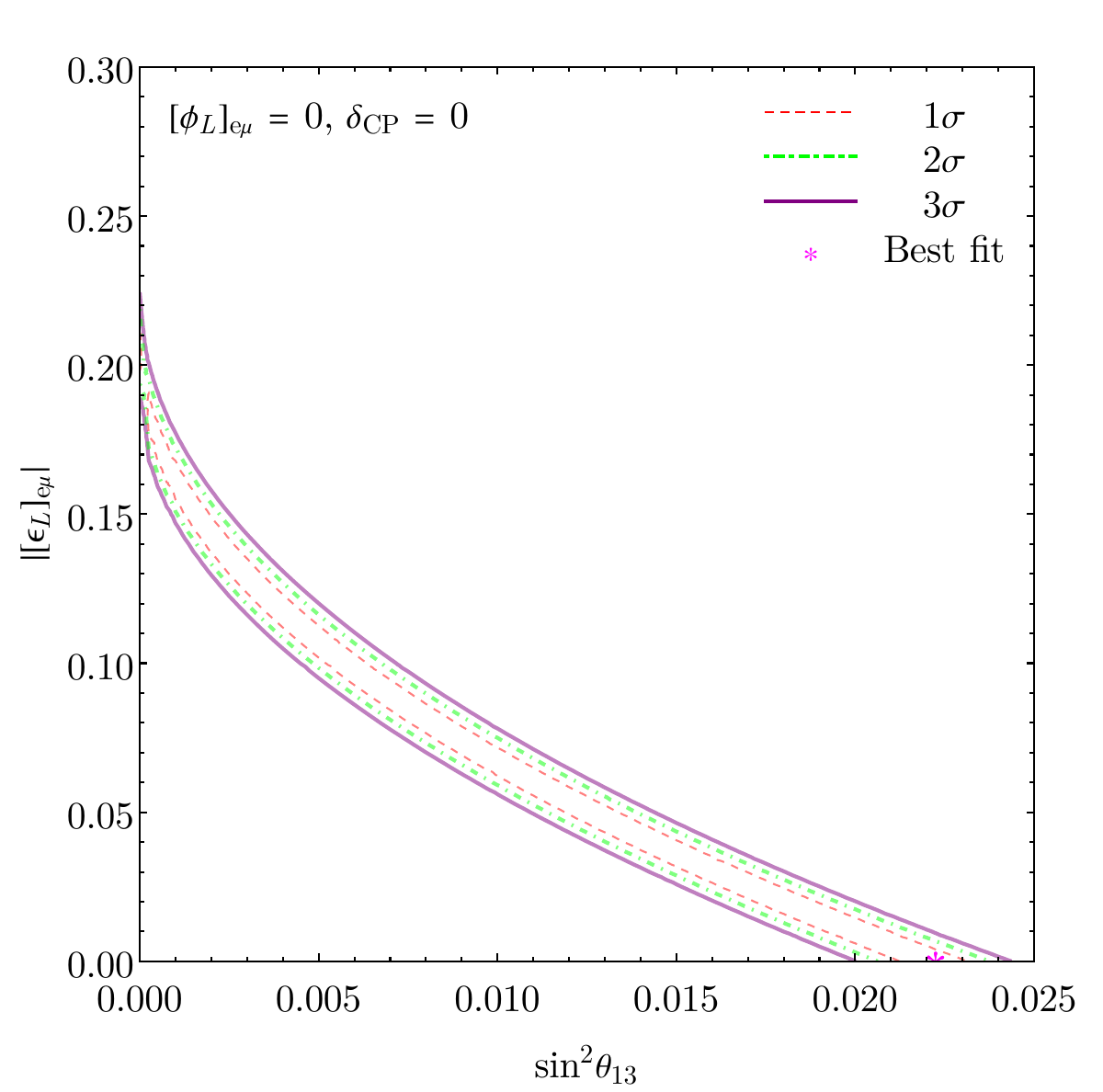}
\par\end{centering}
\caption{Allowed region in the $(\sin^{2}\theta_{13},\left|[\varepsilon_{L}]_{e\mu}\right|)$
plane for $[\phi_{L}]_{e\mu}=\delta_{\text{CP}}=0$. Details of the analysis can be found in section \ref{subsec:Constraints-on-EFT-epsilon_L}.
\label{fig:Lemu}}
\end{figure}

We first consider the effect of the new physics represented by the
term $[\varepsilon_{L}]_{e\mu}$ which describes interactions of the
structure of $V-A$ as in the SM CC weak interactions. But differing
from that in the SM, it couples two leptons of  $e$ and $\bar{\nu}_{\mu}$ instead of $e$ and $\bar{\nu}_{e}$. To first order in $[\varepsilon_{L}]_{e\mu}$ \citep{Falkowski:2019xoe}, the survival probability has the standard form of eq.\,(\ref{eq:Pee_std}) when the small contribution
from the term depending on $\Delta m_{21}^{2}L_\nu/E_\nu$ is ignored:
\begin{equation}
P_{\bar{\nu}_{e}\rightarrow\bar{\nu}_{e}}^{\textrm{WEFT-NSI}}=1-\sin^{2}\left(\frac{\Delta m_{31}^{2}L_\nu}{4E_\nu}\right)\sin^{2}\left(2\tilde{\theta}_{13}\right)+\mathcal{O}(\varepsilon_{L}^{2}),
\end{equation}
with the effective mixing angle
\begin{equation}
\tilde{\theta}_{13}=\theta_{13}+\sin\theta_{23}\left|[\varepsilon_{L}]_{e\mu}\right|\cos([\phi_{L}]_{e\mu}+\delta_{\text{CP}}).
\end{equation}
For $[\phi_{L}]_{e\mu}=\delta_{\text{CP}}=0$, $\tilde{\theta}_{13}=\theta_{13}+\sin\theta_{23}\left|[\varepsilon_{L}]_{e\mu}\right|$.
The effect of the mixing angle $\theta_{13}$ is compensated by the
effect of $\left|[\varepsilon_{L}]_{e\mu}\right|$. Such a behavior remains
when the higher order effects are included, see figure \ref{fig:WEFT-NSI-effect-shape}
for the example of $\left|[\varepsilon_{L}]_{e\mu}\right|=0.01$. The
allowed region is shown in figure \ref{fig:Lemu}. As for the case of
$\epsilon^{s/d}$, the Daya Bay experimental data is still consistent
with the standard oscillation framework (i.e., $\left|[\varepsilon_{L}]_{e\mu}\right|=0$)
at 1$\sigma$ C.L. for the presence of the new $V-A$ type interaction.
In the case of $\delta_{\text{CP}}=0$ and $[\phi_{L}]_{e\mu}=\pi$, however,
the allowed NSI parameter $\left|[\varepsilon_{L}]_{e\mu}\right|$ increases
with $\theta_{13}$ as can be seen from the first order relation $\tilde{\theta}_{13}=\theta_{13}-\sin\theta_{23}\left|[\varepsilon_{L}]_{e\mu}\right|$.
Higher order contributions do not change the trend and the allowed
value of $\left|[\varepsilon_{L}]_{e\mu}\right|$ tends to become infinite
at $\sin^{2}\theta_{13}\approx0.96$. No bound can be put on $\left|[\varepsilon_{L}]_{e\mu}\right|$
in this case nor in the case that $[\phi_{L}]_{e\mu}$ and $\delta_{\text{CP}}$ are allowed to vary freely from the reactor neutrino oscillation experiments.

We note at this point that the identification of the allowed regions in the $(\sin^{2}\theta_{13},\left|[\varepsilon_{L}]_{e\mu}\right|)$ plane and the $(\sin^{2}\theta_{13},\left|\epsilon_{e\mu}\right|)$ plane in figures \ref{fig:s=d_emu_a} and \ref{fig:Lemu} for $\phi=\delta_{\text{CP}}=0$. Such an identification is expected from the relationship between the WEFT-NSI and QM-NSI parameters \citep{Falkowski:2019xoe,Falkowski:2019kfn} which leads to $[\varepsilon_{L}]_{e\mu}=\epsilon_{e\mu}^{*}$ at first order in these NSI parameters.  We also note that an improvement on the uncertainty of the reactor flux normalization has little effect on the constraint on $\left|[\varepsilon_{L}]_{e\mu}\right|$ for $[\phi_L]_{e\mu}=\delta_{\textrm{CP}}=0$. This is similar to the case of $\epsilon_{e\mu}$ as discussed in section \ref{subsec:Constraint-epsilon_emu}.

\subsubsection{Constraints on right-handed NSI coupling $[\varepsilon_{R}]_{e\mu}$
\label{subsec:Constraints-on-EFT-epsilon_R}}

\begin{figure}
\begin{centering}
\includegraphics[width=7.0cm]{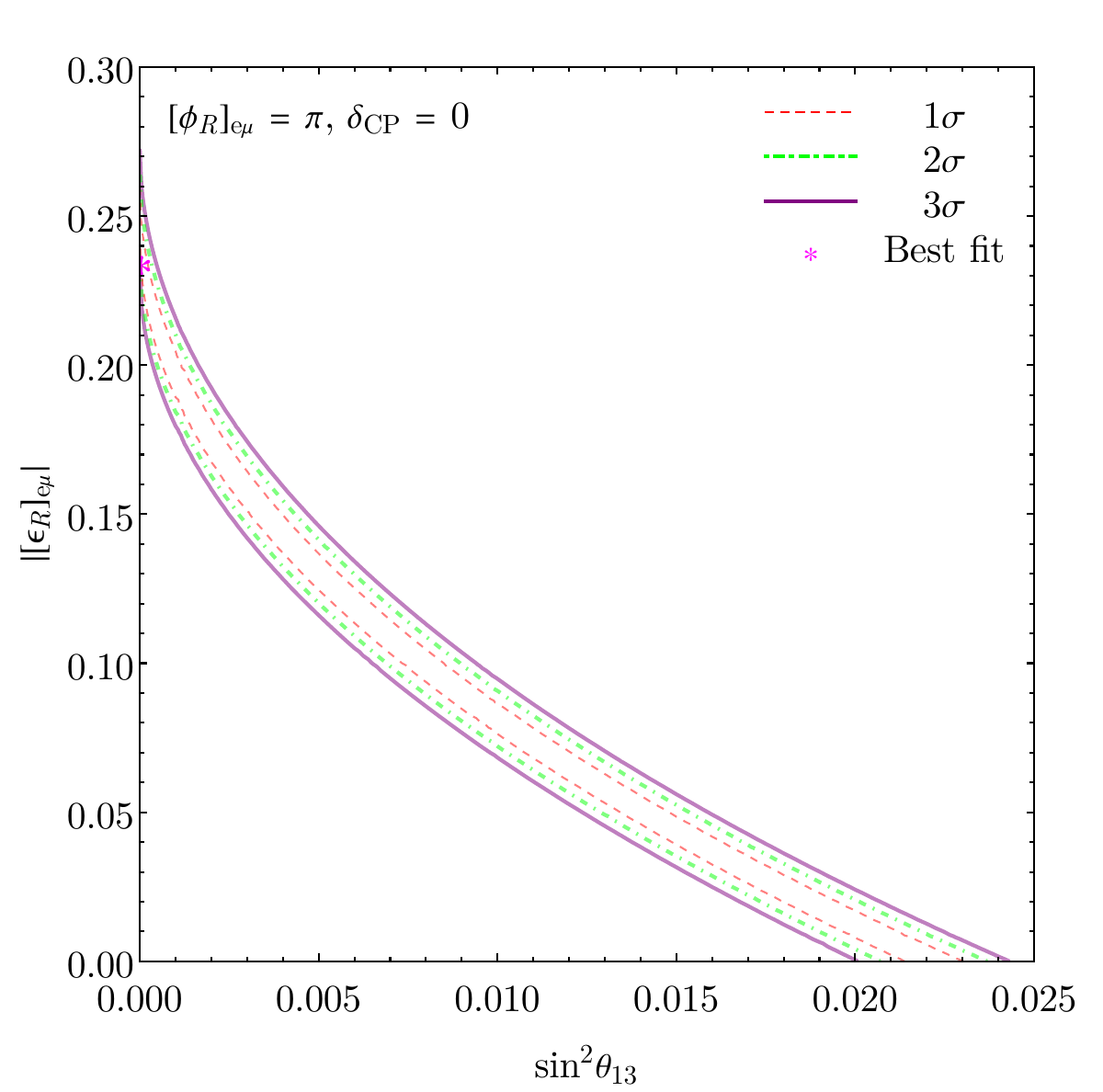}
\par\end{centering}
\caption{The allowed region in the $(\sin^{2}\theta_{13},\left|[\varepsilon_{R}]_{e\mu}\right|)$
plane for, e.g., $\delta_{\text{CP}}=0$ and $[\phi_{R}]_{e\mu}=\pi$.
Details of the analysis can be found in section \ref{subsec:Constraints-on-EFT-epsilon_R}.
\label{fig:Remu}}
\end{figure}

The new interaction represented by the term of $[\varepsilon_{R}]_{e\mu}$
is of the $V+A$ type for the coupling of $u$ and $d$ quarks. The
first order survival probability reads 
\begin{align}
P_{\bar{\nu}_{e}\rightarrow\bar{\nu}_{e}}^{\textrm{WEFT-NSI}} & =1-\sin^{2}\left(\frac{\Delta m_{31}^{2}L_\nu}{4E_\nu}\right)\sin^{2}\left(2\tilde{\theta}_{13}\right)\nonumber \\
 & \quad-\left(\frac{2}{3g_{A}^{2}+1}\sin\theta_{23}\left|[\epsilon_{R}]_{e\mu}\right|\sin([\phi_{R}]_{e\mu}+\delta_{CP})\right)\sin\left(\frac{\Delta m_{31}^{2}L_\nu}{2E_\nu}\right)\sin(2\tilde{\theta}_{13})+\mathcal{O}(\epsilon_{R}^{2}),
\end{align}
where $\tilde{\theta}_{13}=\theta_{13}-(3g_{A}^{2}/(3g_{A}^{2}+1))\sin\theta_{23}\left|[\varepsilon_{R}]_{e\mu}\right|$ $\cos([\phi_{R}]_{e\mu}+\delta_{\text{CP}})$.
This expression reduces to the standard form
\begin{equation}
P_{\bar{\nu}_{e}\rightarrow\bar{\nu}_{e}}^{\textrm{WEFT-NSI}}=1-\sin^{2}\left(\frac{\Delta m_{31}^{2}L_\nu}{4E_{\nu}}\right)\sin^{2}\left(2\tilde{\theta}_{13}\right)+\mathcal{O}(\varepsilon_{R}^{2}),
\end{equation}
when $\sin([\phi_{R}]_{e\mu}+\delta_{\text{CP}})=0$. The situation now becomes
the same to that of $[\varepsilon_{L}]_{e\mu}$ except for the minus
sign before $\cos([\phi_{R}]_{e\mu}+\delta_{\text{CP}})$. For $[\phi_{R}]_{e\mu}=\delta_{\text{CP}}=0$,
$\tilde{\theta}_{13}=\theta_{13}-(3g_{A}^{2}/(3g_{A}^{2}+1))\sin\theta_{23}\left|[\varepsilon_{R}]_{e\mu}\right|$,
corresponding to the case of $\delta_{\text{CP}}=0$ and $[\phi_{L}]_{e\mu}=\pi$
for $[\varepsilon_{L}]_{e\mu}$. For the same reason, the constraint
on $\left|[\varepsilon_{R}]_{e\mu}\right|$ is not possible for $[\phi_{R}]_{e\mu}=\delta_{\text{CP}}=0$ and
thus for the case that both phases are marginalized over. Constraints
may exist for other choices of the phases. For instance, when $\delta_{\text{CP}}=0$
and $[\phi_{R}]_{e\mu}=\pi,$ $\tilde{\theta}_{13}=\theta_{13}+(3g_{A}^{2}/(3g_{A}^{2}+1))\sin\theta_{23}\left|[\varepsilon_{R}]_{e\mu}\right|$.
The situation is similar to that of $[\varepsilon_{L}]_{e\mu}$ when
$[\phi_{L}]_{e\mu}=\delta_{\text{CP}}=0$. The bound on $[\varepsilon_{R}]_{e\mu}$
in this case is thus a factor of $((3g_{A}^{2}+1)/(3g_{A}^{2})\approx1.21$
larger than that on $[\varepsilon_{L}]_{e\mu}$, as can be seen from figure \ref{fig:Remu}. As to the effect of an improvement on the uncertainty of the normalization, the situation is the same as to the case of $\left|[\varepsilon_{L}]_{e\mu}\right|$. The correspondence between $[\varepsilon_{L}]_{e\mu}$
and $[\varepsilon_{R}]_{e\mu}$ discussed here originates from their opposite effects on the effective mixing angle as can be seen from the relation $\tilde{\theta}_{13}=\theta_{13}+\text{Re}[L]-3g_{A}^{2}\text{Re}[R]/(3g_{A}^{2}+1)$
when $\sin([\phi_{R}]_{e\mu}+\delta_{\text{CP}})=0$.  The parameter $[X]$ is defined as $[X]\equiv e^{i\delta_{\text{CP}}}$ $\left(\sin\theta_{23}[\varepsilon_{X}]_{e\mu}+\right.$
$\left.\cos\theta_{23}[\varepsilon_{X}]_{e\tau}\right)$ in ref. \citep{Falkowski:2019xoe}.

\subsubsection{Constraints on scalar NSI coupling $[\varepsilon_{S}]_{e\mu}$ 
\label{subsec:Constraints-on-EFT-epsilon_S}}

\begin{figure*}
\begin{centering}
\subfigure[]{\includegraphics[width=7.0cm]{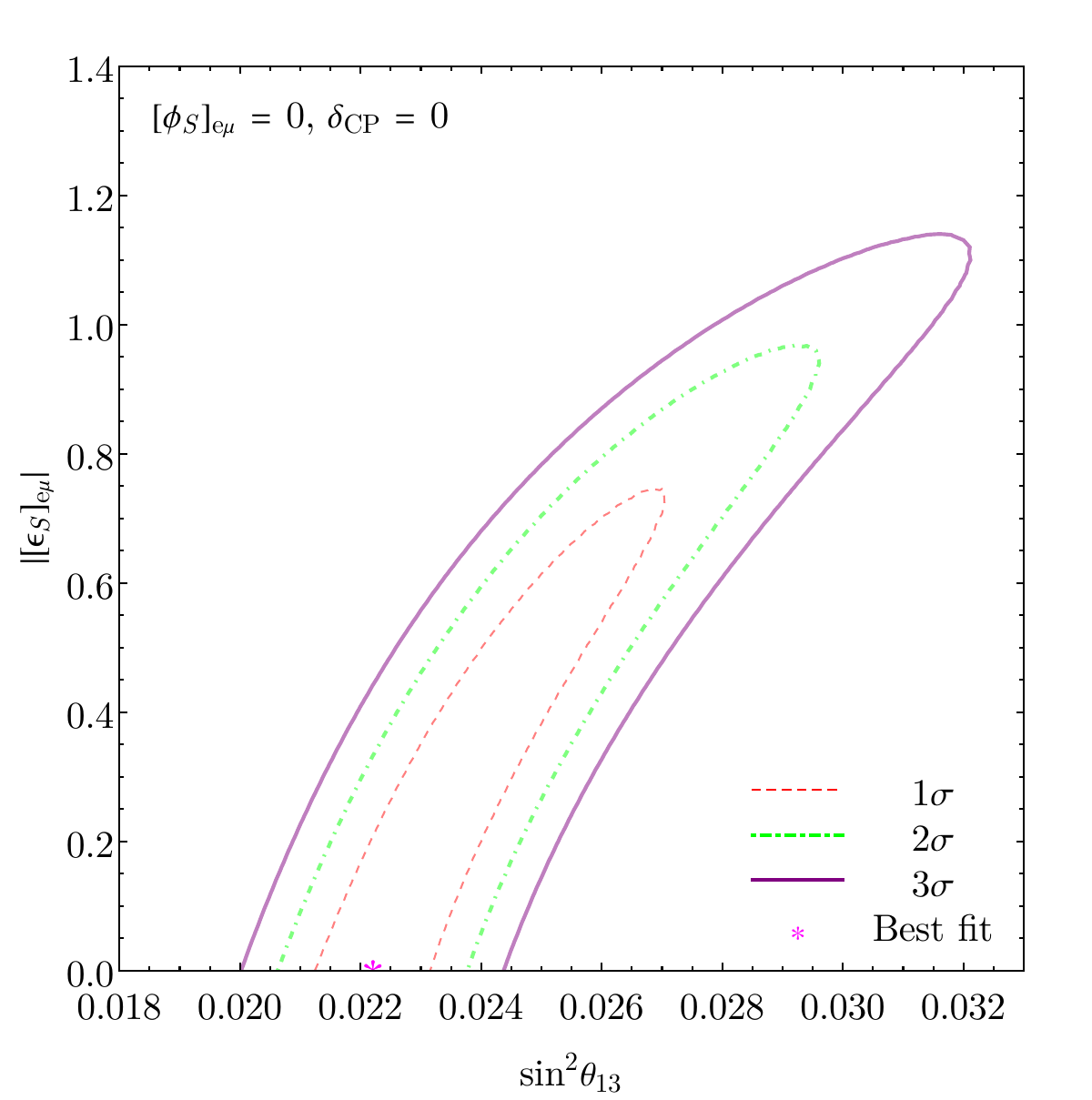}
\label{fig:Semu_a}}
\subfigure[]{\includegraphics[width=7.0cm]{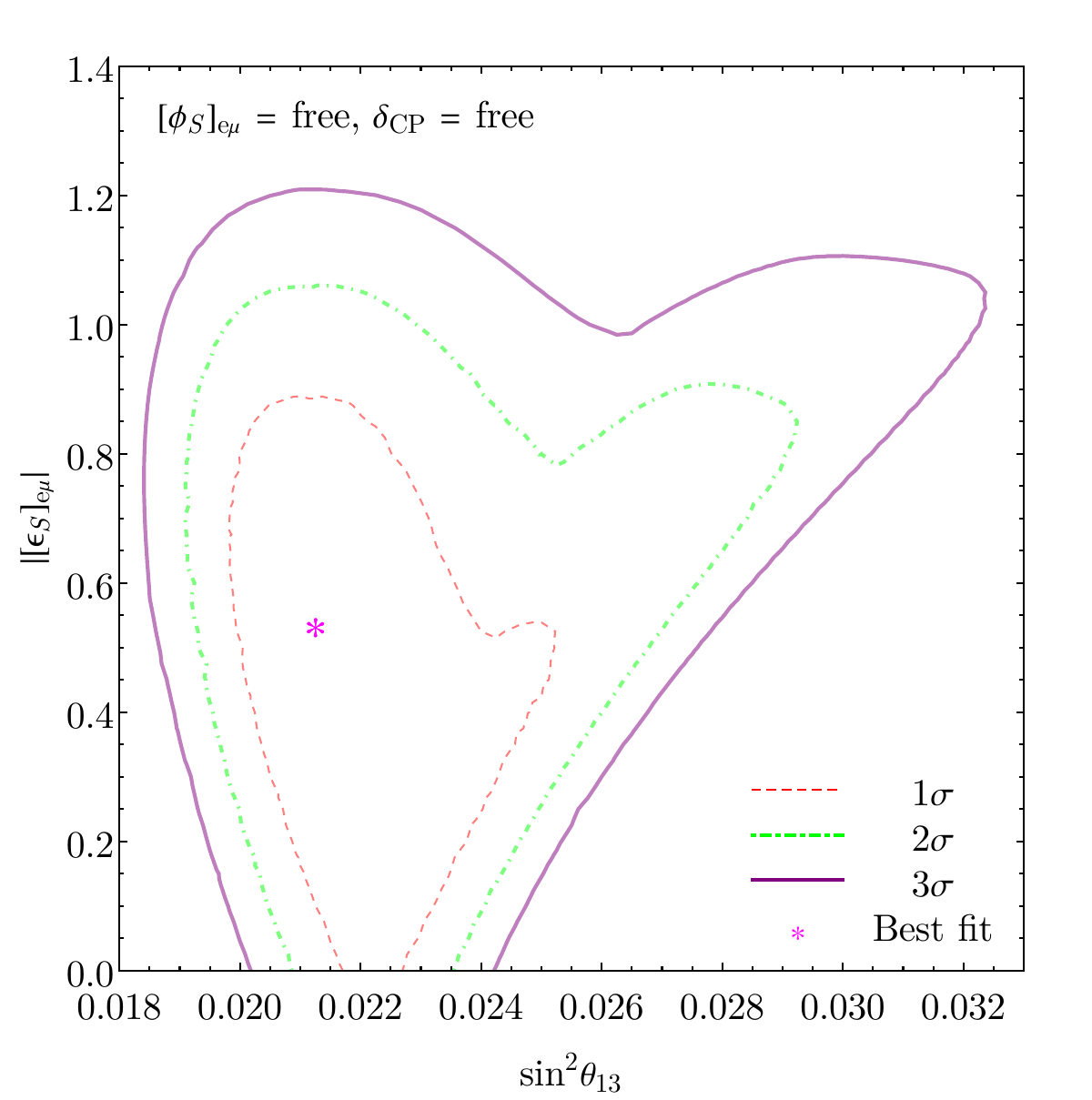}\label{fig:Semu_b}}
\par\end{centering}
\caption{Allowed region in the $(\sin^{2}\theta_{13},\left|[\varepsilon_{S}]_{e\mu}\right|)$
plane for $[\phi_{S}]_{e\mu}=\delta_{\text{CP}}=0$ (left) and for them being
marginalized over ($[\phi_{S}]_{e\mu}=$free and $\delta_{\text{CP}}=$free,
right). Details of the analysis can be found in section \ref{subsec:Constraints-on-EFT-epsilon_S}.
\label{fig:Semu}}
\end{figure*}

If the effect of the new physics is of the scalar type, only $[\varepsilon_{S}]_{e\mu}$ term is present. The first order
survival probability can be written as 
\begin{align}
P_{\bar{\nu}_{e}\rightarrow\bar{\nu}_{e}}^{\textrm{WEFT-NSI}} & =1-\sin^{2}\left(\frac{\Delta m_{31}^{2}L_\nu}{4E_\nu}\right)\sin^{2}\left(2\theta_{13}-\alpha_{D}\frac{m_{e}}{E_\nu-\Delta}\right)\nonumber \\
 & \quad+\sin\left(\frac{\Delta m_{31}^{2}L_\nu}{2E_\nu}\right)\sin(2\theta_{13})\left(\beta_{D}\frac{m_{e}}{E_\nu-\Delta}\right)+\mathcal{O}(\varepsilon_{S}^{2}),
\end{align}
where $m_{e}$ is the electron mass, $\alpha_{D}=(g_{S}/(3g_{A}^{2}+1))\sin\theta_{23}\left|[\varepsilon_{S}]_{e\mu}\right|\cos([\phi_{S}]_{e\mu}+\delta_{\text{CP}})$, $\beta_{D}=(g_{S}/(3g_{A}^{2}+1))$ $\sin\theta_{23}\left|[\varepsilon_{S}]_{e\mu}\right|\sin([\phi_{S}]_{e\mu}+\delta_{\text{CP}})$ and $\Delta\equiv m_{n}-m_{p}$ is the neutron and proton mass difference.
For $\delta_{\text{CP}}=0$ and $[\phi_{S}]_{e\mu}=0$ or $\pi$, $\sin([\phi_{S}]_{e\mu}+\delta_{\text{CP}})=0$.
The survival probability reduces to 
\begin{equation}
P_{\bar{\nu}_{e}\rightarrow\bar{\nu}_{e}}^{\textrm{WEFT-NSI}} =1-\sin^{2}\left(\frac{\Delta m_{31}^{2}L_\nu}{4E_\nu}\right)\sin^{2}\left(2\theta_{13}-\alpha_{D}\frac{m_{e}}{E_\nu-\Delta}\right)+\mathcal{O}(\varepsilon_{S}^{2}).
\end{equation}
Thus 
\begin{equation}
\tilde{\theta}_{13}\approx\theta_{13}\mp\frac{g_{S}/2}{3g_{A}^{2}+1}\sin\theta_{23}\left|[\varepsilon_{S}]_{e\mu}\right|\frac{m_{e}}{E_\nu-\Delta},
\end{equation}
where the $-$ sign is for $[\phi_{S}]_{e\mu}=\delta_{\text{CP}}=0$ and
the $+$ sign for $[\phi_{S}]_{e\mu}=\pi$ and $\delta_{\text{CP}}=0$.
We see that $\left|[\varepsilon_{S}]_{e\mu}\right|$ has to increase and
decrease with $\theta_{13}$ in these two cases, respectively. When
the two phases are marginalized over in the analysis, the allowed
regions of these two cases extend to the left and right wings of the final allowed region as shown in figure \ref{fig:Semu_b}. These constraints are not sensitive to the neutrino flux uncertainty as for the cases of $\left|[\varepsilon_{L}]_{e\mu}\right|$ and $\left|[\varepsilon_{R}]_{e\mu}\right|$.

\subsubsection{Constraints on tensor NSI coupling $[\varepsilon_{T}]_{e\mu}$
\label{subsec:Constraints-on-EFT-epsilon_T}}

\begin{figure*}
\begin{centering}
\subfigure[]{\includegraphics[width=7.0cm]{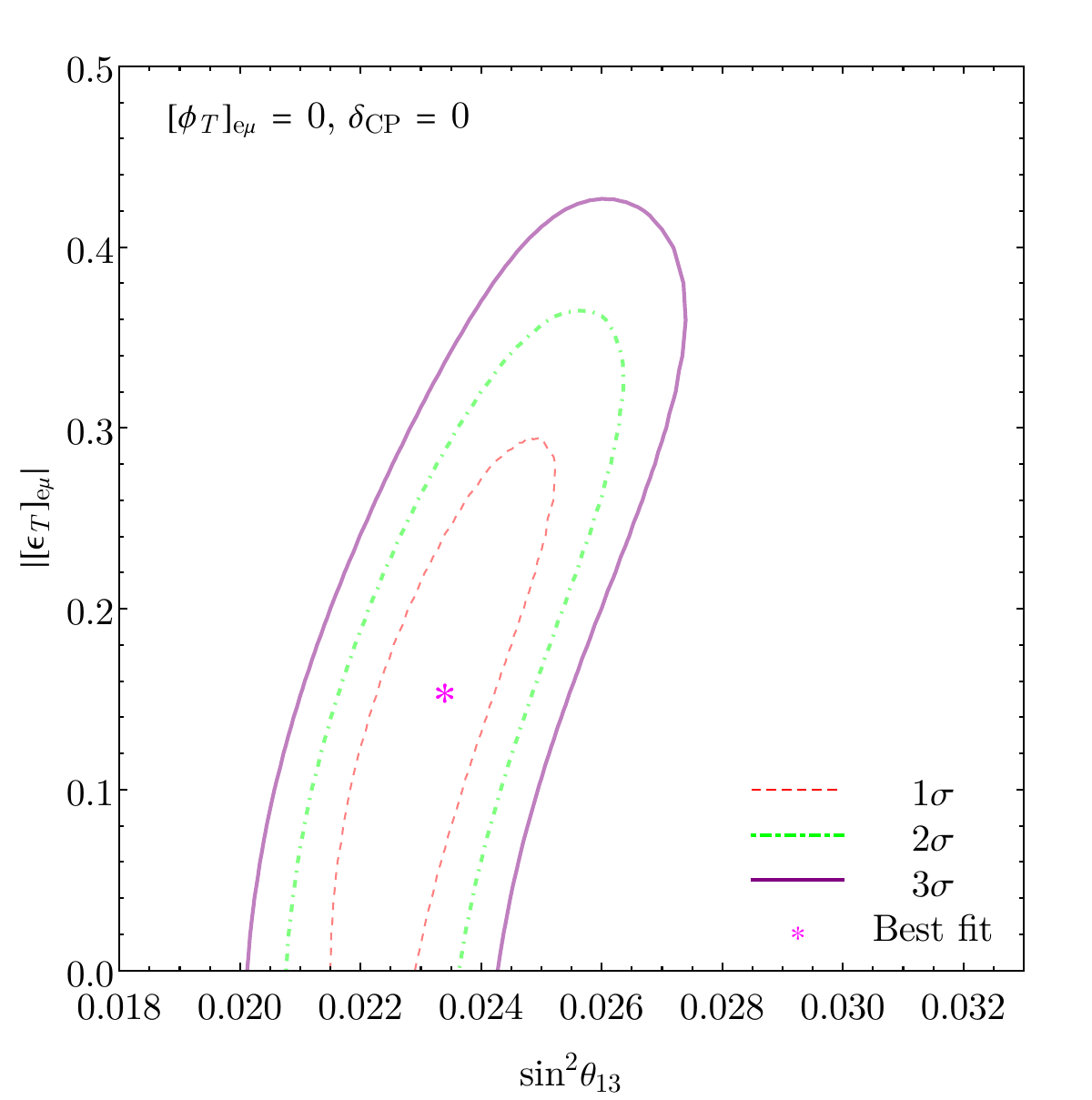}
\label{fig:Temu_a}}
\subfigure[]{\includegraphics[width=7.0cm]{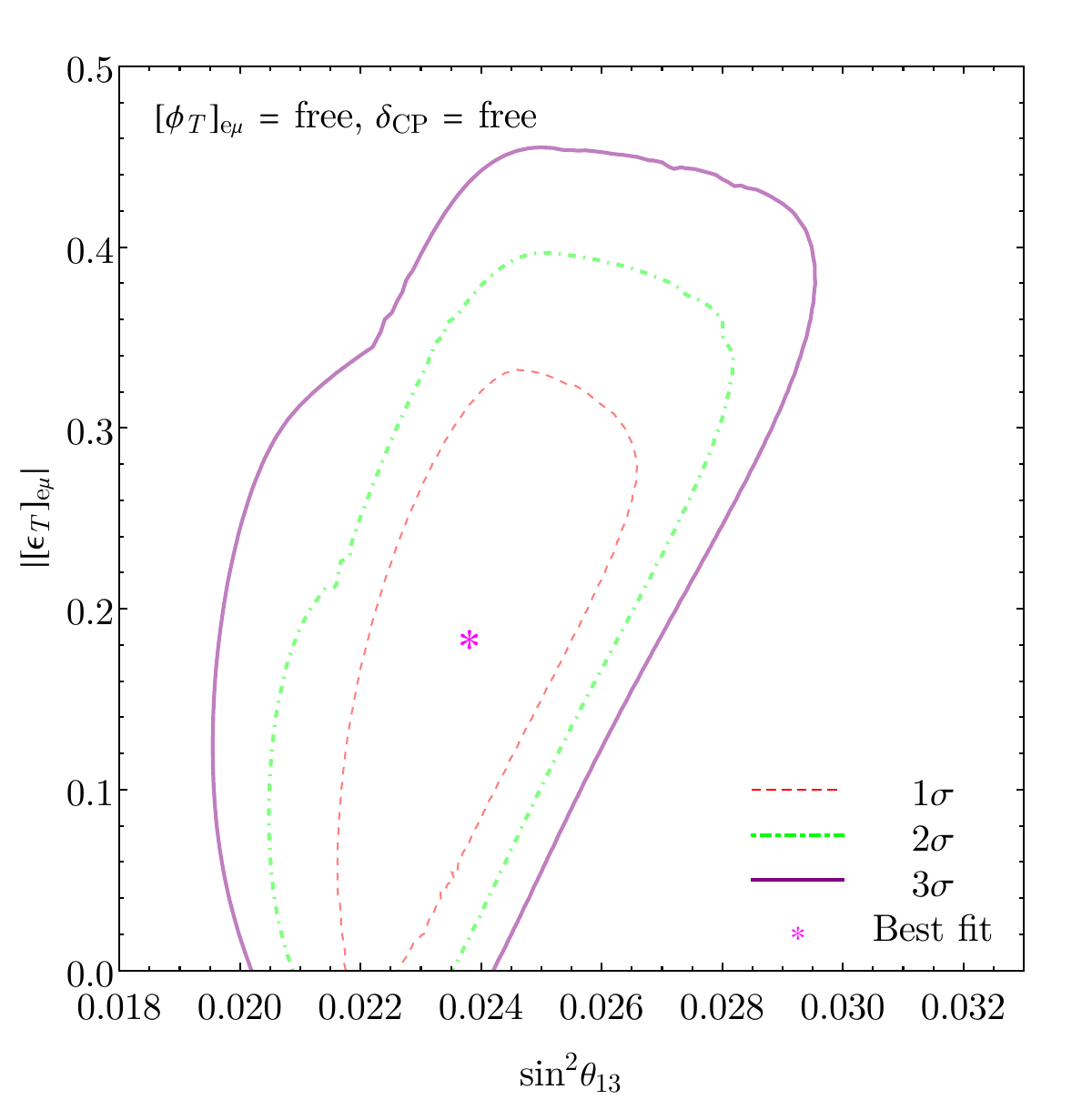}\label{fig:Temu_b}}
\par\end{centering}
\caption{Allowed region in the $(\sin^{2}\theta_{13},\left|[\varepsilon_{T}]_{e\mu}\right|)$
plane for $[\phi_{T}]_{e\mu}=\delta_{\text{CP}}=0$ (left) and for it being
marginalized over ($[\phi_{T}]_{e\mu}=$free and $\delta_{\text{CP}}=$free,
right). Details of the analysis can be found in section \ref{subsec:Constraints-on-EFT-epsilon_T}.
\label{fig:Temu}}
\end{figure*}

The situation with the tensor type interaction is similar to that
with the scalar type interaction, but the expressions are more complicated
with all four coefficients $\alpha_{D}$, $\alpha_{P}$, $\beta_{D}$
and $\beta_{P}$ and the energy dependence of $m_{e}/(E_\nu-\Delta)$
and $m_{e}/f_{T}(E_\nu)$ all present. The form factor $f_{T}(E_\nu)$ is from the production coefficients $p_{TL}$ and $p_{TR}$ and its explicit expression can be found in \citep{Falkowski:2019xoe}. A simple analysis is not possible
even for the case of $[\phi_{T}]_{e\mu}=\delta_{\text{CP}}=0$. We show in
figure \ref{fig:WEFT-NSI-effect-shape} the effect of $[\varepsilon_{T}]_{e\mu}$
on the shape of the survival probability for the case of $[\phi_{T}]_{e\mu}=\delta_{\text{CP}}=0$
for a typical choice of $E_{\nu}=4$ MeV and $\left|[\varepsilon_{T}]_{e\mu}\right|=0.1$.
The behavior of $\left|[\varepsilon_{T}]_{e\mu}\right|$ increasing with
$\sin^{2}\theta_{13}$ is implied. The allowed regions
determined by Daya Bay data are shown in figure \ref{fig:Temu} for $[\phi_{T}]_{e\mu}=\delta_{\text{CP}}=0$ and for both phases to vary freely.

\subsubsection{Constraints in $([\phi_{X}]_{e\mu},\left|[\varepsilon_{X}]_{e\mu}\right|)$ plane
\label{subsec:Constraints-on-EFT-phi-epsilon}}

\begin{figure*}
\begin{centering}
\subfigure[]{\includegraphics[width=7.0cm]{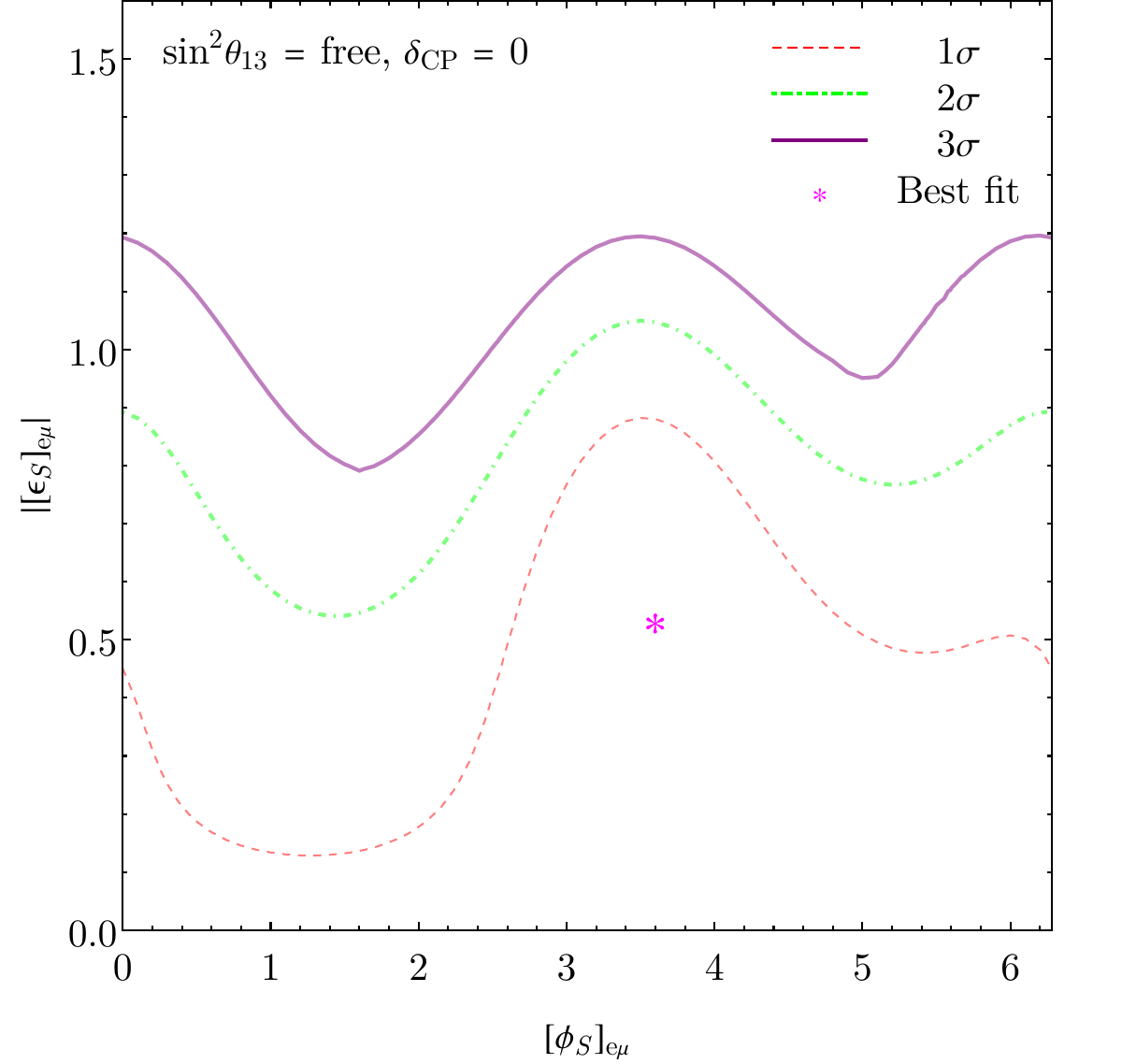}\label{fig:EFT-NSI-phi-epsilon12_a}}
\subfigure[]{\includegraphics[width=7.0cm]{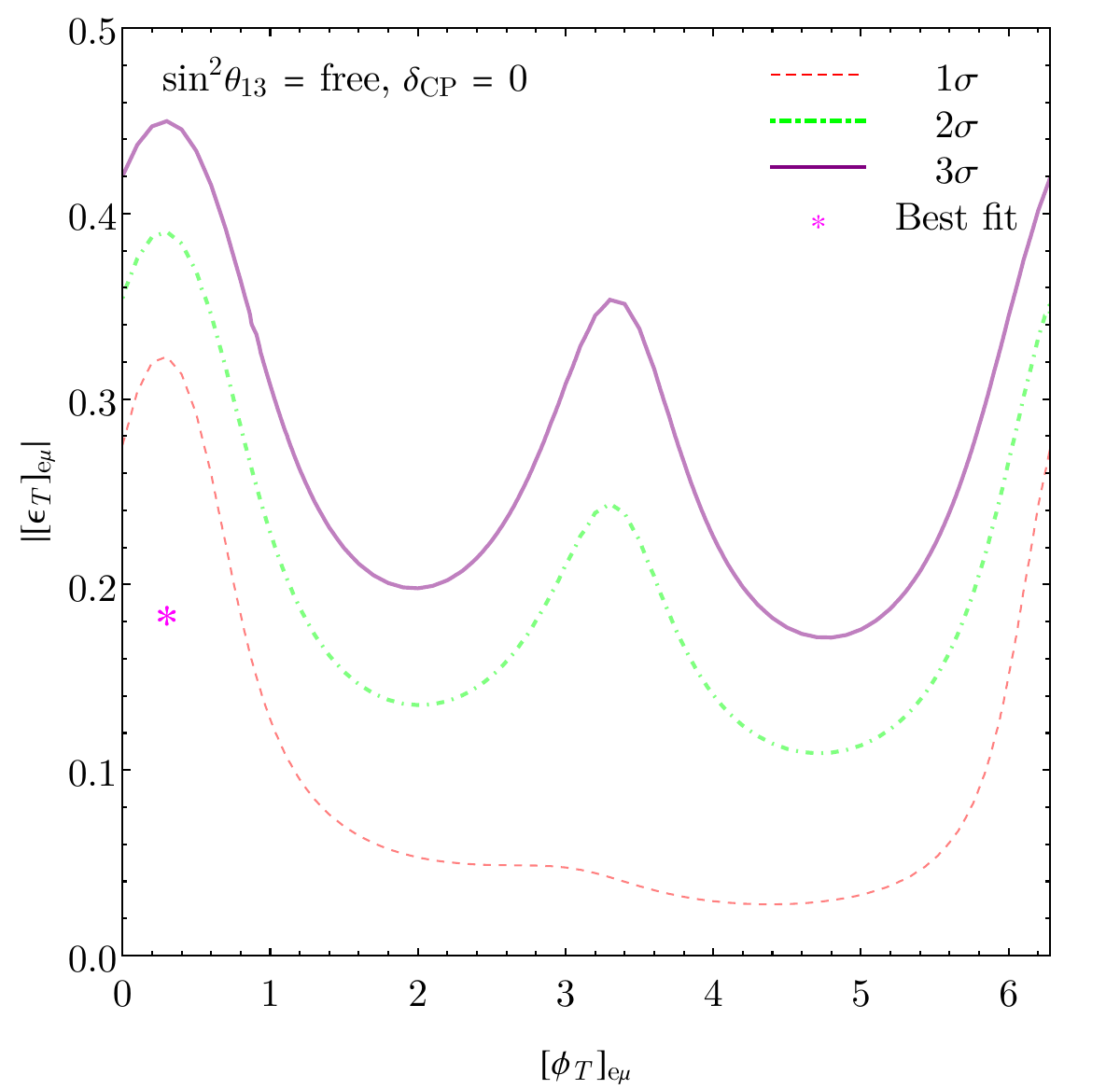}\label{fig:EFT-NSI-phi-epsilon12_b}}
\par\end{centering}
\caption{Allowed region in the $([\phi_{X}]_{e\mu},\left|[\varepsilon_{X}]_{e\mu}\right|)$
plane marginalizing over $\sin^{2}\theta_{13}$ for $\delta_{\text{CP}}=0$
for $X=S$ (left) and $T$ (right), respectively. Details of the analysis can be found in section \ref{subsec:Constraints-on-EFT-phi-epsilon}.
\label{fig:EFT-NSI-phi-epsilon12}}
\end{figure*}

As for QM-NSI, we show in figure \ref{fig:EFT-NSI-phi-epsilon12}
the allowed region plots in the $([\phi_{X}]_{e\mu},\left|[\varepsilon_{X}]_{e\mu}\right|)$
plane for $X=S$ and $T$. As before, we take $\delta_{\text{CP}}=0$ and
let $\sin^{2}\theta_{13}$ vary freely. The corresponding allowed regions can not be set properly for $\left|[\varepsilon_{L}]_{e\mu}\right|$ and $\left|[\varepsilon_{R}]_{e\mu}\right|$. These plots can be understood in the same way
as in QM-NSI with the help of the discussion in, e.g., the subsection
\ref{subsec:Constraints-on-EFT-epsilon_S}. Also as for QM-NSI, the phases $[\phi_{S}]_{e\mu}$ and $[\phi_{T}]_{e\mu}$ are not constrained by the Daya Bay data and can take values in the full range of $[0,2\pi)$. 
\begin{table*}
\centering{}
\begin{tabular}{|c|c|c|c|c|}
\hline 
$([\phi_{X}]_{e\mu},\delta_{\text{CP}})$ & $\left|[\varepsilon_{L}]_{e\mu}\right|$ & $\left|[\varepsilon_{R}]_{e\mu}\right|$ & $\left|[\varepsilon_{S}]_{e\mu}\right|$ & $\left|[\varepsilon_{T}]_{e\mu}\right|$\tabularnewline
\hline 
\multirow{1}{*}{$(0,0)$ } & $\left|[\varepsilon_{L}]_{e\mu}\right|<0.214$ & no limit & $\left|[\varepsilon_{S}]_{e\mu}\right|<0.783$ & $\left|[\varepsilon_{T}]_{e\mu}\right|<0.306$\tabularnewline
(free, free) & no limit & no limit & $\left|[\varepsilon_{S}]_{e\mu}\right|<0.911$ & $\left|[\varepsilon_{T}]_{e\mu}\right|<0.341$\tabularnewline
\hline 
\end{tabular}
\caption{$90\%$ C.L. constraints (1 d.o.f) on the WEFT-NSI parameters $\left|[\varepsilon_{X}]_{e\mu}\right|$
projected from the $(\sin^{2}\theta_{13},\left|[\varepsilon_{X}]_{e\mu}\right|)$
planes for the phases $\delta_{\text{CP}}=[\phi_{X}]_{e\mu}=0$ and being
marginalized over ($(\delta_{\text{CP}},[\phi_{X}]_{e\mu})$= (free, free))
for $X=L,R,S$ and $T$, respectively. \label{tab:Constraint-WEFT-NSI_L_R_S_T}}
\end{table*}

\section{Summary \label{sec:Summary}}

In this paper, we have investigated charged current non-standard neutrino interactions with two different approaches, QM-NSI and WEFT-NSI, using the full IBD data set of Daya Bay. The Huber-Mueller reactor neutrino flux model has
been used with an enlarged $5\%$ rate uncertainty. The effects of CC-NSI are introduced at the quantum state level in QM-NSI, as can be seen from eqs. (\ref{eq:QM-NSI-epsilon-s}) and (\ref{eq:QM-NSI-epsilon-d}), while for WEFT-NSI, they are encoded at the Lagrangian level as in eq.\,(\ref{eq:Lagrangian_WEFT}). It turns out that the effect of the CC-NSI on the reactor neutrino oscillation experiments depends on both the magnitude and the phase of each CC-NSI parameter, as well as on the standard oscillation parameters. For a large number of NSI parameters, we have first considered the effect of one NSI parameter at a time for each approach. In the case of QM-NSI, the two situations, $\epsilon_{e\alpha}^{s}=\epsilon_{\alpha e}^{d*}$ and $\epsilon_{e\alpha}^{s}\neq\epsilon_{\alpha e}^{d*}$, have been studied. For both QM-NSI and WEFT-NSI approaches, the analytical expressions of eq.\,(\ref{eq:QM-NSI-s_d_eff}) and eq.\,(\ref{eq:prob_EFT_complete}) for the effective survival probability are used in analyses. Both of the effective survival probability expressiones are approximately symmetric under the exchange of $\theta_{13} \leftrightarrow \pi/2-2\tilde{\theta}_{13}+\theta_{13}$ or $\theta_{13} \leftrightarrow \pi/2-\theta_{13}$ depending on the values of the Dirac CP-violating phase and the NSI phases if the magnitude of the NSI parameters are small. We focus our discussion in the small $\theta_{13}$ region when we explore the the allowed regions in the $(\sin^{2}\theta_{13},\left|\epsilon\right|)$ plane.

There is no evidence of CC-NSI found in either approach. Bounds
on the magnitude of each CC-NSI parameter have been extracted
under different assumptions on the corresponding CC-NSI phase and/or the Dirac CP-violating phase, especially for the case that these phases are marginalized over. No bounds can be placed on the NSI phases themselves, as shown in figures \ref{fig:QM-NSI-phi-epsilon}, \ref{fig:QM-NSI-phi-epsilon-s} and \ref{fig:EFT-NSI-phi-epsilon12}. The CC-NSI parameters associated with the tau neutrino (e.g., $\epsilon_{e\tau}$) play similar roles as the corresponding CC-NSI parameters with the muon neutrino (e.g., $\epsilon_{e\mu}$) in both approaches, thus the constraints on these parameters are similar. For $\epsilon_{e\alpha}^{s}\neq\epsilon_{\alpha e}^{d*}$ in QM-NSI, the constraints on $\left|\epsilon_{e\alpha}^{s}\right|$ and $\left|\epsilon_{\alpha e}^{d}\right|$ are closely related through eq.\,(\ref{eq:QM-NSI-s-d-symmetry}) since
we consider one NSI parameter at a time. 

For the constraints under different assumptions on the phases, better constraints have been obtained when the phases are fixed to zero or other special values, e.g., $\pi/2,\pi$ and/or $3\pi/2$. We have found $\left|\epsilon_{ex}^{s}\right|<0.013$ (90\% C.L.) for $\phi_{ex}^{s}=3\pi/2$, for example. In other cases, the bounds cannot be set by the Daya Bay experiment when the phase takes such values. For instance,
$\left|[\varepsilon_{L}]_{e\mu}\right|$ is unconstrained in the case $[\phi_{L}]_{e\mu}=\pi$ and $\delta_{\text{CP}}=0$. The upper bounds usually grow enormously when the phases are treated as free parameters. Taking $\left|\epsilon_{ex}^{s}\right|$
as an example, the allowed range of $\left|\epsilon_{ex}^{s}\right|$ increases to $\left|\epsilon_{ex}^{s}\right|<2.02$ for  both $\phi_{ex}^{s}$ and $\delta_{\text{CP}}$ being allowed to vary freely. While a much stringent constraint $\left|\epsilon_{ex}^{s}\right|<0.0296$ is found for $\phi_{ex}^{s}=\delta_{\text{CP}}=0$. Our constraints on the CC-NSI parameters $\left|\epsilon_{e\alpha}\right|$ are consistent with those obtained in ref.\,\citep{Agarwalla:2014bsa} where the special case of $\epsilon_{e\alpha}^{s}=\epsilon_{\alpha e}^{d*}$ and $\phi_{e\alpha}=\delta_{\text{CP}}=0$ for QM-NSI with the $5\%$ total normalization error included is studied with the effective survival probality valid up to second order in $\epsilon_{e\alpha}$. 

For Daya Bay experiment, the effect of $\epsilon_{ee}$
or $\epsilon_{ee}^{s}$ is directly related to the reactor flux normalization. The constraints on $\left|\epsilon_{ee}\right|$ or $\left|\epsilon^s_{ee}\right|$ are thus  sensitive to the normalization uncertainty when the phases are fixed at some special values. If the neutrino flux can be accurately predicted in the future, the constraints on these parameters can be further improved in these cases. Unlike for the case of $\epsilon_{ee}$ or $\epsilon_{ee}^{s}$, the non-zero parameter $\epsilon_{e\alpha}$ or $\epsilon_{e\alpha}^{s/d}$ with $\alpha\neq e$ usually gives rise to an effective mixing angle $\tilde{\theta}_{13}$ and affect the measurement of the true value of $\theta_{13}$. The constraints on these parameters depend on both the systematical and statistical uncertainties, and are not so sensitive to the normalization uncertainty. The constraints on the WEFT-NSI parameters  $\left|[\varepsilon_{X}]_{e\alpha}\right|$ with $\alpha\neq e$ are not so sensitive to the normalization uncertainty either.

In summary, the constraints on the magnitude of the QM-NSI parameters $\epsilon_{ee}$, $\epsilon_{ex}$, $\epsilon_{ee}^{s}$ and $\epsilon_{ex}^{s}$ can reach $\mathcal{O}(0.01)$ with the phases set to zero or other special values, while they get relaxed to $\mathcal{O}(1)$ for the phases being allowed to vary freely. For $\left|\epsilon_{e\mu}\right|$ or $\left|\epsilon_{e\tau}\right|$, the constraints can reach $\mathcal{O}(0.1)$ in both cases. The constraints on $\left|\epsilon_{e\mu}^s\right|$ or $\left|\epsilon_{e\tau}^s\right|$ cannot be set by the Daya Bay experiment alone when the phases are allowed to vary freely. The WEFT-NSI parameters $[\epsilon_{L}]_{e\alpha}$ and $[\epsilon_{R}]_{e\alpha}$ are unconstrained when the phases are free, but constraints of ${\cal O}(0.1)$ can be set for certain value of the phases. For $\left|[\varepsilon_{S}]_{e\alpha}\right|$ and $\left|[\varepsilon_{T}]_{e\alpha}\right|$ for $\alpha=\mu,\tau$, the constraints can reach $\mathcal{O}(0.1)$ whether or not the phases are fixed.

\section*{Acknowledgments }

We are grateful to Professor Jiajun Liao for useful discussions and
suggestions on various aspects of the NSI effect in reactor neutrino
oscillation experiments. The Daya Bay experiment is supported in part by the Ministry of Science and Technology of China, the U.S. Department of Energy, the Chinese Academy of Sciences, the National Natural Science Foundation of China, the New Cornerstone Science Foundation, the Guangdong provincial government, the Shenzhen municipal government,
the China General Nuclear Power Group, the Research Grants Council
of the Hong Kong Special Administrative Region of China, the Ministry
of Education in Taiwan, the U.S. National Science Foundation, the
Ministry of Education, Youth, and Sports of the Czech Republic, the
Charles University Research Centre UNCE, and the Joint Institute of
Nuclear Research in Dubna, Russia. We acknowledge Yellow River Engineering
Consulting Co., Ltd., and China Railway 15th Bureau Group Co., Ltd.,
for building the underground laboratory. We are grateful for the cooperation
from the China Guangdong Nuclear Power Group and China Light $\&$
Power Company.

\bibliography{DYBref}

\end{document}

%% file: name_JHEP.tex
\author[1]{F.~P.~An,}
\author[1]{W.~D.~Bai,}
\author[2]{A.~B.~Balantekin,}
\author[3]{M.~Bishai,}
\author[4]{S.~Blyth,}
\author[5]{G.~F.~Cao,}
\author[5,6]{J.~Cao,}
\author[5]{J.~F.~Chang,}
\author[7]{Y.~Chang,}
\author[5]{H.~S.~Chen,}
% ...
\affiliation[1]{Sun Yat-Sen (Zhongshan) University, Guangzhou}
\affiliation[2]{University~of~Wisconsin, Madison, Wisconsin 53706}
\affiliation[3]{Brookhaven~National~Laboratory, Upton, New York 11973}
\affiliation[4]{Department of Physics, National~Taiwan~University, Taipei}
\affiliation[5]{Institute~of~High~Energy~Physics, Beijing}
\affiliation[6]{New Cornerstone Science Laboratory, Institute of High Energy Physics, Beijing}
\affiliation[7]{National~United~University, Miao-Li}
% ...

%% file: CC_NSI_at_Daya_Bay.bbl
\providecommand{\href}[2]{#2}\begingroup\raggedright\begin{thebibliography}{10}

\bibitem{Workman:2022ynf}
{\scshape Particle Data Group} collaboration, \emph{{Review of Particle
  Physics}}, \href{https://doi.org/10.1093/ptep/ptac097}{\emph{PTEP} {\bfseries
  2022} (2022) 083C01}.

\bibitem{Proceedings:2019qno}
\emph{{Neutrino Non-Standard Interactions: A Status Report}}, vol.~2, 2019.
\newblock 10.21468/SciPostPhysProc.2.001.

\bibitem{PhysRevD.17.2369}
L.~Wolfenstein, \emph{Neutrino oscillations in matter},
  \href{https://doi.org/10.1103/PhysRevD.17.2369}{\emph{Phys. Rev. D}
  {\bfseries 17} (1978) 2369}.

\bibitem{Guzzo:1991hi}
M.M.~Guzzo, A.~Masiero and S.T.~Petcov, \emph{{On the MSW effect with massless
  neutrinos and no mixing in the vacuum}},
  \href{https://doi.org/10.1016/0370-2693(91)90984-X}{\emph{Phys. Lett. B}
  {\bfseries 260} (1991) 154}.

\bibitem{Biggio:2009nt}
C.~Biggio, M.~Blennow and E.~Fernandez-Martinez, \emph{{General bounds on
  non-standard neutrino interactions}},
  \href{https://doi.org/10.1088/1126-6708/2009/08/090}{\emph{JHEP} {\bfseries
  08} (2009) 090} [\href{https://arxiv.org/abs/0907.0097}{{\ttfamily
  0907.0097}}].

\bibitem{ANTUSCH2009369}
S.~Antusch, J.~Baumann and E.~Fernández-Martínez, \emph{Non-standard neutrino
  interactions with matter from physics beyond the standard model},
  \href{https://doi.org/https://doi.org/10.1016/j.nuclphysb.2008.11.018}{\emph{Nuclear
  Physics B} {\bfseries 810} (2009) 369}.

\bibitem{Ohlsson:2012kf}
T.~Ohlsson, \emph{{Status of non-standard neutrino interactions}},
  \href{https://doi.org/10.1088/0034-4885/76/4/044201}{\emph{Rept. Prog. Phys.}
  {\bfseries 76} (2013) 044201}
  [\href{https://arxiv.org/abs/1209.2710}{{\ttfamily 1209.2710}}].

\bibitem{Miranda:2015dra}
O.G.~Miranda and H.~Nunokawa, \emph{{Non standard neutrino interactions:
  current status and future prospects}},
  \href{https://doi.org/10.1088/1367-2630/17/9/095002}{\emph{New J. Phys.}
  {\bfseries 17} (2015) 095002}
  [\href{https://arxiv.org/abs/1505.06254}{{\ttfamily 1505.06254}}].

\bibitem{Farzan:2017xzy}
Y.~Farzan and M.~Tortola, \emph{{Neutrino oscillations and Non-Standard
  Interactions}}, \href{https://doi.org/10.3389/fphy.2018.00010}{\emph{Front.
  in Phys.} {\bfseries 6} (2018) 10}
  [\href{https://arxiv.org/abs/1710.09360}{{\ttfamily 1710.09360}}].

\bibitem{Esteban:2018ppq}
I.~Esteban, M.C.~Gonzalez-Garcia, M.~Maltoni, I.~Martinez-Soler and J.~Salvado,
  \emph{{Updated constraints on non-standard interactions from global analysis
  of oscillation data}},
  \href{https://doi.org/10.1007/JHEP08(2018)180}{\emph{JHEP} {\bfseries 08}
  (2018) 180} [\href{https://arxiv.org/abs/1805.04530}{{\ttfamily
  1805.04530}}].

\bibitem{Kopp:2007ne}
J.~Kopp, M.~Lindner, T.~Ota and J.~Sato, \emph{{Non-standard neutrino
  interactions in reactor and superbeam experiments}},
  \href{https://doi.org/10.1103/PhysRevD.77.013007}{\emph{Phys. Rev. D}
  {\bfseries 77} (2008) 013007}
  [\href{https://arxiv.org/abs/0708.0152}{{\ttfamily 0708.0152}}].

\bibitem{Davidson:2003ha}
S.~Davidson, C.~Pena-Garay, N.~Rius and A.~Santamaria, \emph{{Present and
  future bounds on nonstandard neutrino interactions}},
  \href{https://doi.org/10.1088/1126-6708/2003/03/011}{\emph{JHEP} {\bfseries
  03} (2003) 011} [\href{https://arxiv.org/abs/hep-ph/0302093}{{\ttfamily
  hep-ph/0302093}}].

\bibitem{Falkowski:2019xoe}
A.~Falkowski, M.~Gonz\'alez-Alonso and Z.~Tabrizi, \emph{{Reactor neutrino
  oscillations as constraints on Effective Field Theory}},
  \href{https://doi.org/10.1007/JHEP05(2019)173}{\emph{JHEP} {\bfseries 05}
  (2019) 173} [\href{https://arxiv.org/abs/1901.04553}{{\ttfamily
  1901.04553}}].

\bibitem{Fornengo:2001pm}
N.~Fornengo, M.~Maltoni, R.~Tomas and J.W.F.~Valle, \emph{{Probing neutrino
  nonstandard interactions with atmospheric neutrino data}},
  \href{https://doi.org/10.1103/PhysRevD.65.013010}{\emph{Phys. Rev. D}
  {\bfseries 65} (2002) 013010}
  [\href{https://arxiv.org/abs/hep-ph/0108043}{{\ttfamily hep-ph/0108043}}].

\bibitem{Li:2014mlo}
Y.-F.~Li and Y.-L.~Zhou, \emph{{Shifts of neutrino oscillation parameters in
  reactor antineutrino experiments with non-standard interactions}},
  \href{https://doi.org/10.1016/j.nuclphysb.2014.09.013}{\emph{Nucl. Phys. B}
  {\bfseries 888} (2014) 137}
  [\href{https://arxiv.org/abs/1408.6301}{{\ttfamily 1408.6301}}].

\bibitem{Liao:2016orc}
J.~Liao, D.~Marfatia and K.~Whisnant, \emph{{Nonstandard neutrino interactions
  at DUNE, T2HK and T2HKK}},
  \href{https://doi.org/10.1007/JHEP01(2017)071}{\emph{JHEP} {\bfseries 01}
  (2017) 071} [\href{https://arxiv.org/abs/1612.01443}{{\ttfamily
  1612.01443}}].

\bibitem{Liao:2017awz}
J.~Liao, D.~Marfatia and K.~Whisnant, \emph{{Nonstandard interactions in solar
  neutrino oscillations with Hyper-Kamiokande and JUNO}},
  \href{https://doi.org/10.1016/j.physletb.2017.05.054}{\emph{Phys. Lett. B}
  {\bfseries 771} (2017) 247}
  [\href{https://arxiv.org/abs/1704.04711}{{\ttfamily 1704.04711}}].

\bibitem{ANTARES:2021crm}
{\scshape ANTARES} collaboration, \emph{{Search for non-standard neutrino
  interactions with 10 years of ANTARES data}},
  \href{https://doi.org/10.1007/JHEP07(2022)048}{\emph{JHEP} {\bfseries 07}
  (2022) 048} [\href{https://arxiv.org/abs/2112.14517}{{\ttfamily
  2112.14517}}].

\bibitem{IceCube:2022ubv}
{\scshape IceCube} collaboration, \emph{{Strong Constraints on Neutrino
  Nonstandard Interactions from TeV-Scale $\nu_u$ Disappearance at IceCube}},
  \href{https://doi.org/10.1103/PhysRevLett.129.011804}{\emph{Phys. Rev. Lett.}
  {\bfseries 129} (2022) 011804}
  [\href{https://arxiv.org/abs/2201.03566}{{\ttfamily 2201.03566}}].

\bibitem{Leitner:2011aa}
R.~Leitner, M.~Malinsky, B.~Roskovec and H.~Zhang, \emph{{Non-standard
  antineutrino interactions at Daya Bay}},
  \href{https://doi.org/10.1007/JHEP12(2011)001}{\emph{JHEP} {\bfseries 12}
  (2011) 001} [\href{https://arxiv.org/abs/1105.5580}{{\ttfamily 1105.5580}}].

\bibitem{Girardi:2014gna}
I.~Girardi and D.~Meloni, \emph{{Constraining new physics scenarios in neutrino
  oscillations from Daya Bay data}},
  \href{https://doi.org/10.1103/PhysRevD.90.073011}{\emph{Phys. Rev. D}
  {\bfseries 90} (2014) 073011}
  [\href{https://arxiv.org/abs/1403.5507}{{\ttfamily 1403.5507}}].

\bibitem{Du:2020dwr}
Y.~Du, H.-L.~Li, J.~Tang, S.~Vihonen and J.-H.~Yu, \emph{{Non-standard
  interactions in SMEFT confronted with terrestrial neutrino experiments}},
  \href{https://doi.org/10.1007/JHEP03(2021)019}{\emph{JHEP} {\bfseries 03}
  (2021) 019} [\href{https://arxiv.org/abs/2011.14292}{{\ttfamily
  2011.14292}}].

\bibitem{Falkowski:2019kfn}
A.~Falkowski, M.~Gonz\'alez-Alonso and Z.~Tabrizi, \emph{{Consistent QFT
  description of non-standard neutrino interactions}},
  \href{https://doi.org/10.1007/JHEP11(2020)048}{\emph{JHEP} {\bfseries 11}
  (2020) 048} [\href{https://arxiv.org/abs/1910.02971}{{\ttfamily
  1910.02971}}].

\bibitem{Pontecorvo:1957cp}
B.~Pontecorvo, \emph{{Mesonium and anti-mesonium}}, {\emph{Sov. Phys. JETP}
  {\bfseries 6} (1957) 429}.

\bibitem{Pontecorvo:1957qd}
B.~Pontecorvo, \emph{{Inverse beta processes and nonconservation of lepton
  charge}}, {\emph{Zh. Eksp. Teor. Fiz.} {\bfseries 34} (1957) 247}.

\bibitem{Maki:1960ut}
Z.~Maki, M.~Nakagawa, Y.~Ohnuki and S.~Sakata, \emph{{A unified model for
  elementary particles}},
  \href{https://doi.org/10.1143/PTP.23.1174}{\emph{Prog. Theor. Phys.}
  {\bfseries 23} (1960) 1174}.

\bibitem{Maki:1962mu}
Z.~Maki, M.~Nakagawa and S.~Sakata, \emph{{Remarks on the unified model of
  elementary particles}}, \href{https://doi.org/10.1143/PTP.28.870}{\emph{Prog.
  Theor. Phys.} {\bfseries 28} (1962) 870}.

\bibitem{Pontecorvo:1967fh}
B.~Pontecorvo, \emph{{Neutrino Experiments and the Problem of Conservation of
  Leptonic Charge}}, {\emph{Zh. Eksp. Teor. Fiz.} {\bfseries 53} (1967) 1717}.

\bibitem{Grossman:1995wx}
Y.~Grossman, \emph{{Nonstandard neutrino interactions and neutrino oscillation
  experiments}},
  \href{https://doi.org/10.1016/0370-2693(95)01069-3}{\emph{Phys. Lett. B}
  {\bfseries 359} (1995) 141}
  [\href{https://arxiv.org/abs/hep-ph/9507344}{{\ttfamily hep-ph/9507344}}].

\bibitem{Gonzalez-Garcia:2001snt}
M.C.~Gonzalez-Garcia, Y.~Grossman, A.~Gusso and Y.~Nir, \emph{{New CP violation
  in neutrino oscillations}},
  \href{https://doi.org/10.1103/PhysRevD.64.096006}{\emph{Phys. Rev. D}
  {\bfseries 64} (2001) 096006}
  [\href{https://arxiv.org/abs/hep-ph/0105159}{{\ttfamily hep-ph/0105159}}].

\bibitem{Ohlsson:2008gx}
T.~Ohlsson and H.~Zhang, \emph{{Non-Standard Interaction Effects at Reactor
  Neutrino Experiments}},
  \href{https://doi.org/10.1016/j.physletb.2008.12.005}{\emph{Phys. Lett. B}
  {\bfseries 671} (2009) 99} [\href{https://arxiv.org/abs/0809.4835}{{\ttfamily
  0809.4835}}].

\bibitem{Meloni:2009cg}
D.~Meloni, T.~Ohlsson, W.~Winter and H.~Zhang, \emph{{Non-standard interactions
  versus non-unitary lepton flavor mixing at a neutrino factory}},
  \href{https://doi.org/10.1007/JHEP04(2010)041}{\emph{JHEP} {\bfseries 04}
  (2010) 041} [\href{https://arxiv.org/abs/0912.2735}{{\ttfamily 0912.2735}}].

\bibitem{Ohlsson:2013nna}
T.~Ohlsson, H.~Zhang and S.~Zhou, \emph{{Nonstandard interaction effects on
  neutrino parameters at medium-baseline reactor antineutrino experiments}},
  \href{https://doi.org/10.1016/j.physletb.2013.11.052}{\emph{Phys. Lett. B}
  {\bfseries 728} (2014) 148}
  [\href{https://arxiv.org/abs/1310.5917}{{\ttfamily 1310.5917}}].

\bibitem{Agarwalla:2014bsa}
S.K.~Agarwalla, P.~Bagchi, D.V.~Forero and M.~T\'ortola, \emph{{Probing
  Non-Standard Interactions at Daya Bay}},
  \href{https://doi.org/10.1007/JHEP07(2015)060}{\emph{JHEP} {\bfseries 07}
  (2015) 060} [\href{https://arxiv.org/abs/1412.1064}{{\ttfamily 1412.1064}}].

\bibitem{Langacker:1988up}
P.~Langacker and D.~London, \emph{{Lepton Number Violation and Massless
  Nonorthogonal Neutrinos}},
  \href{https://doi.org/10.1103/PhysRevD.38.907}{\emph{Phys. Rev. D} {\bfseries
  38} (1988) 907}.

\bibitem{Antusch:2006vwa}
S.~Antusch, C.~Biggio, E.~Fernandez-Martinez, M.B.~Gavela and J.~Lopez-Pavon,
  \emph{{Unitarity of the Leptonic Mixing Matrix}},
  \href{https://doi.org/10.1088/1126-6708/2006/10/084}{\emph{JHEP} {\bfseries
  10} (2006) 084} [\href{https://arxiv.org/abs/hep-ph/0607020}{{\ttfamily
  hep-ph/0607020}}].

\bibitem{T2K:2019bcf}
{\scshape T2K} collaboration, \emph{{Constraint on the
  matter\textendash{}antimatter symmetry-violating phase in neutrino
  oscillations}},
  \href{https://doi.org/10.1038/s41586-020-2177-0}{\emph{Nature} {\bfseries
  580} (2020) 339} [\href{https://arxiv.org/abs/1910.03887}{{\ttfamily
  1910.03887}}].

\bibitem{Buchmuller:1985jz}
W.~Buchmuller and D.~Wyler, \emph{{Effective Lagrangian Analysis of New
  Interactions and Flavor Conservation}},
  \href{https://doi.org/10.1016/0550-3213(86)90262-2}{\emph{Nucl. Phys. B}
  {\bfseries 268} (1986) 621}.

\bibitem{Grzadkowski:2010es}
B.~Grzadkowski, M.~Iskrzynski, M.~Misiak and J.~Rosiek, \emph{{Dimension-Six
  Terms in the Standard Model Lagrangian}},
  \href{https://doi.org/10.1007/JHEP10(2010)085}{\emph{JHEP} {\bfseries 10}
  (2010) 085} [\href{https://arxiv.org/abs/1008.4884}{{\ttfamily 1008.4884}}].

\bibitem{Chaves:2021kxe}
M.E.~Chaves, P.C.~de~Holanda and O.L.G.~Peres, \emph{{Testing non-standard
  neutrino interactions in (anti)-electron neutrino disappearance
  experiments}}, \href{https://doi.org/10.1007/JHEP03(2023)180}{\emph{JHEP}
  {\bfseries 03} (2023) 180}
  [\href{https://arxiv.org/abs/2106.15725}{{\ttfamily 2106.15725}}].

\bibitem{Gonzalez-Alonso:2018omy}
M.~Gonz\'alez-Alonso, O.~Naviliat-Cuncic and N.~Severijns, \emph{{New physics
  searches in nuclear and neutron $\beta$ decay}},
  \href{https://doi.org/10.1016/j.ppnp.2018.08.002}{\emph{Prog. Part. Nucl.
  Phys.} {\bfseries 104} (2019) 165}
  [\href{https://arxiv.org/abs/1803.08732}{{\ttfamily 1803.08732}}].

\bibitem{YEH2007329}
M.~Yeh, A.~Garnov and R.~Hahn, \emph{Gadolinium-loaded liquid scintillator for
  high-precision measurements of antineutrino oscillations and the mixing
  angle, θ13},
  \href{https://doi.org/https://doi.org/10.1016/j.nima.2007.03.029}{\emph{Nuclear
  Instruments and Methods in Physics Research Section A: Accelerators,
  Spectrometers, Detectors and Associated Equipment} {\bfseries 578} (2007)
  329}.

\bibitem{DING2008238}
Y.~Ding, Z.~Zhang, J.~Liu, Z.~Wang, P.~Zhou and Y.~Zhao, \emph{A new
  gadolinium-loaded liquid scintillator for reactor neutrino detection},
  \href{https://doi.org/https://doi.org/10.1016/j.nima.2007.09.044}{\emph{Nuclear
  Instruments and Methods in Physics Research Section A: Accelerators,
  Spectrometers, Detectors and Associated Equipment} {\bfseries 584} (2008)
  238}.

\bibitem{BERIGUETE201482}
W.~Beriguete, J.~Cao, Y.~Ding, S.~Hans, K.M.~Heeger, L.~Hu et~al.,
  \emph{Production of a gadolinium-loaded liquid scintillator for the daya bay
  reactor neutrino experiment},
  \href{https://doi.org/https://doi.org/10.1016/j.nima.2014.05.119}{\emph{Nuclear
  Instruments and Methods in Physics Research Section A: Accelerators,
  Spectrometers, Detectors and Associated Equipment} {\bfseries 763} (2014)
  82}.

\bibitem{DayaBay:2014cmr}
{\scshape Daya Bay} collaboration, \emph{{The muon system of the Daya Bay
  Reactor antineutrino experiment}},
  \href{https://doi.org/10.1016/j.nima.2014.09.070}{\emph{Nucl. Instrum. Meth.
  A} {\bfseries 773} (2015) 8}
  [\href{https://arxiv.org/abs/1407.0275}{{\ttfamily 1407.0275}}].

\bibitem{DayaBay:2015kir}
{\scshape Daya Bay} collaboration, \emph{{The Detector System of The Daya Bay
  Reactor Neutrino Experiment}},
  \href{https://doi.org/10.1016/j.nima.2015.11.144}{\emph{Nucl. Instrum. Meth.
  A} {\bfseries 811} (2016) 133}
  [\href{https://arxiv.org/abs/1508.03943}{{\ttfamily 1508.03943}}].

\bibitem{DayaBay:2022orm}
{\scshape Daya Bay} collaboration, \emph{{Precision Measurement of Reactor
  Antineutrino Oscillation at Kilometer-Scale Baselines by Daya Bay}},
  \href{https://doi.org/10.1103/PhysRevLett.130.161802}{\emph{Phys. Rev. Lett.}
  {\bfseries 130} (2023) 161802}
  [\href{https://arxiv.org/abs/2211.14988}{{\ttfamily 2211.14988}}].

\bibitem{DayaBay:2016ggj}
{\scshape Daya Bay} collaboration, \emph{{Measurement of electron antineutrino
  oscillation based on 1230 days of operation of the Daya Bay experiment}},
  \href{https://doi.org/10.1103/PhysRevD.95.072006}{\emph{Phys. Rev. D}
  {\bfseries 95} (2017) 072006}
  [\href{https://arxiv.org/abs/1610.04802}{{\ttfamily 1610.04802}}].

\bibitem{PhysRevC.84.024617}
P.~Huber, \emph{Determination of antineutrino spectra from nuclear reactors},
  \href{https://doi.org/10.1103/PhysRevC.84.024617}{\emph{Phys. Rev. C}
  {\bfseries 84} (2011) 024617}.

\bibitem{PhysRevC.83.054615}
T.A.~Mueller, D.~Lhuillier, M.~Fallot, A.~Letourneau, S.~Cormon, M.~Fechner
  et~al., \emph{Improved predictions of reactor antineutrino spectra},
  \href{https://doi.org/10.1103/PhysRevC.83.054615}{\emph{Phys. Rev. C}
  {\bfseries 83} (2011) 054615}.

\bibitem{Hayes:2013wra}
A.C.~Hayes, J.L.~Friar, G.T.~Garvey, G.~Jungman and G.~Jonkmans,
  \emph{{Systematic Uncertainties in the Analysis of the Reactor Neutrino
  Anomaly}}, \href{https://doi.org/10.1103/PhysRevLett.112.202501}{\emph{Phys.
  Rev. Lett.} {\bfseries 112} (2014) 202501}
  [\href{https://arxiv.org/abs/1309.4146}{{\ttfamily 1309.4146}}].

\bibitem{Hayes:2015ctl}
A.C.~Hayes, \emph{{Uncertainties in Reactor Neutrino Fluxes and in the
  Anomaly}},  in \emph{{50th Rencontres de Moriond on EW Interactions and
  Unified Theories}}, pp.~241--248, 2015.

\bibitem{VanegasForero:2019mqo}
D.~Vanegas~Forero, \emph{{Standard and non-standard neutrino physics at reactor
  experiments}}, \href{https://doi.org/10.22323/1.341.0148}{\emph{PoS}
  {\bfseries NuFACT2018} (2019) 148}.

\bibitem{Esteban:2020cvm}
I.~Esteban, M.C.~Gonzalez-Garcia, M.~Maltoni, T.~Schwetz and A.~Zhou,
  \emph{{The fate of hints: updated global analysis of three-flavor neutrino
  oscillations}}, \href{https://doi.org/10.1007/JHEP09(2020)178}{\emph{JHEP}
  {\bfseries 09} (2020) 178}
  [\href{https://arxiv.org/abs/2007.14792}{{\ttfamily 2007.14792}}].

\end{thebibliography}\endgroup
